\documentclass[12pt]{iopart}

\usepackage{iopams}
\usepackage{graphicx}
\usepackage{bm}
\usepackage{amssymb}
\usepackage{color}
\usepackage[breaklinks=true,colorlinks=true,linkcolor=blue,urlcolor=blue,citecolor=blue]{hyperref}

\newcommand{\discard}[1]{{}}

\newcommand{\suppl}{\url{https://arxiv.org/e-print/2003.10990}}

\newcommand{\tfrac}[2]{{\textstyle{\frac{#1}{#2}}}}

\newcommand{\bfr}{\mathbf r}

\newcommand{\bfk}{\mathbf k}
\newcommand{\bfnab}{\boldsymbol\nabla}
\newcommand{\bfsig}{\boldsymbol\sigma}

\newcommand{\text}{\rm}

\begin{document}

\title[Properties of spherical and deformed nuclei\ldots]
{Properties of spherical and deformed nuclei using regularized pseudopotentials in nuclear DFT}

\author{K~Bennaceur$^1$, J~Dobaczewski$^{2,3,4}$, T~Haverinen$^{4,5}$, and M~Kortelainen$^{5,4}$}
\address{$^1$Univ Lyon, Universit\'e Claude Bernard Lyon 1, CNRS, IPNL, UMR 5822,
4 rue E. Fermi, F-69622 Villeurbanne Cedex, France}
\address{$^2$Department of Physics, University of York, Heslington, York YO10 5DD, United Kingdom}
\address{$^3$Institute of Theoretical Physics, Faculty of Physics, University of Warsaw,
Pasteura 5, 02-093 Warszawa, Poland}
\address{$^4$Helsinki Institute of Physics, P.O. Box 64, 00014 University of Helsinki, Finland}
\address{$^5$Department of Physics, University of Jyv\"askyl\"a,
P.O. Box 35 (YFL), 40014 University of Jyv\"askyl\"a, Finland}

\begin{abstract}
We developed new parameterizations of local regularized finite-range
pseudopotentials up to next-to-next-to-next-to-leading order
(N$^3$LO), used as generators of nuclear density functionals. When
supplemented with zero-range spin-orbit and density-dependent terms,
they provide a correct single-reference description of binding
energies and radii of spherical and deformed nuclei. We compared the
obtained results to experimental data and discussed benchmarks
against the standard well-established Gogny D1S functional.
\end{abstract}

\submitto{\JPG}

\section{Introduction}

The nuclear density functional theory (DFT) offers one of the most flexible frameworks
to microscopically describe structure of atomic nuclei~\cite{(Ben03e),(Sch19)}.
A key element in the nuclear DFT is the energy density functional (EDF), which is
usually obtained by employing effective forces as its generators.
A long-standing goal of nuclear DFT is to construct an EDF with high precision
of describing existing data and high predictive power.

The Skyrme and Gogny EDFs~\cite{(Ben03e),(Rob18)} are the most utilized non-relativistic EDFs
in nuclear structure calculations.
The Skyrme EDF is based on a zero-range generator, combined with a momentum expansion up to second order,
whereas the Gogny EDF is based on the generator constructed with two Gaussian terms.
While zero-range potentials are computationally simpler and less demanding, they lack in flexibility
of their exchange terms. In addition, in the pairing channel they
manifest the well-known problem of nonconvergent pairing energy, which needs to be regularized,
see Refs.~\cite{(Bul02a),(Bor06a)} and references cited therein.
While Skyrme-type EDFs can reproduce various nuclear bulk properties relatively well, their
limits have been reached~\cite{(Kor14a)}, and proposed extensions of
zero-range generators~\cite{(Car08a),(Dav13a)} did not prove
efficient enough~\cite{(Szp16)}. New approaches are, therefore,
required.

To improve present EDFs, a possible route
is to use EDFs based on regularized finite-range pseudopotentials~\cite{(Dob12b)}.
Such EDFs stem from a momentum expansion around a finite-range regulator and thus have a form compatible with
powerful effective-theory methods~\cite{(Cas02),(Lep97a)}. Here, as well as in our
earlier studies~\cite{(Rai14a),(Ben17a)}, we chose a Gaussian regulator, which
offers numerically simple treatment, particularly when combined with the harmonic oscillator basis.
The momentum expansion can be built order-by-order, resulting in an EDF with increasing
precision. Due to its finite-range nature, treatment of the pairing channel does not require any
particular regularization or renormalization.

The ultimate goal of building EDFs based on regularized finite-range pseudopotentials is to apply them to
beyond mean-field multi-reference calculations.
However, before that, to evaluate expected performance and detect possible pitfalls, their
predictive power should be benchmarked at the single-reference level.
The goal of this work is to adjust the single-reference parameters of pseudopotentials up to
next-to-next-to-next-to-leading order (N$^3$LO) and to compare the obtained results
to experimental data on the one hand and to those obtained for the Gogny
D1S EDF~\cite{(Ber91d)} on the other. The D1S EDF offers an excellent reference to compare to, because it
contains finite-range terms of a similar nature, although its possible extensions to more than two
Gaussians~\cite{(Dav17)},
cannot be cast in the form of an effective-theory expansion.
Because EDFs adjusted in present work are intended to be used solely at the single-reference
level, they include a density-dependent term. This term significantly improves
infinite nuclear matter properties, with the drawback that such EDFs become unsuitable for
multi-reference calculations, see, e.g., Refs.~\cite{(Dob07),(She19b)}.

This article is organized as follows. In Sec.~\ref{Pseudopotential}, we briefly recall the formalism of
the regularized finite-range pseudopotential and in Sec.~\ref{sec:fit} we present details of
adjusting its parameters. Then, in Secs.~\ref{sec:res} and~\ref{sec:conclusion}, we present results
and conclusions of our study, respectively. In Appendices A--D, we give specific details of our approach and
in the supplemental material (\suppl) we collected files with numerical
results given in a machine readable format.

\section{Pseudopotential}
\label{Pseudopotential}

In this study, we use the local regularized pseudopotential with terms at $n$th order
introduced in~\cite{(Rai14a)},
\begin{eqnarray}
\label{eq:locpot}
&&{\cal V}^{(n)}_{\mathrm{loc}}(\bfr_{1},\bfr_{2};\bfr_{3},\bfr_{4})=\left(W^{(n)}_1\hat 1_\sigma\hat 1_\tau
+B^{(n)}_1\hat 1_\tau\hat P^\sigma
-H^{(n)}_1\hat 1_\sigma\hat P^\tau
-M^{(n)}_1\hat P^\sigma\hat P^\tau\right)  \nonumber \\
&&\hskip 4cm\times\delta(\bfr_{13})\delta(\bfr_{24})
\left(\frac{1}{2}\right)^{n/2}\bfk_{12}^n g_a(\bfr_{12})\,,
\end{eqnarray}
where the Gaussian regulator is defined as
\begin{equation}
g_a(\bfr)=\frac{1}{(a\sqrt{\pi})^3}\,\rme^{-\frac{\bfr^2}{a^2}}\,,
\label{eq:g}
\end{equation}
and $\hat 1_\sigma$ and $\hat 1_\tau$ are respectively the identity
operators in spin and isospin space and $\hat P^\sigma$ and $\hat
P^\tau$ the spin and isospin exchange operators. Standard
relative-momentum operators are defined as
$\bfk_{ij}=\tfrac{1}{2i}\left(\bfnab_{i}-\bfnab_{j}\right)$
and relative positions as $\bfr_{ij}=\bfr_{i}-\bfr_{j}$.

Up to the $n$th order (N$^n$LO), this pseudopotential depends on the following parameters,
\begin{itemize}
\item 8 parameters up to the next-to-leading-order (NLO):
$W_1^{(0)}$, $B_1^{(0)}$, $H_1^{(0)}$, $M_1^{(0)}$,
$W_1^{(2)}$, $B_1^{(2)}$, $H_1^{(2)}$ and $M_1^{(2)}$;
\item 4 additional parameters up to N$^2$LO:
$W_1^{(4)}$, $B_1^{(4)}$, $H_1^{(4)}$ and $M_1^{(4)}$;
\item and 4 additional parameters up to N$^3$LO:
$W_1^{(6)}$, $B_1^{(6)}$, $H_1^{(6)}$ and $M_1^{(6)}$.
\end{itemize}

In the present study, we determined coupling constants of
pseudopotentials that are meant to be used at the single-reference
level. Therefore, we complemented pseudopotentials (\ref{eq:locpot})
with standard zero-range spin-orbit and density-dependent terms,
\begin{equation}
{\cal V}_{\text{SO}}(\bfr_{1},\bfr_{2};\bfr_{3},\bfr_{4})
=iW_\mathrm{SO}\left(\bfsig_1+\bfsig_2\right)\cdot\left(\bfk_{12}^*\times\bfk_{34}\right)
\delta(\bfr_{13})\delta(\bfr_{24})\delta(\bfr_{12}) \,,
\label{eq:so}
\end{equation}
\begin{equation}
{\cal V}_{\text{DD}}(\bfr_{1},\bfr_{2};\bfr_{3},\bfr_{4})
=\frac{1}{6}\,t_3\left(1+\hat P_\sigma\right)\rho_0^{1/3}(\mathbf r_1)
\delta(\bfr_{13})\delta(\bfr_{24})\delta(\bfr_{12}) \,,
\label{eq:dd}
\end{equation}
which carry two additional parameters $W_\mathrm{SO}$ and $t_3$. The
density-dependent term, which has the same form as in the Gogny D1S
interaction~\cite{(Ber91d)}, represents a convenient way to adjust the
nucleon effective mass in infinite nuclear matter to any reasonable
value in the interval $0.70\lesssim
\frac{m^*}{m}\lesssim0.90$~\cite{(Dav18)}. This term contributes neither
to the binding of the neutron matter nor to nuclear pairing in
time-even systems. To avoid using a zero-range term in the pairing
channel, we neglect contribution of the spin-orbit term
to pairing.

Expressions giving the contributions to the EDFs of the local
regularized pseudopotential~(\ref{eq:locpot}) can be found in Ref.~\cite{(Ben17a)},
whereas those of the zero-range spin-orbit~(\ref{eq:so}) and
density-dependent term~(\ref{eq:dd}) can be found, for example,
in Refs.~\cite{(Eng75c),(Per04c)}.

\section{Adjustments of parameters}
\label{sec:fit}

As explained in Sec.~\ref{Pseudopotential}, pseudopotentials considered here
contain 10 parameters at NLO, 14 at N$^2$LO,
and 18 at N$^3$LO. In this study, we adjusted 15 series of parameters with effective masses
$m^*/m$ equal to 0.70, 0.75, 0.80, 0.85, and 0.90 at NLO, N$^2$LO, and N$^3$LO. For each series,
the range $a$ of the regulator was varied between 0.8 and 1.6\,fm.

Our previous experience shows that
the use of a penalty function only containing data on finite
nuclei is not sufficient to efficiently constrain parameters of pseudopotentials,
or even to constrain them at all. Typical reasons for these difficulties
are (i) appearance of finite-size instabilities, (ii) phase transitions to unphysical
states (for example those with very large vector pairing), or (iii) numerical problems
due to compensations of large coupling constants with opposite signs. To
avoid these unwanted situations, the penalty function must contain specially
designed empirical constrains. Before performing actual fits, such constrains cannot be easily defined;
therefore, to design the final penalty function, we went through
the steps summarized below.

\begin{itemize}
\item Step 1:

We made some preliminary fits so as
to detect possible pitfalls and devise ways to avoid them.
The main resulting observation was that it seems to be very difficult, if possible at
all, to adjust parameters leading to a value of the slope of the symmetry energy
coefficient $L$ in the range of the commonly accepted values, which is roughly between 40
and 70\,MeV~\cite{(Xu10),(Vin14),(Dan14)}.
Therefore, for all adjustments performed in this study, we set its value to $L=15$\,MeV.
This value is rather low, although it is at a similar lower side as those corresponding
to various Gogny parameterizations: $L=18.4$\,MeV for D1~\cite{(Dec80b)}, $L=22.4$\,MeV for
D1S~\cite{(Ber91d)}, $L=24.8$\,MeV for D1M~\cite{(Gor16a)}, or $L=43.2$\,MeV for
D1M*~\cite{(Gon18)}.

\item Step 2:

With the fixed value of $L=15$\,MeV, we performed a series of exploratory adjustments
with fixed values of other infinite-nuclear-matter properties, that is, for the
saturation density of $\rho_\mathrm{sat}=0.16\,\mathrm{fm}^{-3}$, binding
energy per nucleon in infinite symmetric matter of $E/A=-16$\,MeV, compression
modulus of $K_\infty=230$\,MeV, and symmetry energy coefficient of $J=32$\,MeV.
These initial values were the same as for
the Skyrme interactions of the SLy family~\cite{(Cha97a),(Cha98b)}.
The conclusion drawn from this step was that the favoured values for $\rho_\mathrm{sat}$ and $J$ were
slightly lower than the initial ones. Therefore, we decided to fix $\rho_\mathrm{sat}$
and $J$ at the results corresponding to pseudopotentials giving the lowest values
of the penalty function $\chi^2$, see Fig.~\ref{fig:chi2s} and Table~\ref{tab:inm}.

\begin{figure}[tbp]
\begin{center}
\begin{tabular}{ccc}
 \includegraphics[height=0.305\linewidth,angle=270,viewport=30 10 500 667,clip]%
        {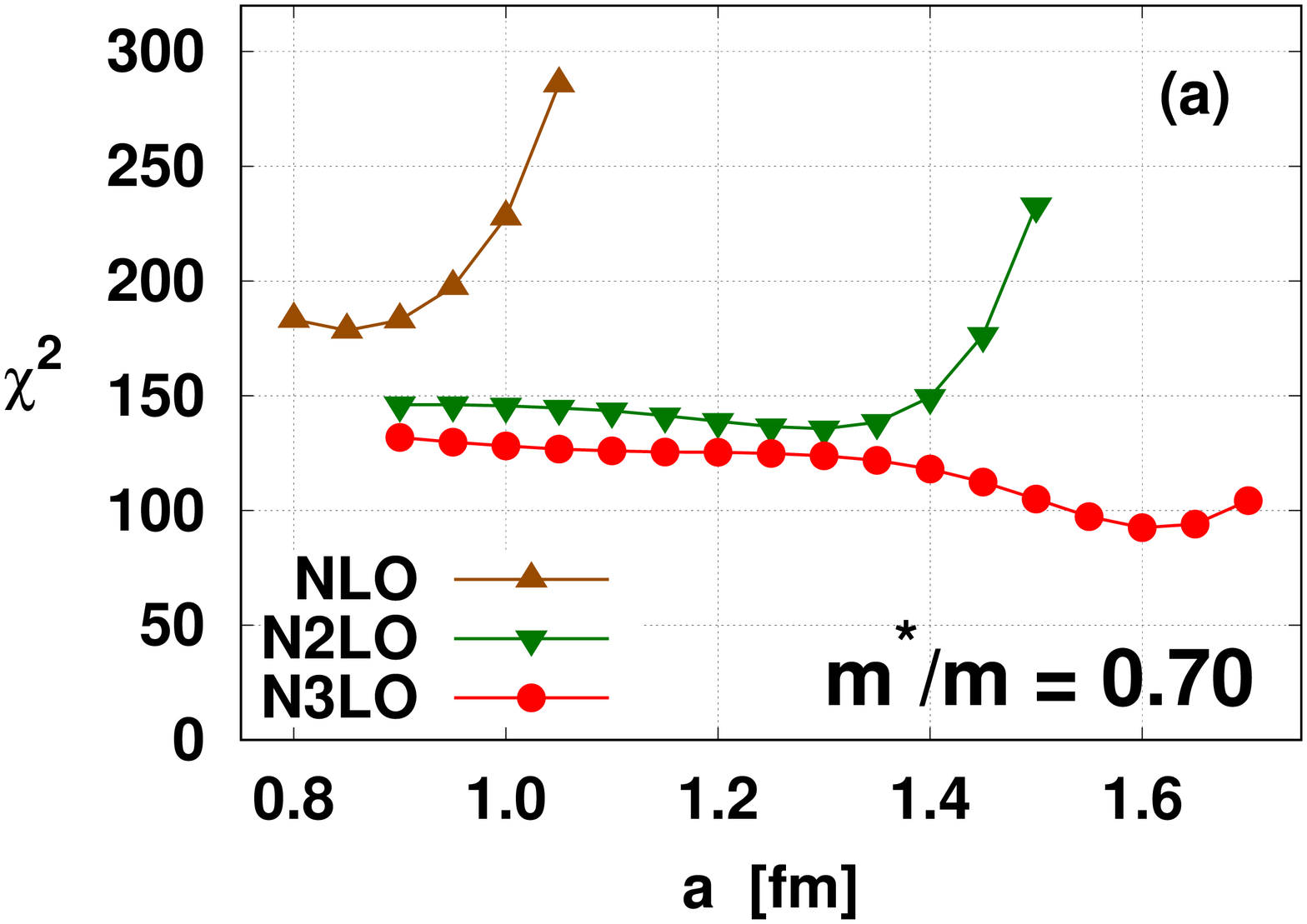}  &
 \includegraphics[height=0.305\linewidth,angle=270,viewport=30 10 500 667,clip]%
        {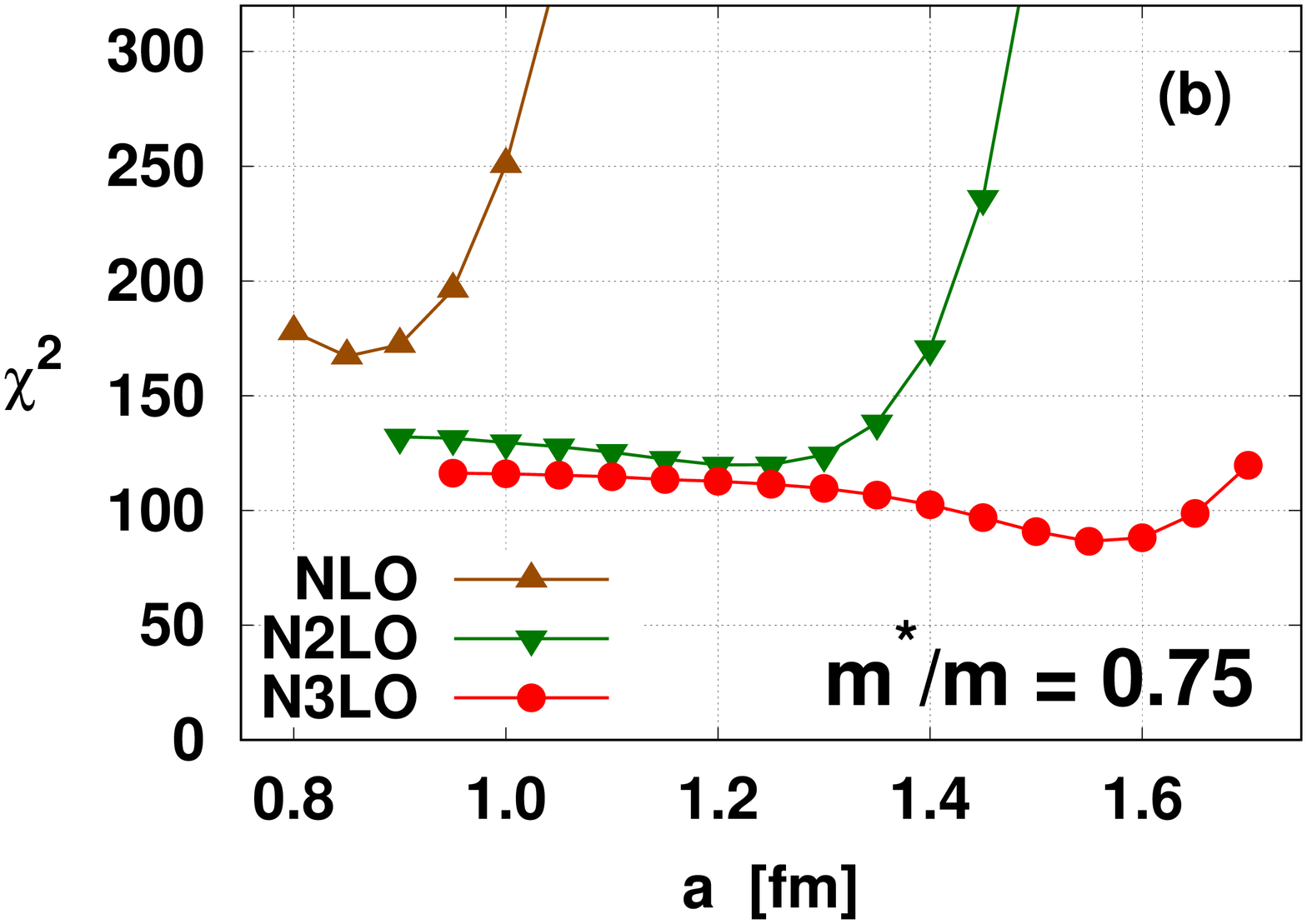}  &
 \includegraphics[height=0.305\linewidth,angle=270,viewport=30 10 500 667,clip]%
        {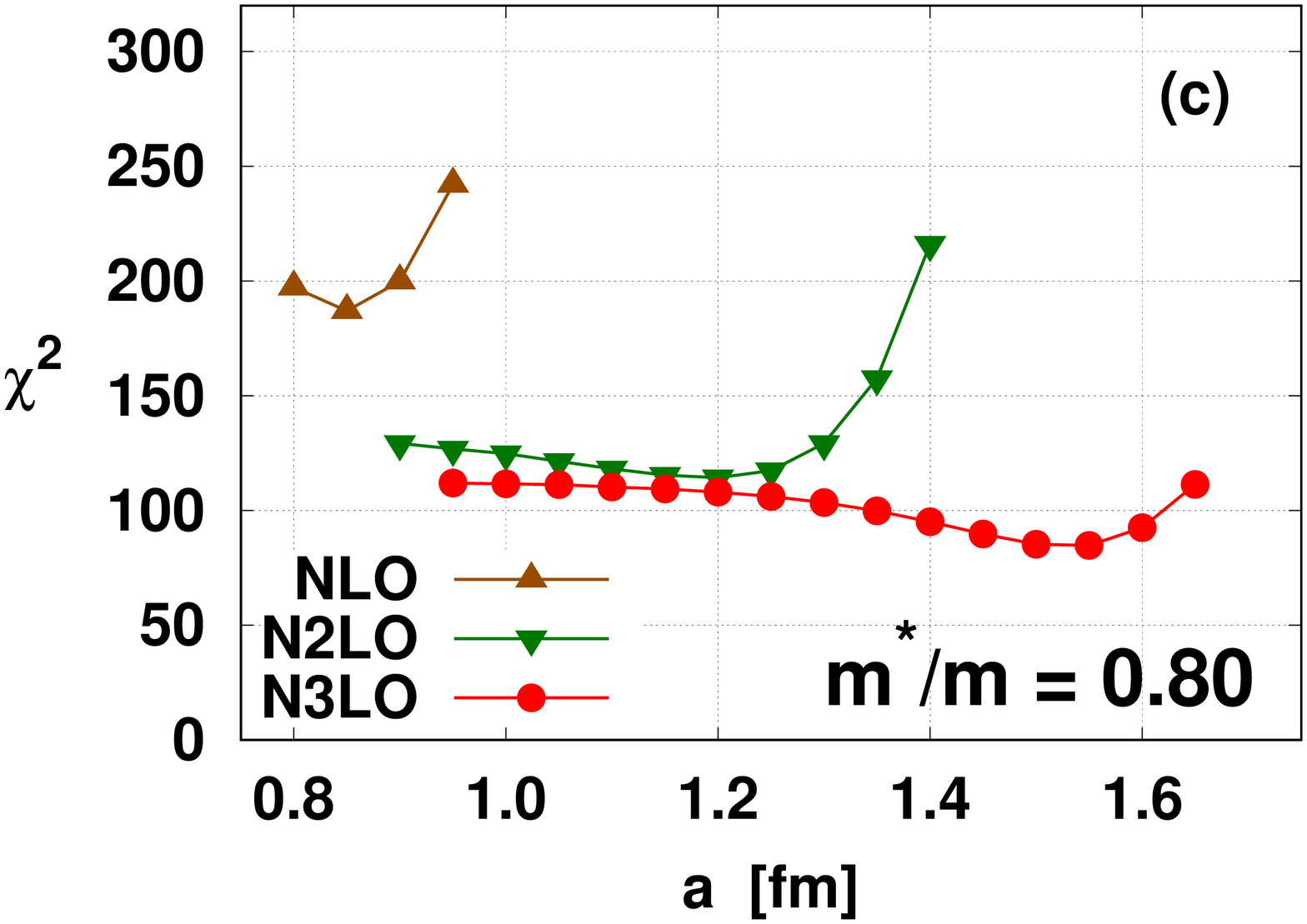}
\end{tabular}
\begin{tabular}{cc}
 \includegraphics[height=0.305\linewidth,angle=270,viewport=30 10 500 667,clip]%
        {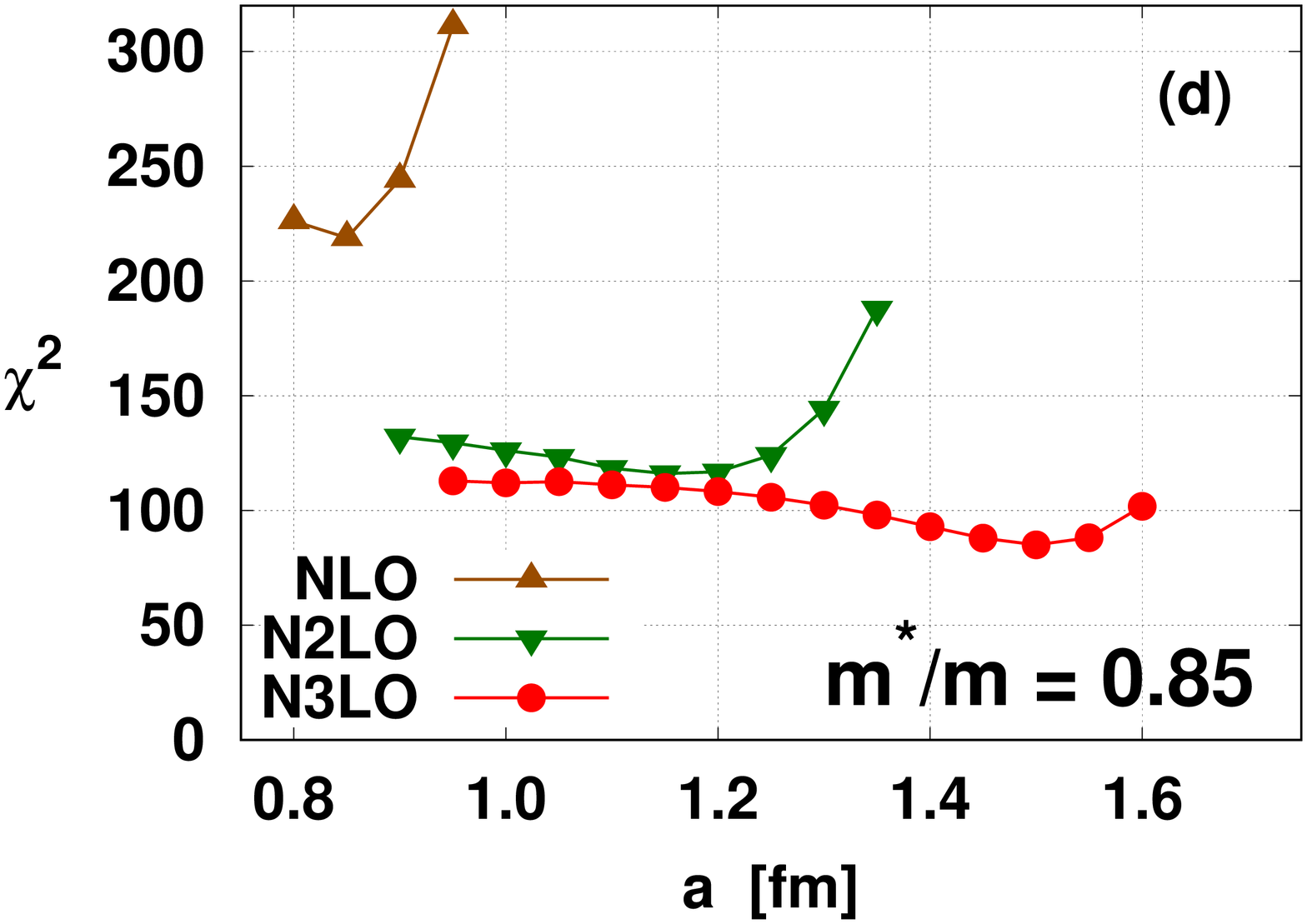}  &
 \includegraphics[height=0.305\linewidth,angle=270,viewport=30 10 500 667,clip]%
        {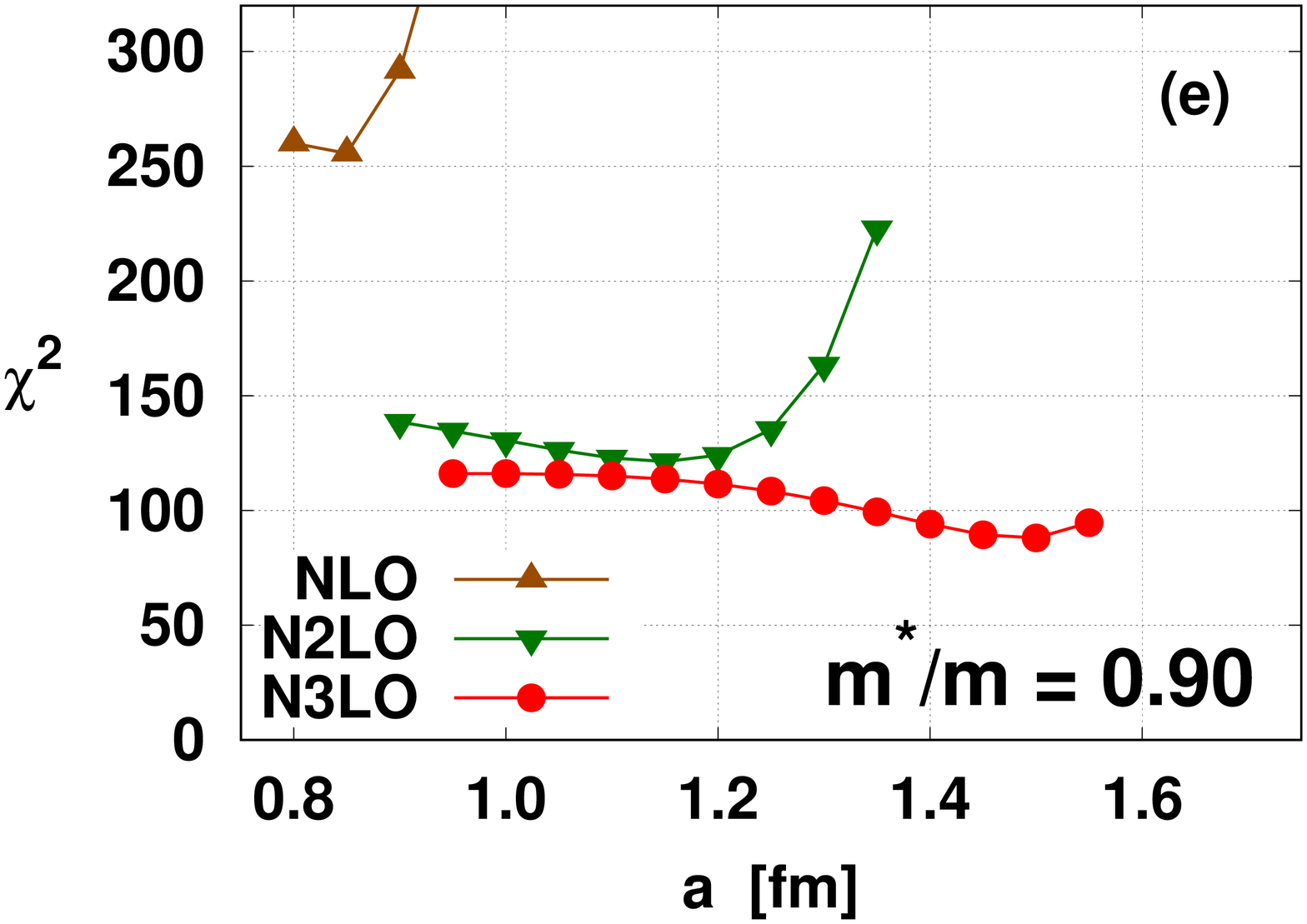}
\end{tabular}
\caption{Penalty functions $\chi^2$ obtained in step 2 of the adjustment (see text) as functions of the
regulator range $a$. Panels (a)-(e) correspond to the five values of the effective mass adopted in this study.
\label{fig:chi2s}}
\end{center}
\end{figure}

\item Step 3:

In a consistent effective theory, with increasing order of expansion,
the dependence of observables on the range $a$ of the regulator
should become weaker and weaker. In our previous
work~\cite{(Ben17a)}, where all terms of the pseudopotential
were regulated with the same range, such a behaviour was clearly
visible. In the present work, the regulated part of the
pseudopotential is combined with two zero-range terms. As a result,
even at N$^3$LO, there remains a significant dependence of the
penalty functions on $a$, see Fig.~\ref{fig:chi2s}. Therefore, in
step 3 we picked for further analyses the parameterizations of
pseudopotentials that correspond to the minimum values of penalty
functions.

Then, for each of the five values of the effective mass and for each
of the three orders of expansion, we optimized the corresponding
parameters of the pseudopotential, but this time with the
infinite-matter properties not rigidly fixed but allowed to change
within small tolerance intervals, see Table~\ref{tab:inm}.

\end{itemize}

In the supplemental material (\suppl), the
corresponding 15 sets of parameters are listed in a machine readable
format. Following the naming convention adopted in
Ref.~\cite{(Ben14b)}, these final sets are named as REG$n$m.190617,
where $n$ stands for the order of the pseudopotential ($n=2$ at NLO,
$n=4$ at N$^2$LO, and $n=6$ at N$^3$LO), and $\mathrm{m}=\mathrm{a}$,
b, c, d, or e stands for one of the five adopted values of the
effective mass $m^*/m$ 0.70, 0.75, 0.80, 0.85, or 0.90, respectively.
For brevity, in the remaining of this paper, we omit the date of the
final adjustment, denoted by 190617, which otherwise is an inherent
part of the name.

\begin{table}
\caption{Target infinite nuclear matter  properties and the corresponding tolerance intervals used in
step 3 of the adjustment (see text).\label{tab:inm}}
\begin{center}
\begin{tabular}{lrrrrrr}
\br
Quantity  & $E/A$~[MeV] &  $\rho_\mathrm{sat}~[\mathrm{fm}^{-3}]$  & $K_\infty$~[MeV] & $ m^*/m$ & $J$~[MeV] & $L$~[MeV] \\
\mr
Value     & -16.0 & 0.158               & 230      & 0.70-0.90 & 29.0 & 15.00 \\
Tolerance & 0.3   & 0.003               &   5      & 0.001     & 0.5 & 0.05 \\
\br
\end{tabular}
\end{center}
\end{table}

We are now in a position to list all contributions to the penalty
function $\chi^2$, which come from the empirical constrains used in step 3 of the adjustment
and from those corresponding to the nuclear data and pseudo-data that we used.
\begin{enumerate}
\item \underline{Empirical properties of the symmetric infinite nuclear matter.}
These correspond to: saturation
density $\rho_\mathrm{sat}$, binding energy per nucleon $E/A$, compression
modulus $K_\infty$, isoscalar effective mass $m^*/m$, symmetry energy
coefficient $J$, and its slope $L$. The target values and the corresponding
tolerance intervals are listed in Table~\ref{tab:inm}.
\item \underline{Potential energies per nucleon in symmetric infinite
nuclear matter.} We used values in four spin-isospin channels $(S,T)$ determined in theoretical
calculations of Refs.~\cite{Bal97,baldo}.
Although it is not clear if these constraints have any significant impact
on the observables calculated in finite nuclei, we observed that they seem to
prevent the aforementioned numerical instabilities due to compensations
of large coupling constants with opposite signs. Explicit formulas
for the decomposition of the potential energy in the $(S,T)$ channels are
given in \ref{app:st}.
\item \underline{Energy per nucleon in infinite neutron matter.}
We used values calculated for
potentials UV14 plus UVII (see Table III in~\cite{(Wir88)}) at
densities below $0.4~\mathrm{fm}^{-3}$ with a tolerance interval of 25~\%.
\item \underline{Energy per nucleon in polarized infinite nuclear matter.}
Adjustment of parameters often leads to the appearance of a bound
state in symmetric polarized matter. To avoid this type of result, we used
the constraint of $E/A=12.52$\,MeV at density $0.1\,\mathrm{fm}^{-3}$ (taken
from Ref.~\cite{(Bor07)}) with a large tolerance interval of 25\,\%.
\item \underline{Average pairing gap in infinite nuclear matter.}
Our goal was to obtain
a reasonable profile for the average gap in symmetric infinite nuclear matter
and to avoid too frequent collapse of pairing for deformed minima (especially
for protons). Therefore, we used as targets the values calculated for the D1S functional at
$k_F=0.4$, $0.8$, and $1.2$\,fm$^{-1}$ with the tolerance intervals
of $0.1$\,MeV.
\item \underline{Binding energies of spherical nuclei.}
We used experimental values of the following 17 spherical (or approximated as
spherical) nuclei $^{36}$Ca, $^{40}$Ca, $^{48}$Ca, $^{54}$Ca,
$^{54}$Ni, $^{56}$Ni, $^{72}$Ni, $^{80}$Zr, $^{90}$Zr, $^{112}$Zr,
$^{100}$Sn, $^{132}$Sn, $^{138}$Sn, $^{178}$Pb,  $^{208}$Pb,  $^{214}$Pb,
and $^{216}$Th. We attributed tolerance intervals of 1\,MeV (2\,MeV) if the binding energy was known
experimentally (extrapolated)~\cite{(Wan17)}. The motivation for this list was to use open-shell nuclei
along with doubly magic ones, so as to better constrain distances between successive shells.
\item \underline{Proton rms radii.} We used values taken from Ref.~\cite{(Kor10c)}
for $^{40}$Ca, $^{48}$Ca, $^{208}$Pb, and $^{214}$Pb with the tolerance intervals of 0.02\,fm and
that for $^{56}$Ni (which is
extrapolated from systematics) with the tolerance interval of 0.03\,fm.
\item \underline{Isovector and isoscalar central densities.}
To avoid finite-size scalar-isovector ({\em i.e.} $S=0$,
$T=1$)
instabilities, we used isovector density at the center of $^{208}$Pb and isoscalar density at the
center of $^{40}$Ca. A use of the linear response methodology (such as in Ref.~\cite{(Hel13)}
for zero-range interactions) would lead to too much time-consuming calculations.
As a proxy, we used the two empirical constraints on central densities, which are
known to grow uncontrollably when the scalar-isovector instabilities develop.
We used the empirical values of $\rho_1(0)<0\,\mathrm{fm}^{-3}$ in $^{208}$Pb and
$\rho_0(0)<0.187\,\mathrm{fm}^{-3}$ in $^{40}$Ca with asymmetric tolerance intervals as
described in Ref.~\cite{(Ben17a)}. For $\rho_0(0)$ in $^{40}$Ca, we have used
the central density obtained with SLy5~\cite{(Cha98b)} as an upper limit.
In the parameter adjustments performed in this study, possible instabilities
in the vector channels ($S=1$) are still not under control.
\item \underline{Surface energy coefficient.}
As it was recently shown~\cite{(Jod16),PhysRevC.99.044315}, a constraint on
the surface energy coefficient is an efficient way to improve properties
of EDFs. For the regularized
pseudopotentials considered here, we calculate a simple estimate of the surface
energy coefficient using a liquid-drop type formula $a_\mathrm{surf}^\mathrm{LDM}$
with target value of 18.5\,MeV and the tolerance interval of 0.2\,MeV. The relevance of this constraint
and the motivation for the target value are discussed in \ref{app:asurf}.
\item \underline{Coupling constants corresponding to vector pairing.}
Terms of the EDF that correspond to this channel are given in Eq.~(36) of
Ref.~\cite{(Ben17a)}.
To avoid transitions to unphysical regions of unrealistically large vector pairing,
we constrain them to be equal to $0\pm 5\,\mathrm{MeV}\,\mathrm{fm}^3$.
\end{enumerate}

\section{Results and discussion}
\label{sec:res}

\subsection{Parameters and statistical uncertainties}
\label{sec:Parameters}

\begin{figure}[tbp]
\begin{center}
\begin{tabular}{cc}
\includegraphics[height=0.47\linewidth,angle=270,viewport=10 10 500 670,clip]{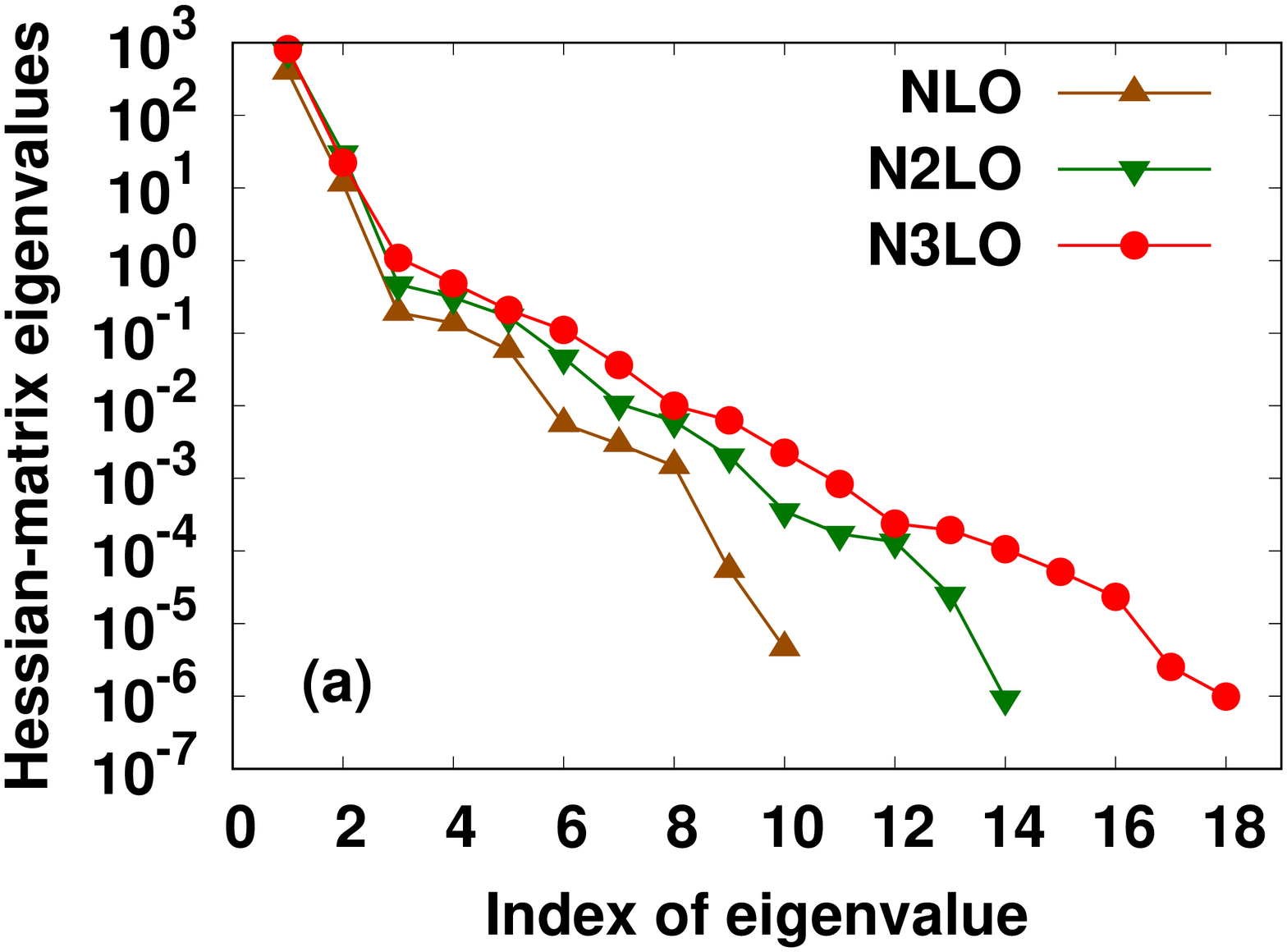}  &
\includegraphics[height=0.47\linewidth,angle=270,viewport=10 10 500 670,clip]{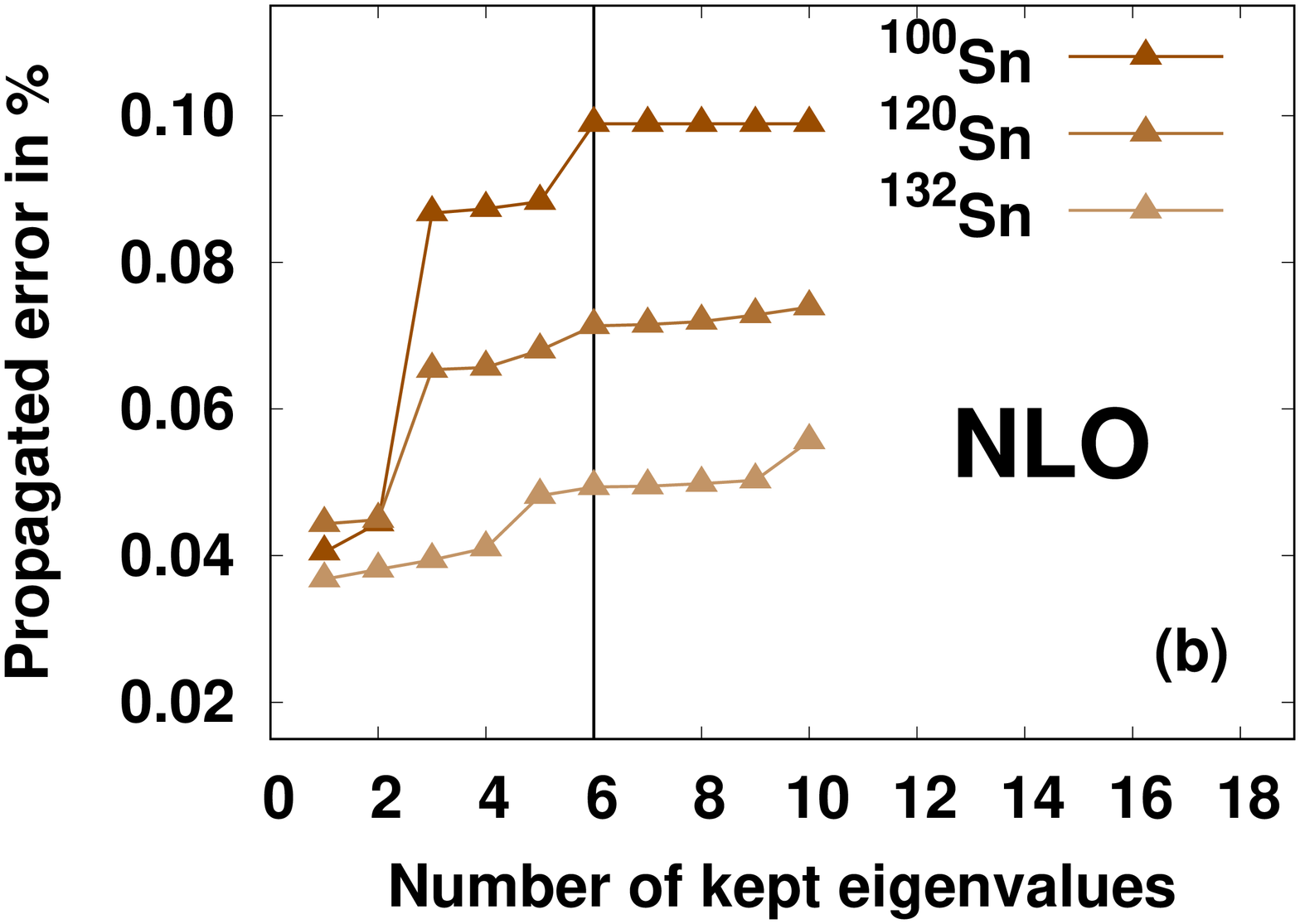} \\
\includegraphics[height=0.47\linewidth,angle=270,viewport=10 10 500 670,clip]{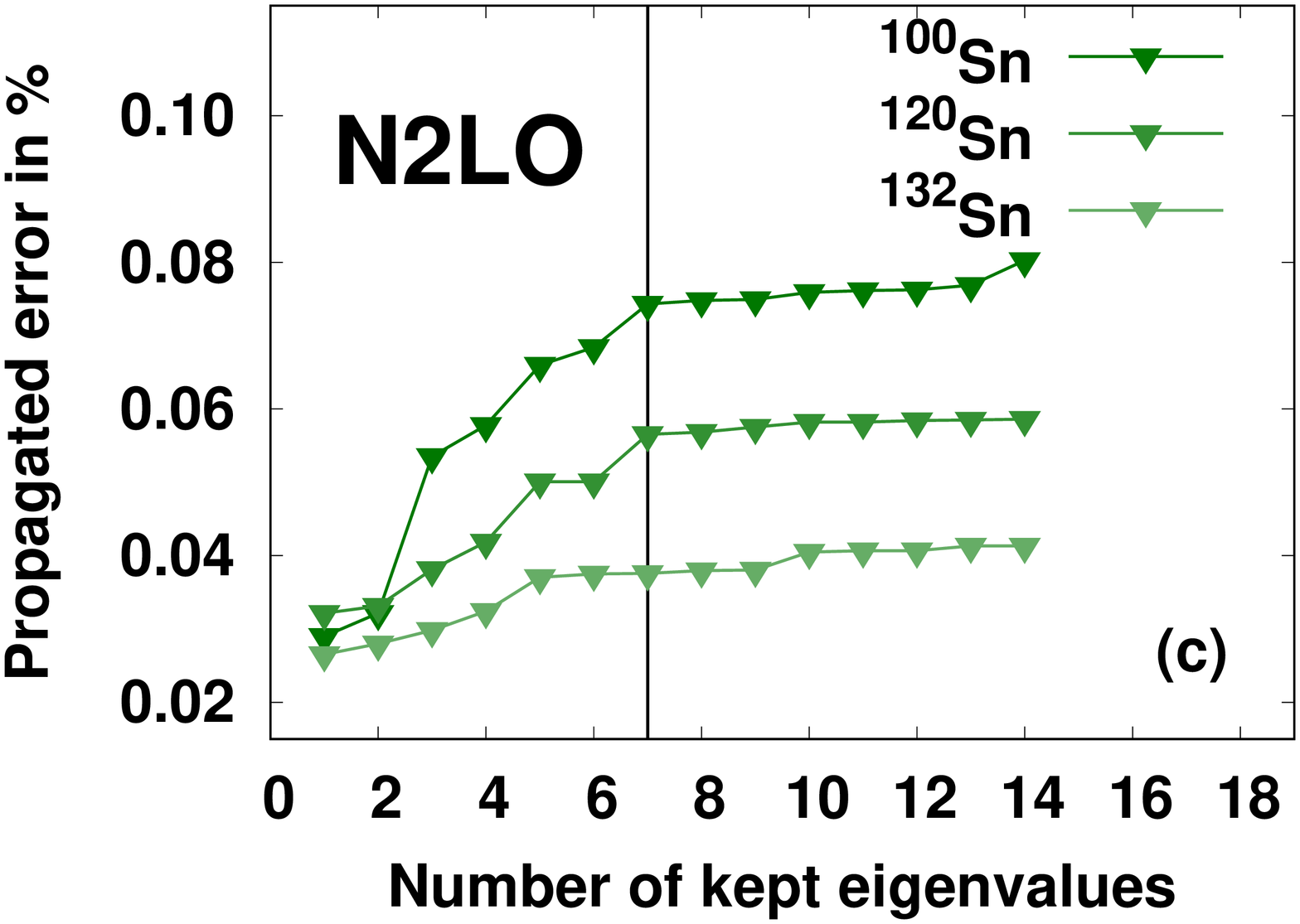} &
\includegraphics[height=0.47\linewidth,angle=270,viewport=10 10 500 670,clip]{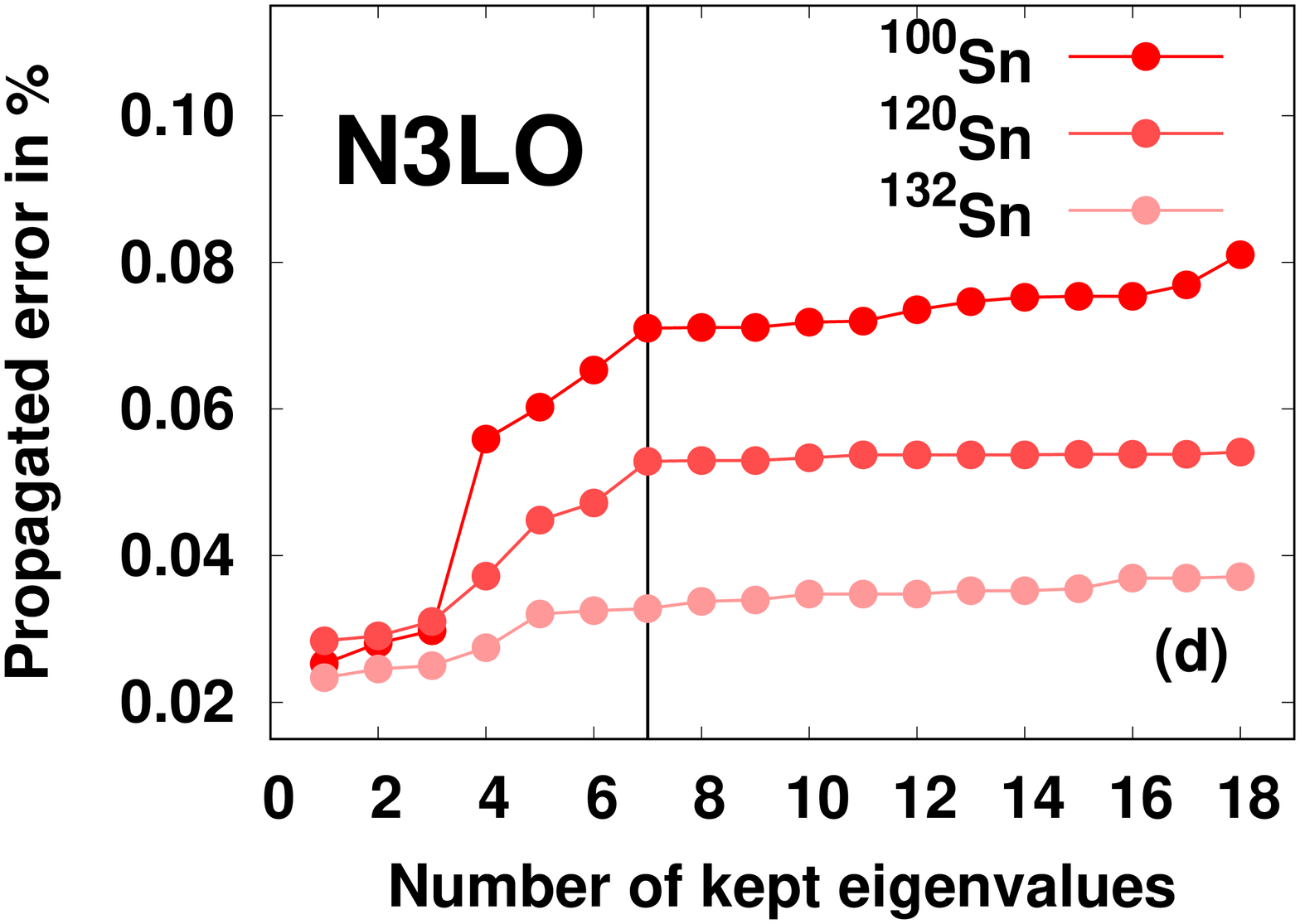}
\end{tabular}
\caption{Results of statistical analyses performed for
parameterizations corresponding to $m^*/m=0.85$. Eigenvalues of the
Hessian matrices (a) are compared with propagated uncertainties of
binding energies of $^{100}$Sn, $^{120}$Sn, and $^{132}$Sn,
determined at NLO REG2d\discard{.190617} (b), N$^2$LO
REG4d\discard{.190617} (c), and N$^3$LO REG2d\discard{.190617} (d) plotted as
functions of the numbers of eigenvalues kept in the Hessian matrices.
\label{fig:eigen}}
\end{center}
\end{figure}

For the purpose of presenting observables calculated in finite
nuclei, we decided to use a criterion of binding energies of
spherical nuclei, see Sec.~\ref{sec:Binding}. It then appears that optimal results are
obtained for $m^*/m=0.85$ at N$^3$LO~\cite{(Ben19)} and $a=1.50$\,fm,
that is, for the pseudopotential named REG6d\discard{.190617}.
Following this guidance, below we also present some results
corresponding to the same effective mass of $m^*/m=0.85$ and lower
orders: REG2d\discard{.190617} (NLO and  $a=0.85$\,fm) and
REG4d\discard{.190617} (N$^2$LO and  $a=1.15$\,fm). For an extended
comparison with the Gogny D1S parameterization~\cite{(Ber91d)}, which
corresponds to $m^*/m=0.697$, we also show results for $m^*/m=0.70$,
that is, for REG6a\discard{.190617} (N$^3$LO and  $a=1.60$\,fm).
Parameters of the four selected pseudopotentials are tabulated
in \ref{app:Parameters}. In the supplemental material
(\suppl) they are collected in a machine readable format.

We performed the standard analysis of statistical uncertainties as presented
in Ref.~\cite{(Dob14b)}. For REG2d\discard{.190617}, REG4d\discard{.190617} and
REG6d\discard{.190617}, eigenvalues of the Hessian matrices corresponding
to penalty functions scaled to $\chi^2=1$ are shown in Fig,~\ref{fig:eigen}(a).
The numbers of eigenvalues correspond to the numbers of parameters
optimized during the adjustments, and, therefore, vary from 10 (NLO) to 18 (N$^3$LO).

The magnitude of the eigenvalues of the Hessian matrices reveals how
well the penalty functions are constrained in the directions of the
corresponding eigenvectors in the parameter space. We observe that
for the three pseudopotentials considered here, there is a rapid
decrease of magnitude from the first to the third eigenvalue and
then a slower and almost regular decrease, where no clear gap can be identified.
This suggests that all parameters of the pseudopotentials are important.

For three tin isotopes of different nature: $^{100}$Sn (closed-shell,
isospin symmetric, unpaired), $^{120}$Sn (open-shell, isospin
asymmetric, paired) and $^{132}$Sn (closed-shell, isospin asymmetric,
unpaired), we calculated the propagated statistical uncertainties of
the total binding energies as functions of the number of kept
eigenvalues of the Hessian matrices, Figs~\ref{fig:eigen}(b)-(d) for
REG2d\discard{.190617}--REG6d\discard{.190617}, respectively. For each
of the considered parameterizations, after a given number of kept
eigenvalues (denoted in Figs~\ref{fig:eigen}(b)-(d) by vertical
lines), we observe a saturation of the propagated statistical
uncertainties. Therefore, we performed the final determination of the
statistical uncertainties by keeping these minimal numbers of
eigenvalues, {\em i.e.} 6 eigenvalues for REG2d\discard{.190617} (NLO)
and 7 for REG4d\discard{.190617} (N$^2$LO) and REG6d\discard{.190617}
(N$^3$LO).

\subsection{Infinite nuclear matter}
\label{sec:Infinite}

\begin{table}
\caption{Infinite nuclear matter properties corresponding to pseudopotentials
REG2d\discard{.190617}, REG4d\discard{.190617}, REG6d\discard{.190617}, and REG6a\discard{.190617}, compared
to those of the Gogny D1S interaction~\protect\cite{(Ber91d)}.\label{tab:inmres}}
\begin{center}
\begin{tabular}{lrrrrrr}
\br
Pseudopotential  & \multicolumn{1}{c}{$E/A$} &  \multicolumn{1}{c}{$\rho_\mathrm{sat}$}  & \multicolumn{1}{c}{$K_\infty$} & \multicolumn{1}{c}{$m^*/m$} & \multicolumn{1}{c}{$J$} & \multicolumn{1}{c}{$L$} \\
  & \multicolumn{1}{c}{[MeV]} &  \multicolumn{1}{c}{$[\mathrm{fm}^{-3}]$}  & \multicolumn{1}{c}{[MeV]} &  & \multicolumn{1}{c}{[MeV]} & \multicolumn{1}{c}{[MeV]} \\
\mr
REG2d\discard{.190617}  & -15.86 & 0.1574  & 235.4 & 0.8499 & 29.24 & 14.99 \\
REG4d\discard{.190617}  & -15.86 & 0.1589  & 225.6 & 0.8492 & 29.17 & 15.00 \\
REG6d\discard{.190617}  & -15.77 & 0.1584  & 232.1 & 0.8496 & 28.56 & 15.00 \\
REG6a\discard{.190617}  & -15.74 & 0.1564  & 233.6 & 0.7014 & 28.23 & 15.00 \\
D1S           & -16.01 & 0.1633  & 202.8 & 0.6970 & 31.13 & 22.44 \\
\br
\end{tabular}
\end{center}
\end{table}

In Table~\ref{tab:inmres}, we list quantities characterizing the
properties of infinite nuclear matter. We present results for
pseudopotentials REG2d\discard{.190617}, REG4d\discard{.190617},
REG6d\discard{.190617}, and REG6a\discard{.190617} compared to those
characterizing the D1S interaction~\cite{(Ber91d)}. For the two
strongly constrained quantities, $m^*/m$ and $L$, the target values
are almost perfectly met, whereas, for the other ones, we observe
some deviations, which, nevertheless, are well within the tolerance
intervals allowed in the penalty function.

\begin{figure}[tbp]
\begin{center}
\begin{tabular}{cc}
\includegraphics[height=0.47\linewidth,angle=270,viewport=10 10 500 670,clip]{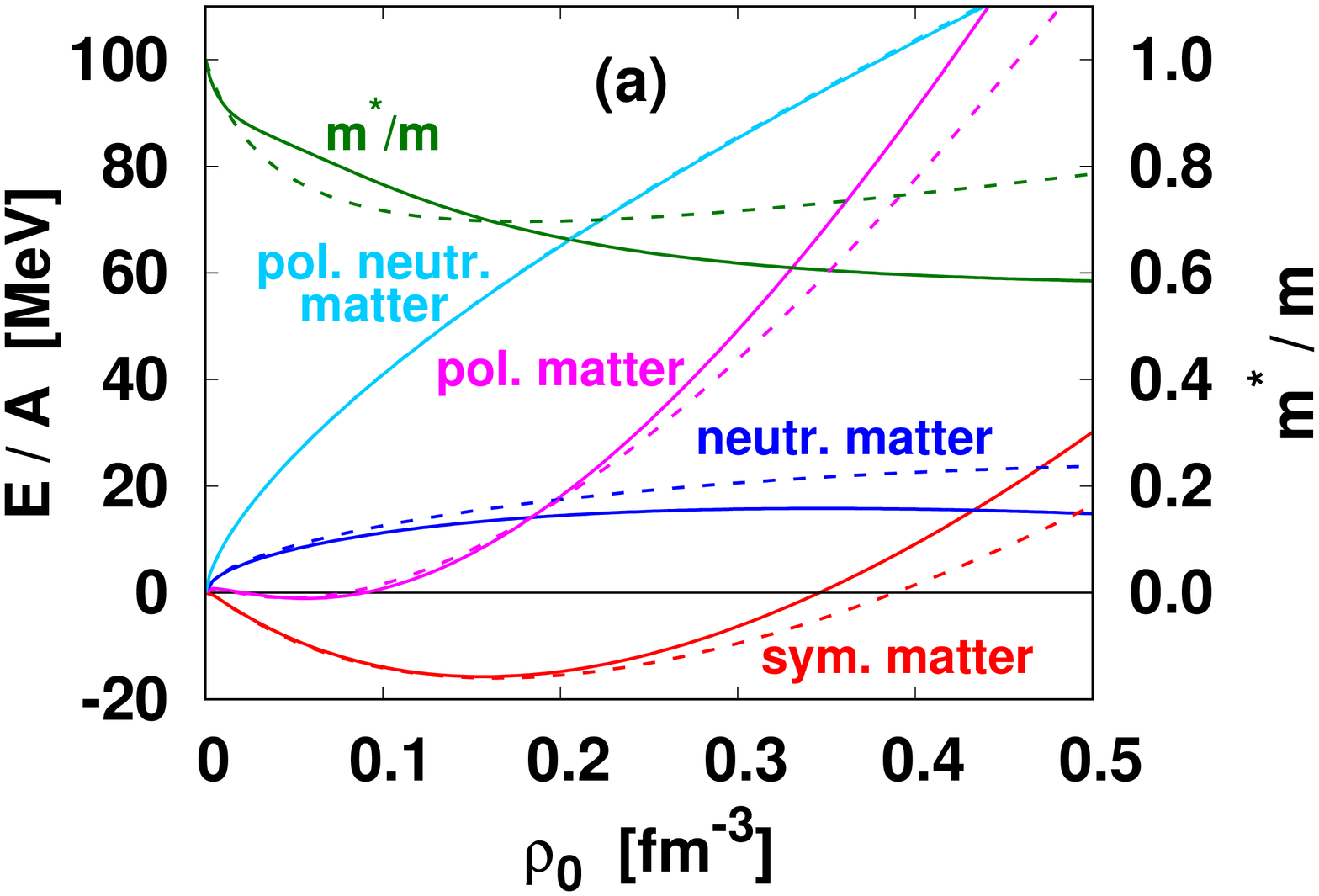}
&
\includegraphics[height=0.47\linewidth,angle=270,viewport=10 10 500 670,clip]{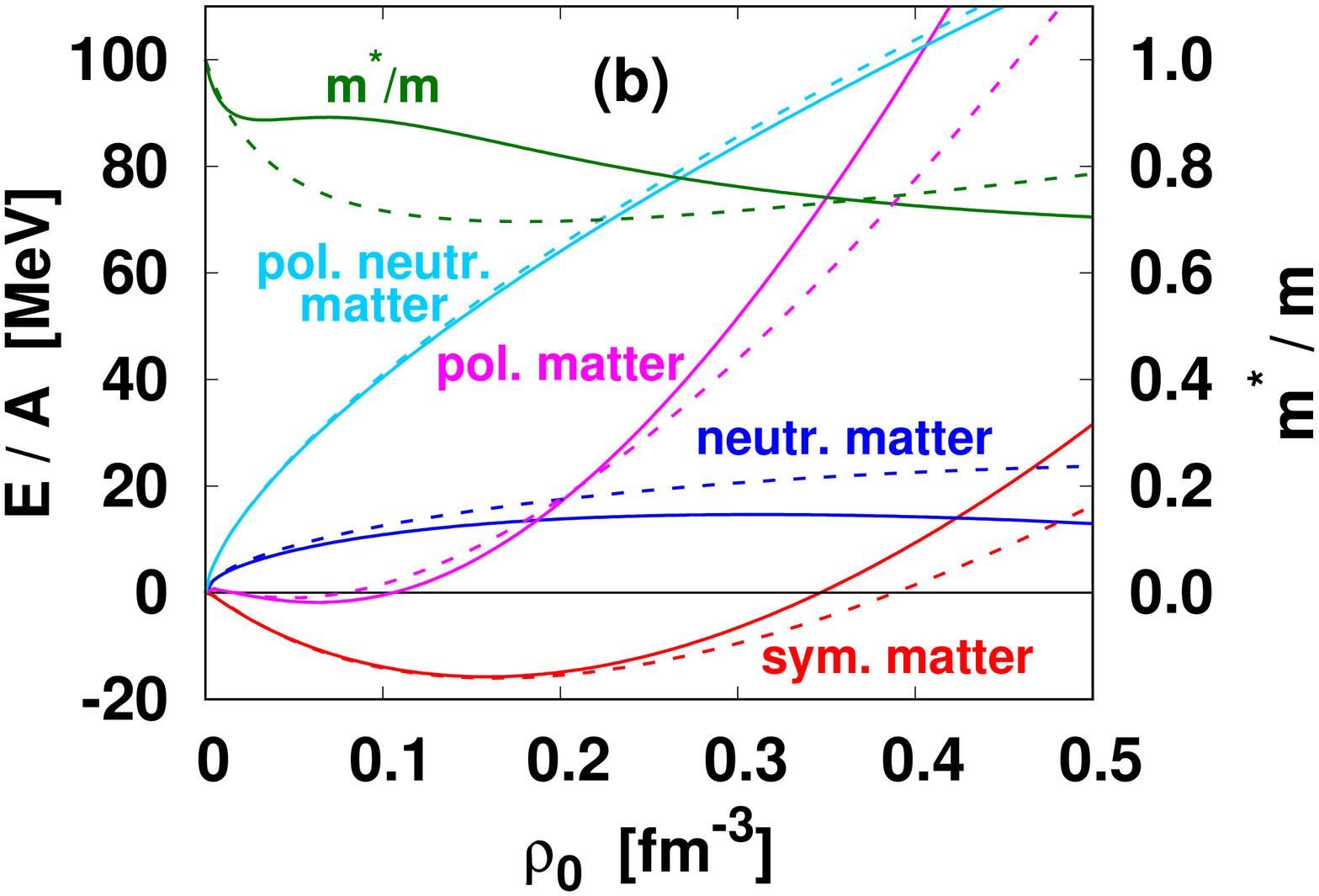}
\end{tabular}
\caption{Infinite-nuclear-matter isoscalar effective mass and
energies per particle in symmetric, neutron, polarized, and polarized neutron matter
as functions of the nuclear density $\rho_0$. Results calculated for the
D1S interaction~\protect\cite{(Ber91d)} (dashed lines) are compared
with those obtained for the two pseudopotentials at N$^3$LO with
$m^*/m=0.70$ (a) and $m^*/m=0.85$ (b) (solid lines).
\label{fig:eos}}
\end{center}
\end{figure}

For pseudopotentials REG6a\discard{.190617} and REG6d\discard{.190617},
the isoscalar effective mass in symmetric matter and energies per
particle (equations of state) for symmetric, neutron, polarized, and polarized neutron
matter are plotted in Fig.~\ref{fig:eos} along with the same
quantities for D1S~\cite{(Ber91d)}. The plotted equations of state
can be obtained from those calculated in four spin-isospin $(S,T)$ channels,
see \ref{app:st}. For these two N$^3$LO pseudopotentials, equations of state of
symmetric matter are somewhat
stiffer than that obtained for D1S. This is because of its slightly larger
compression modulus $K_\infty$. We also can see that for polarized symmetric matter, a shallow bound state
appears at low density. This feature
also affects D1S. The constraint on the equation of state of
polarized symmetric matter introduced in the penalty function has
probably limited the development of this state, but did not totally
avoid its appearance. Further studies are needed to analyze to what
extent it could impact observables calculated in time-odd nuclei and
how this possible flaw might be corrected.

The two main differences that appear when we compare the properties
in infinite nuclear matter of REG6a\discard{.190617} and
REG6d\discard{.190617} on one hand and those of D1S on the other hand
relate to the equation of state of the neutron matter and isoscalar
effective masses. First, near saturation, the regularized
pseudopotentials give equations of state of neutron matter slightly
lower than D1S, which can be attributed to its lower symmetry energy.
Second, for the N$^3$LO pseudopotentials, dependence of the effective
on density is less regular than for D1S. We note, however, that the
N$^3$LO effective masses are monotonically decreasing functions of the
density, and thus the pseudopotentials obtained in this study do not
lead to a surface-peaked effective mass, a feature which
was expected to improve the description of the density of states
around the Fermi energy~\cite{MA1983275}.

\subsection{Binding energies, radii, and pairing gaps of spherical nuclei}
\label{sec:Binding}

\begin{figure*}[tbp]
\noindent
\includegraphics[width=\textwidth]{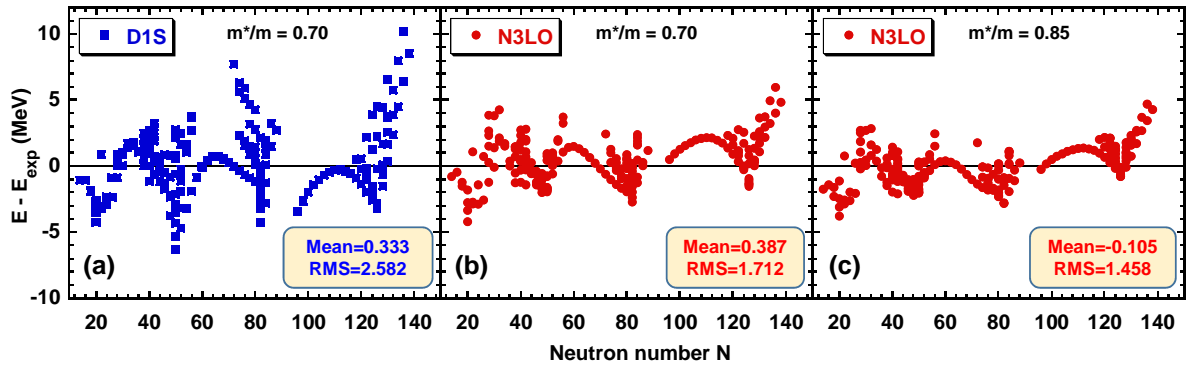}
\caption{Binding-energy residuals as functions of the neutron number, calculated
for a set of spherical nuclei (see text) and plotted for the
D1S~\protect\cite{(Ber91d)} (a), REG6a\discard{.190617} (b), and REG6d\discard{.190617} (c)
pseudopotentials.\label{fig:sphres}}
\end{figure*}

\newlength{\mywidth}
\setlength{\mywidth}{0.91\textwidth}
\begin{figure}
\begin{center}
\begin{tabular}{c@{\hspace*{0.001\mywidth}}c}
\includegraphics[height=0.534\mywidth,angle=270,viewport=40 50  525 765]{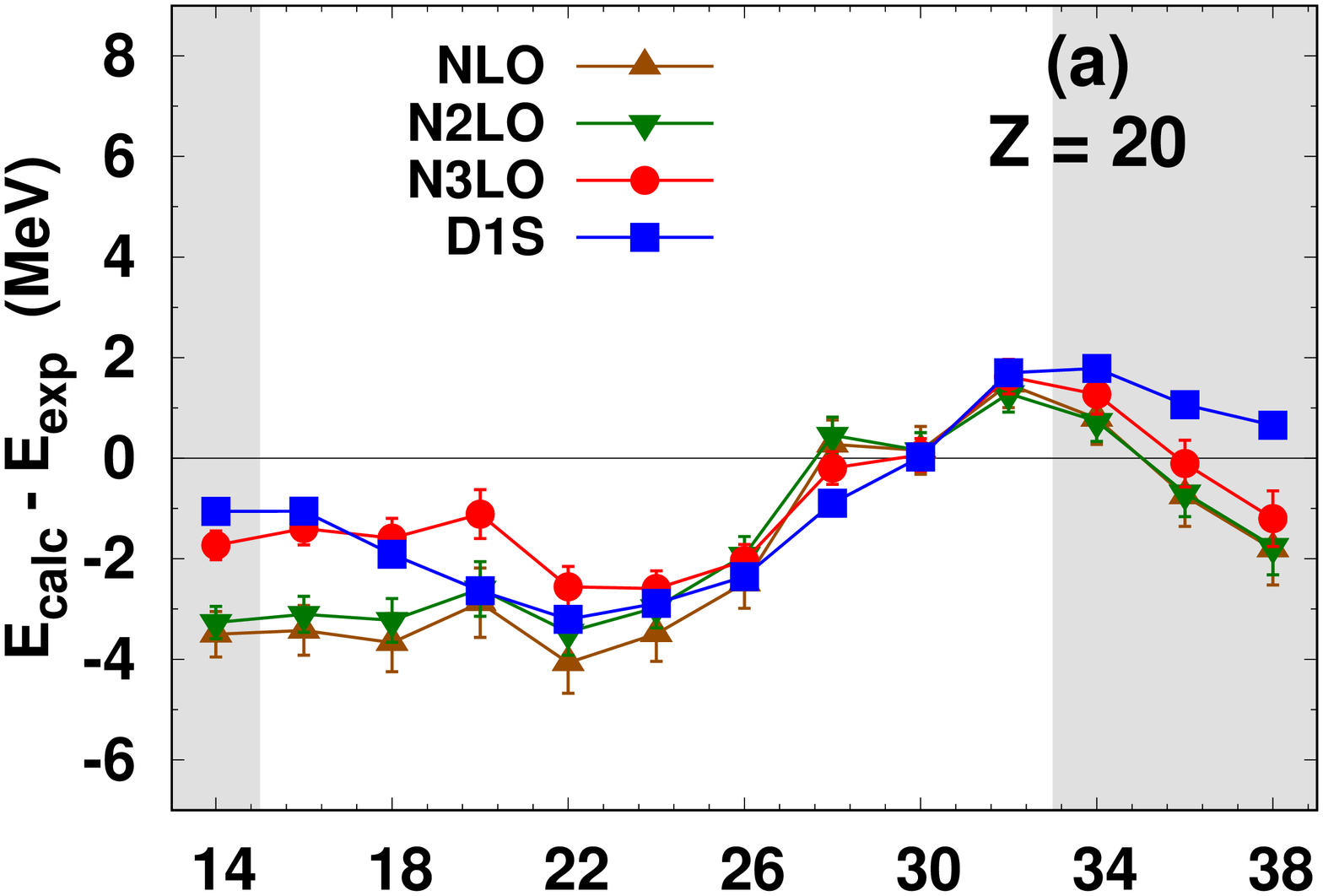} &
\includegraphics[height=0.534\mywidth,angle=270,viewport=40 175 525 890]{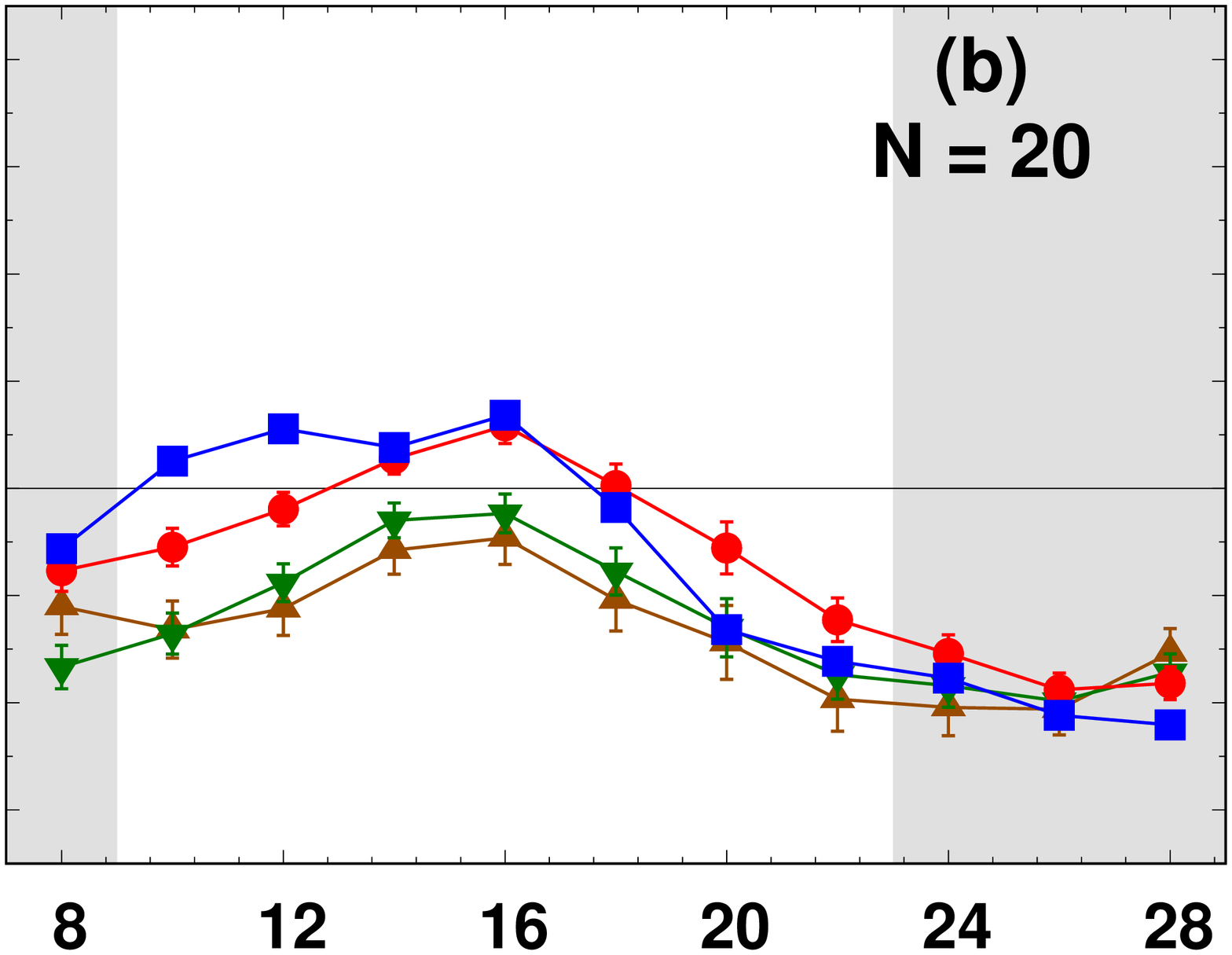} \\[-0.05\mywidth]
\includegraphics[height=0.534\mywidth,angle=270,viewport=40 50  525 765]{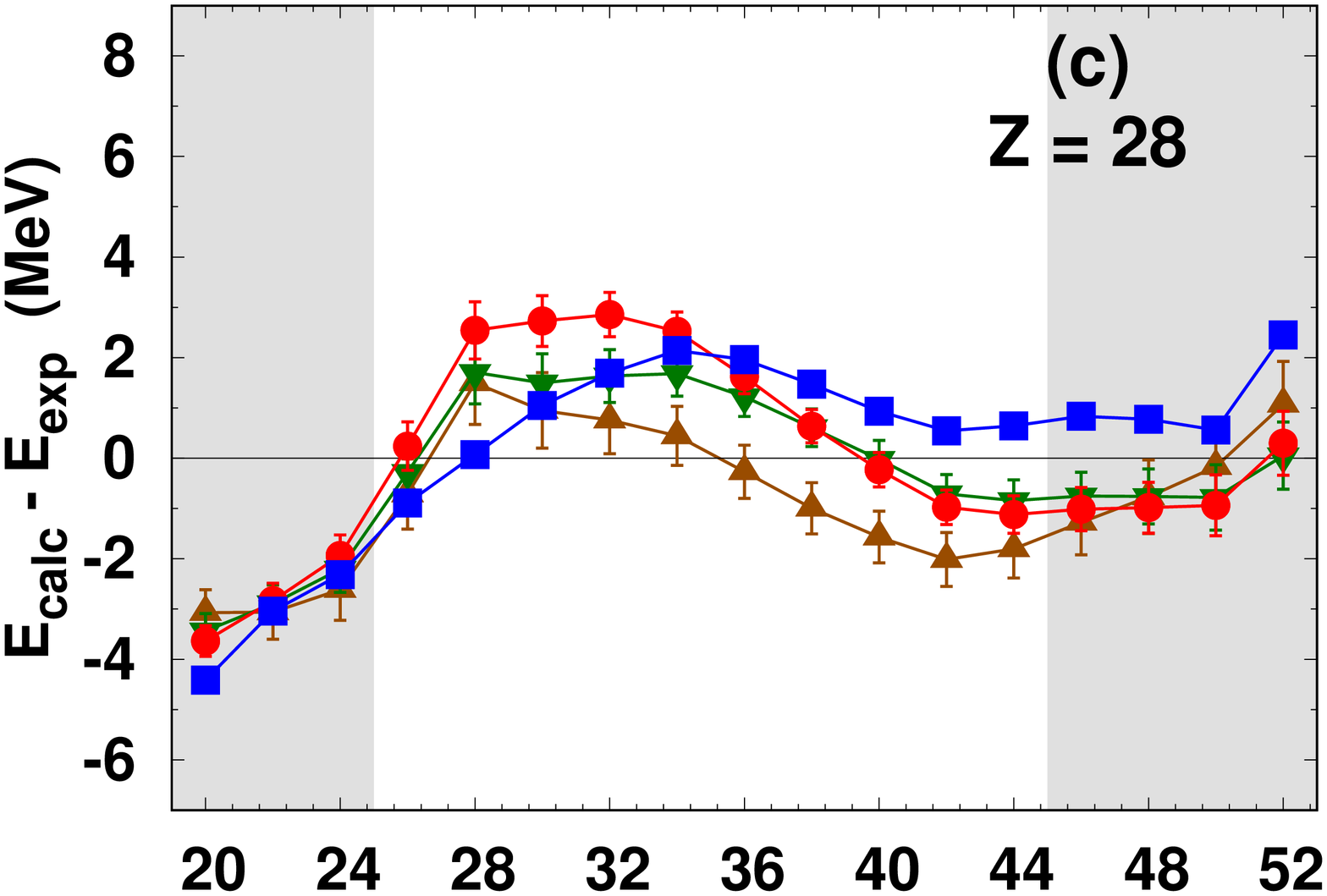} &
\includegraphics[height=0.534\mywidth,angle=270,viewport=40 175 525 890]{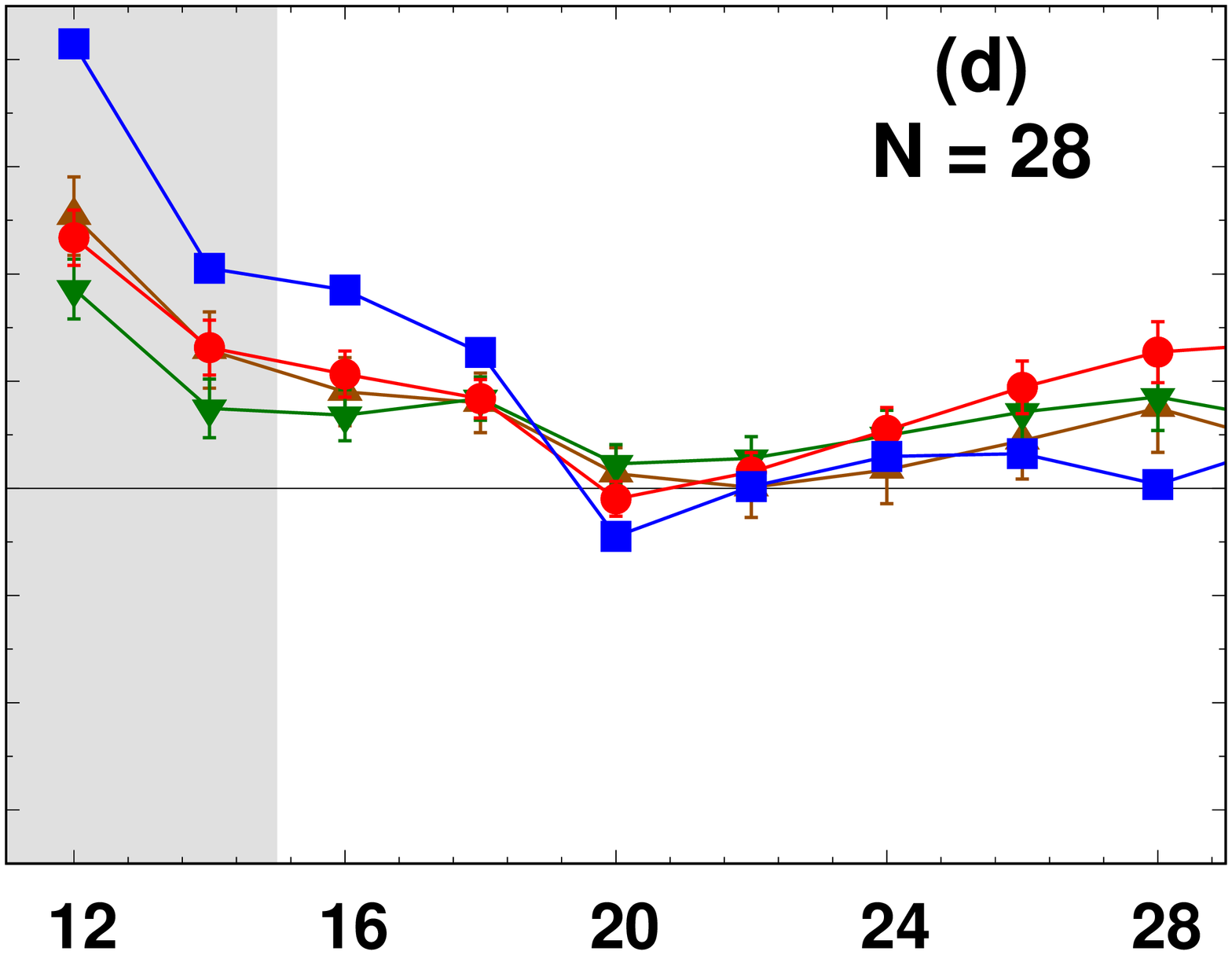} \\[-0.05\mywidth]
\includegraphics[height=0.534\mywidth,angle=270,viewport=40 50  525 765]{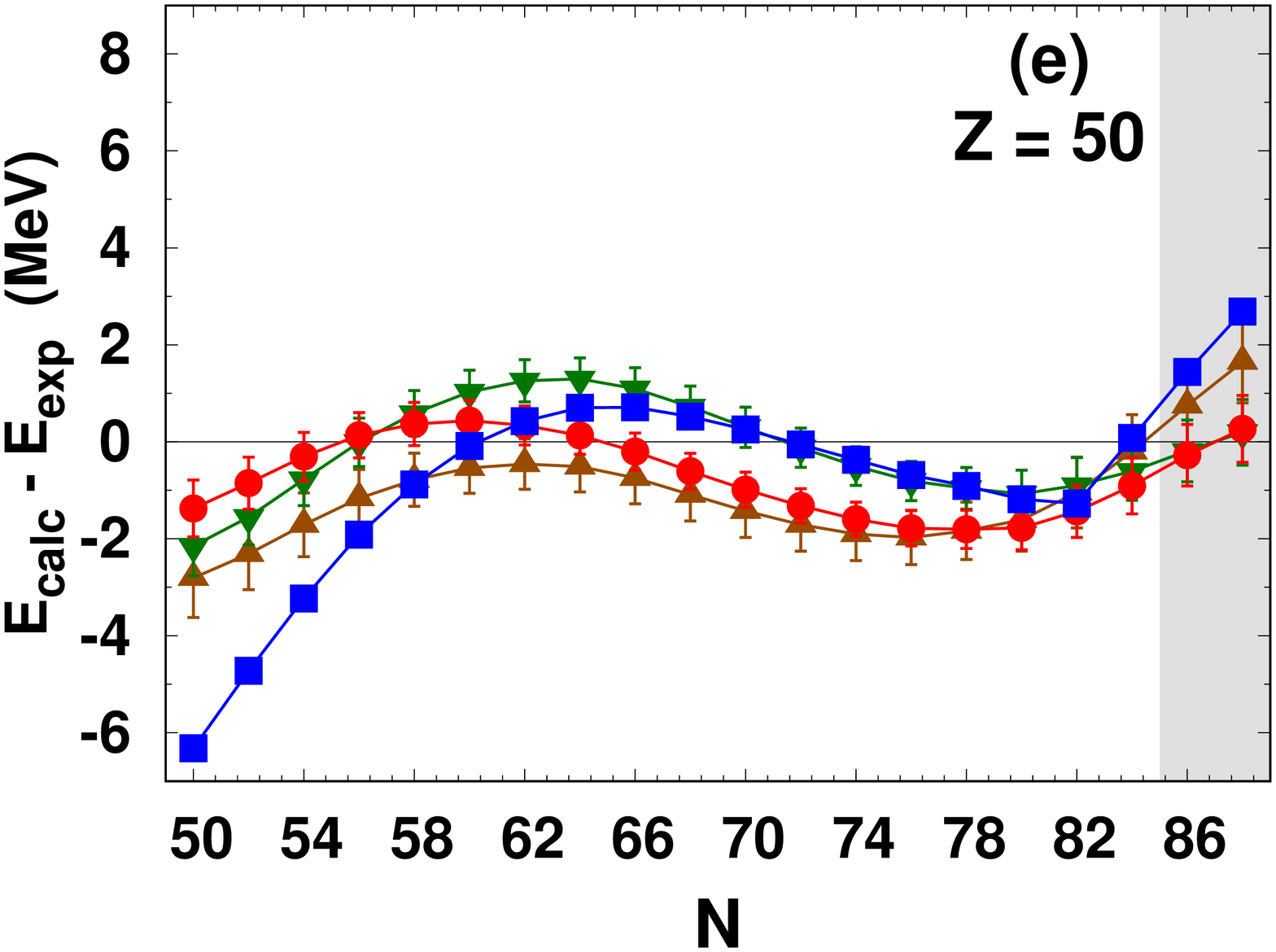} &
\includegraphics[height=0.534\mywidth,angle=270,viewport=40 175 525 890]{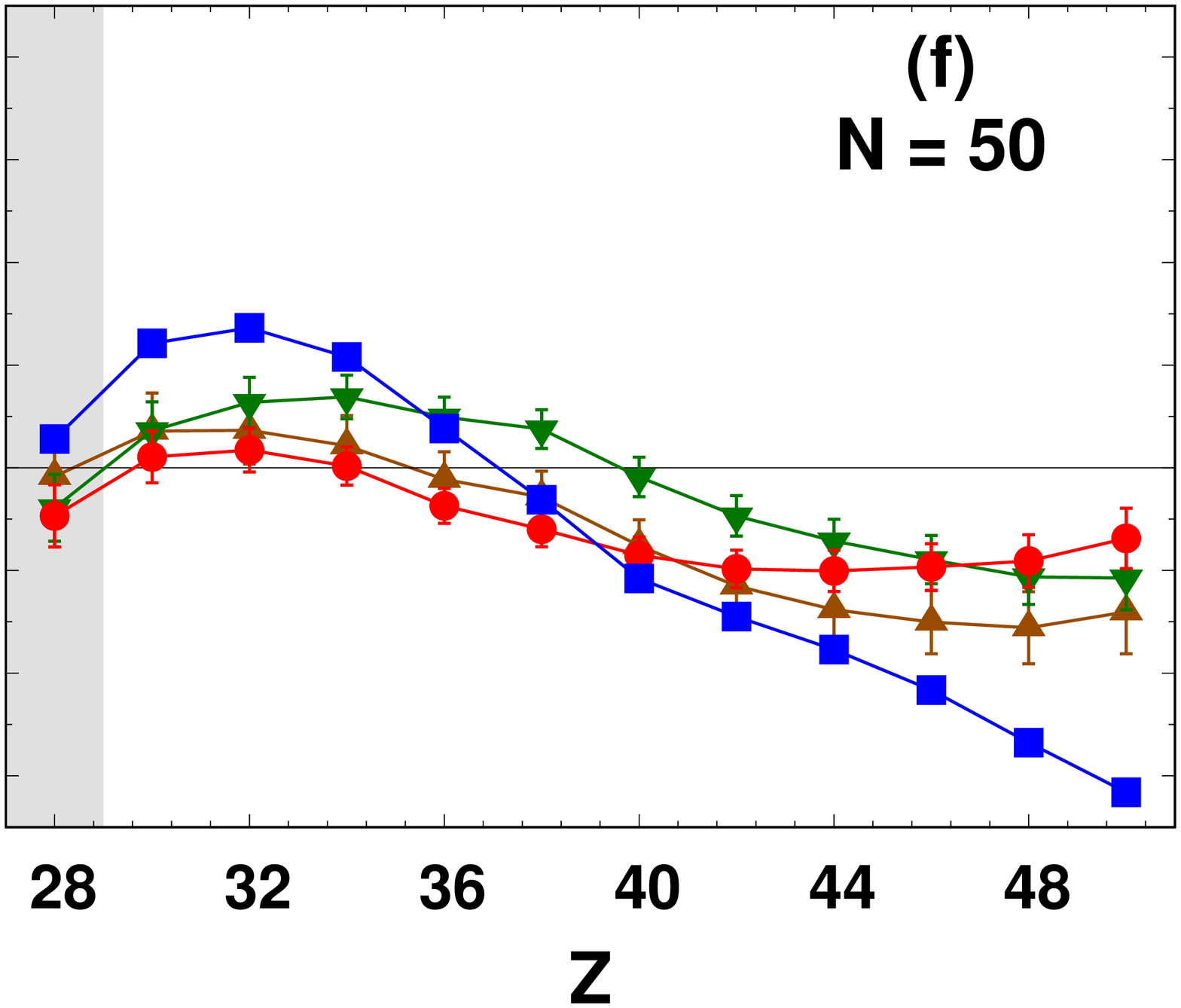}
\end{tabular}
\caption{Binding-energy residuals of proton (neutron) semi-magic
nuclei with $Z$ ($N$) equal to 20, 28, or 50, plotted in the left
(right) panels as functions of the neutron (proton) number,
calculated for the REG2d\discard{.190617} (up triangles),
REG4d\discard{.190617} (down triangles), REG6d\discard{.190617}
(circles), and D1S~\protect\cite{(Ber91d)} (squares)
pseudopotentials. Shaded zones correspond to the AME2016 masses
extrapolated from systematics~\cite{(Wan17)}.
\label{fig:fig1}}
\end{center}
\end{figure}

In this section, we present results of systematic calculations
performed for spherical nuclei and compared with experimental data.
For the purpose of such a comparison, we have selected a set of 214
nuclei that were identified as spherical in the systematic
calculations performed for the D1S functional in
Refs.~\cite{(Del10a),(Del10c)}. In Fig.~\ref{fig:sphres}, we
present an overview of the binding-energy residuals obtained for the D1S,
REG6a\discard{.190617}, and REG6d\discard{.190617} functionals.
Experimental values were taken from the 2016 atomic mass
evaluation~\cite{(Wan17)}. The obtained root-mean-square (RMS)
binding-energy residuals are equal to 2.582\,MeV for
D1S, 1.717\,MeV for REG6a\discard{.190617}, and 1.458\,MeV for
REG6d\discard{.190617}. We also see that for REG6d\discard{.190617}, the
trends of binding-energy residuals along isotopic chains in heavy
nuclei become much better reproduced. As a reference, we have also
determined the analogous RMS value corresponding to the UNEDF0 functional~\cite{(Kor10c),massexplorer},
which turns out to be equal to 1.900\,MeV.

\begin{figure}
\begin{center}
\begin{tabular}{c@{\hspace*{0.001\mywidth}}c}
\includegraphics[height=0.534\mywidth,angle=270,viewport=40 50  525 765]{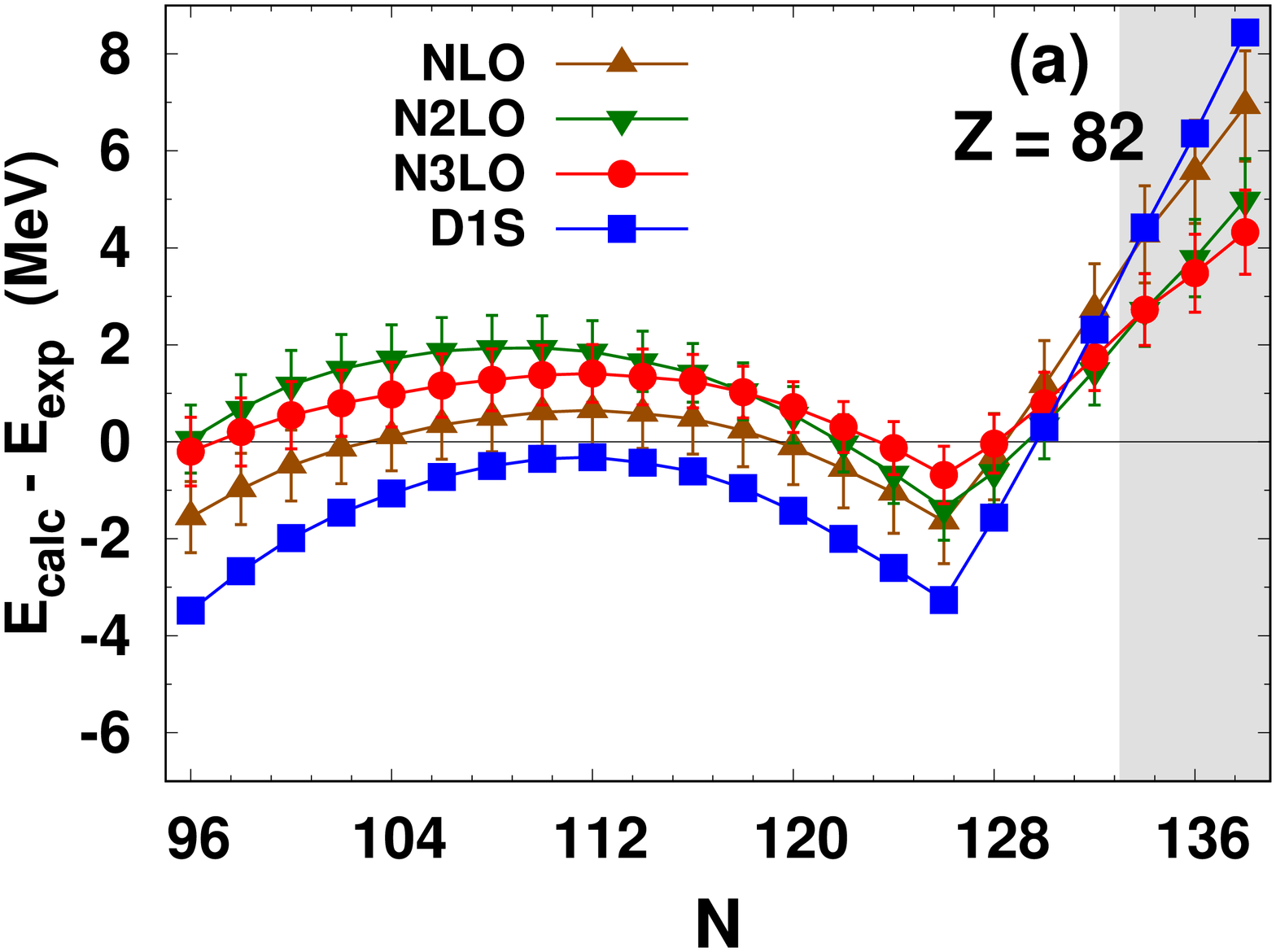} &
\includegraphics[height=0.534\mywidth,angle=270,viewport=40 175 525 890]{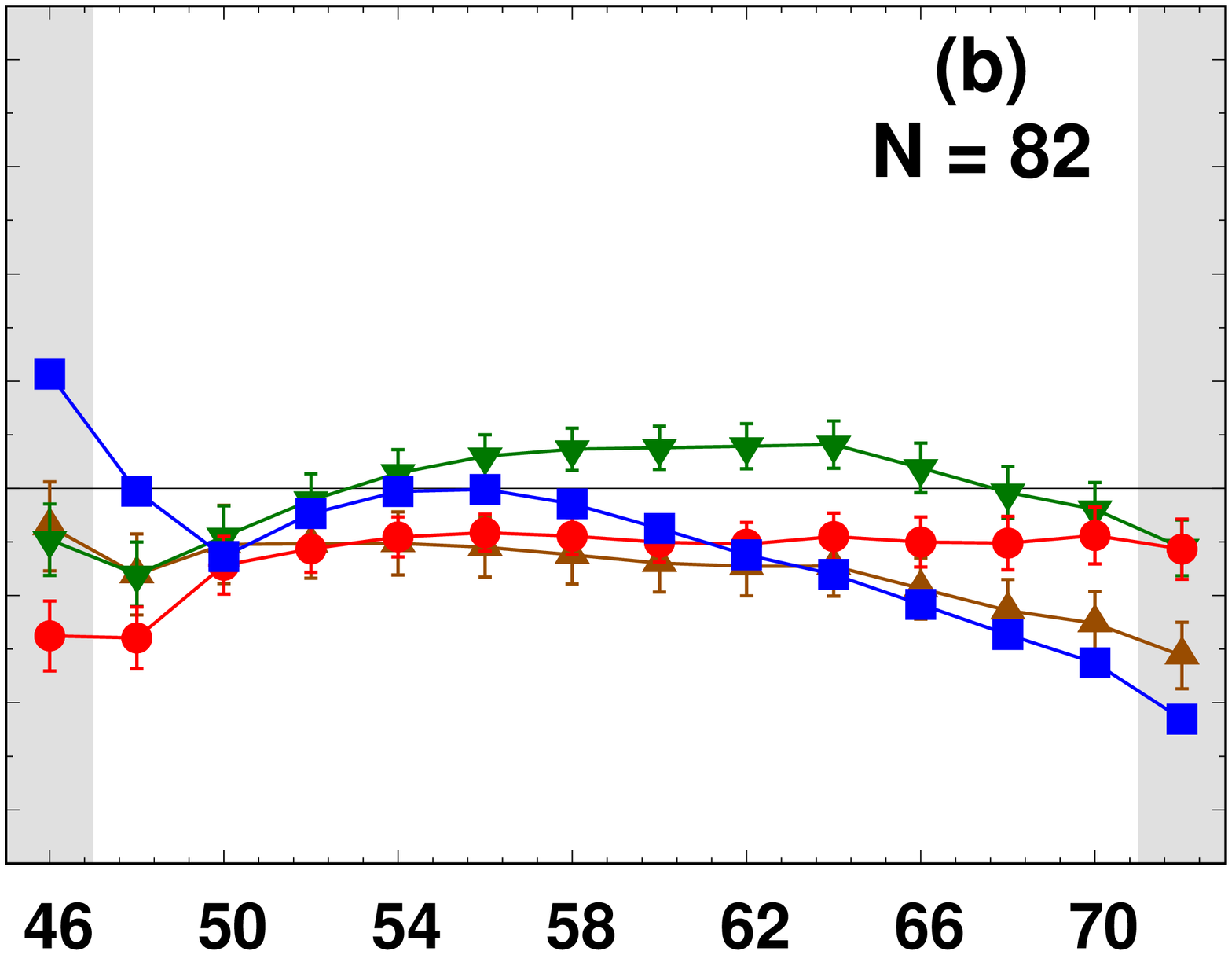} \\[-0.05\mywidth]
 &
\includegraphics[height=0.534\mywidth,angle=270,viewport=40 175 525 890]{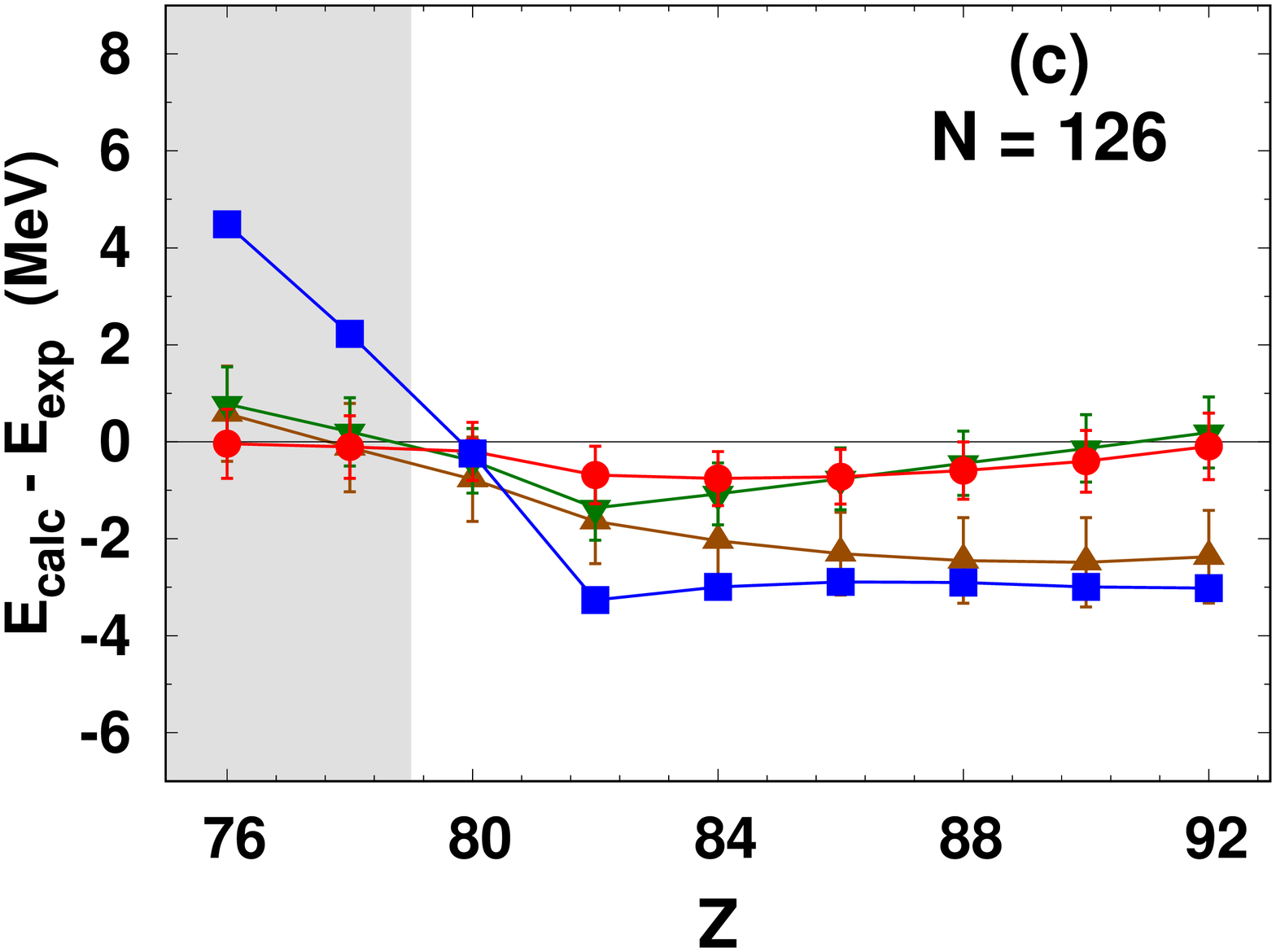}
\end{tabular}
\end{center}
\caption{Same as in Fig.~\ref{fig:fig1} but for the semimagic nuclei with
$Z=82$, $N=82$ and $N=126$.
\label{fig:fig2}}
\end{figure}

In Figs.~\ref{fig:fig1} and~\ref{fig:fig2}, we show detailed values
of binding-energy residuals along the isotopic or isotonic chains of
semi-magic nuclei. In most chains one can see a clear improvement of
the isospin dependence of masses. In particular, in almost all
semi-magic chains, kinks of energy residuals at doubly magic nuclei
either decreased or even vanished completely, like at $N=82$ and 126,
see Figs.~\ref{fig:fig2}(b) and (c), respectively.

\begin{figure*}[tbp]
\noindent
\includegraphics[width=\textwidth]{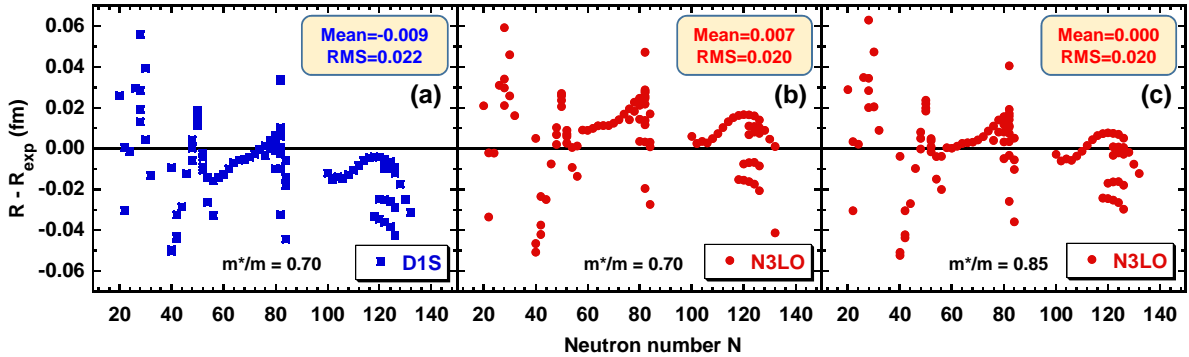}
\caption{Same as in Fig.~\protect\ref{fig:sphres} but for
the charge-radii residuals.\label{fig:rad}}
\end{figure*}

\begin{figure}[tbp]
\noindent
\begin{center}
\includegraphics[height=0.25\textwidth]{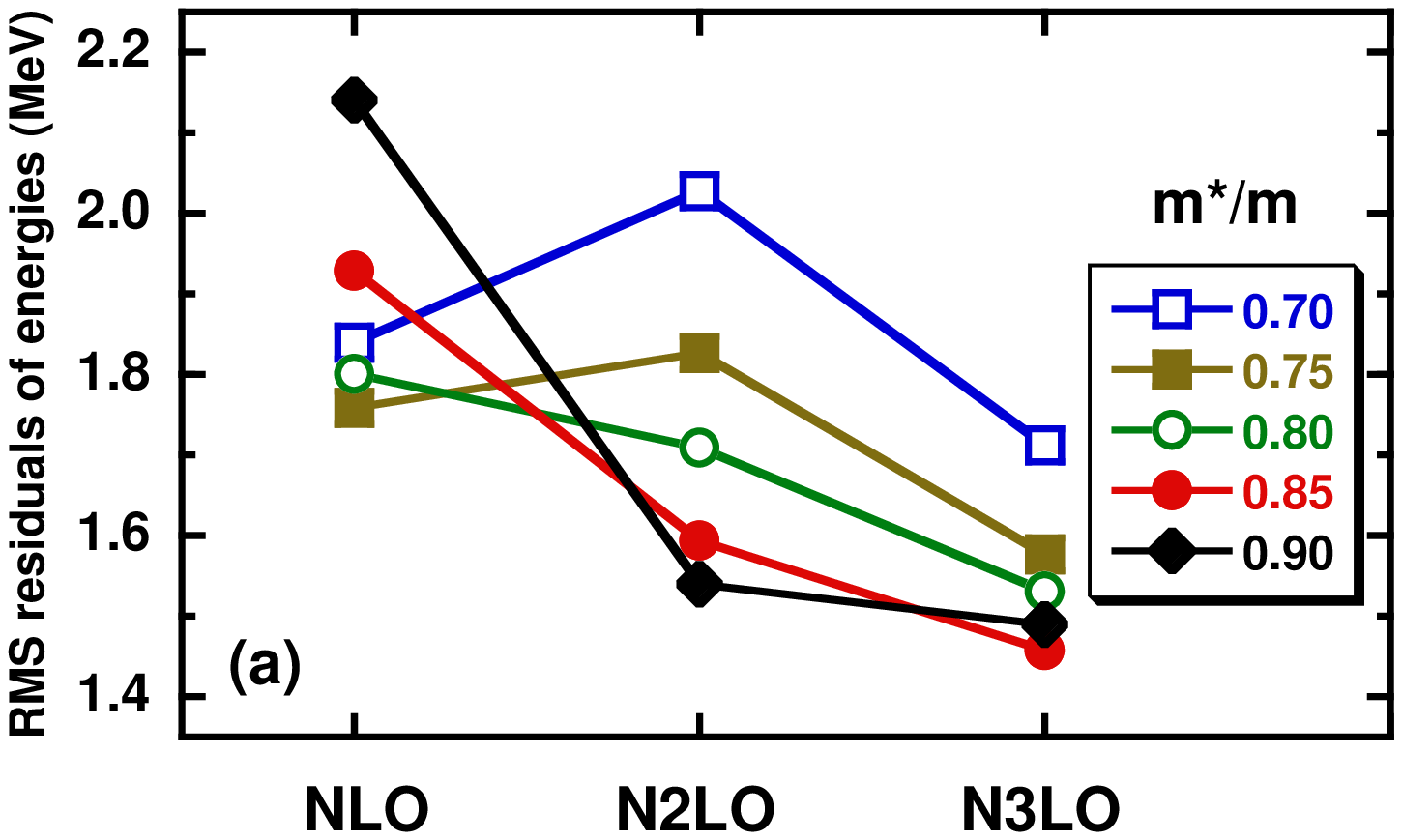}~~~~
\includegraphics[height=0.25\textwidth]{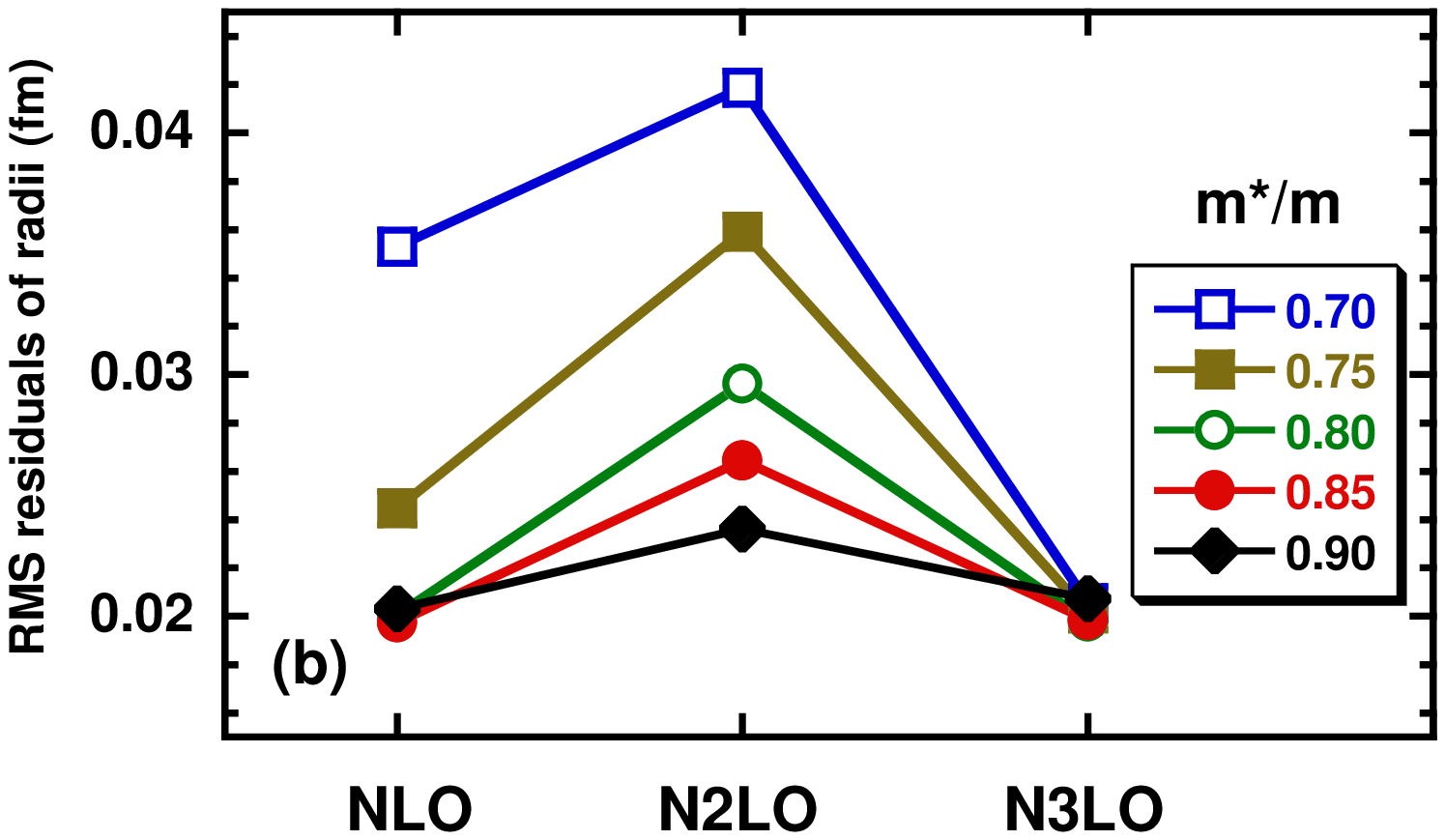}
\caption{RMS residuals of binding energies (a) and charge radii (b)
as functions of the order of pseudopotentials, adjusted in this study
for the five values of the isoscalar effective mass $m^*/m$.\label{fig:rms}}
\end{center}
\end{figure}

In Fig.~\ref{fig:rad}, for the same set of EDFs and nuclei as those used in
Fig.~\ref{fig:sphres}, we show the analogous residuals of the charge
radii of spherical nuclei. The experimental values were taken from
Ref.~\cite{(Nad94)}. Again, the N$^3$LO EDFs provide the smallest
deviations from data. We note that the residuals of the order of
0.02\,fm are typical for many Skyrme-like EDFs, for example, for the
UNEDF family of EDFs~\cite{(Kor14a)}. Figures~\ref{fig:rms}(a) and
(b) present summary of the RMS residuals of binding energies and charge radii,
respectively, which were obtained in this study. We see
that a decrease of the penalty functions when going from NLO to
N$^2$LO, see Fig.~\ref{fig:chi2s}, is often accompanied by an increase
of the RMS residuals. This indicates that the data for 17 spherical
nuclei, which are included in the penalty function, see
Sec.~\ref{sec:fit}, do not automatically lead to a better description
of all spherical nuclei. Only at N$^3$LO a consistently better
description is obtained.

\begin{figure*}[tbp]
\noindent
\includegraphics[width=\textwidth]{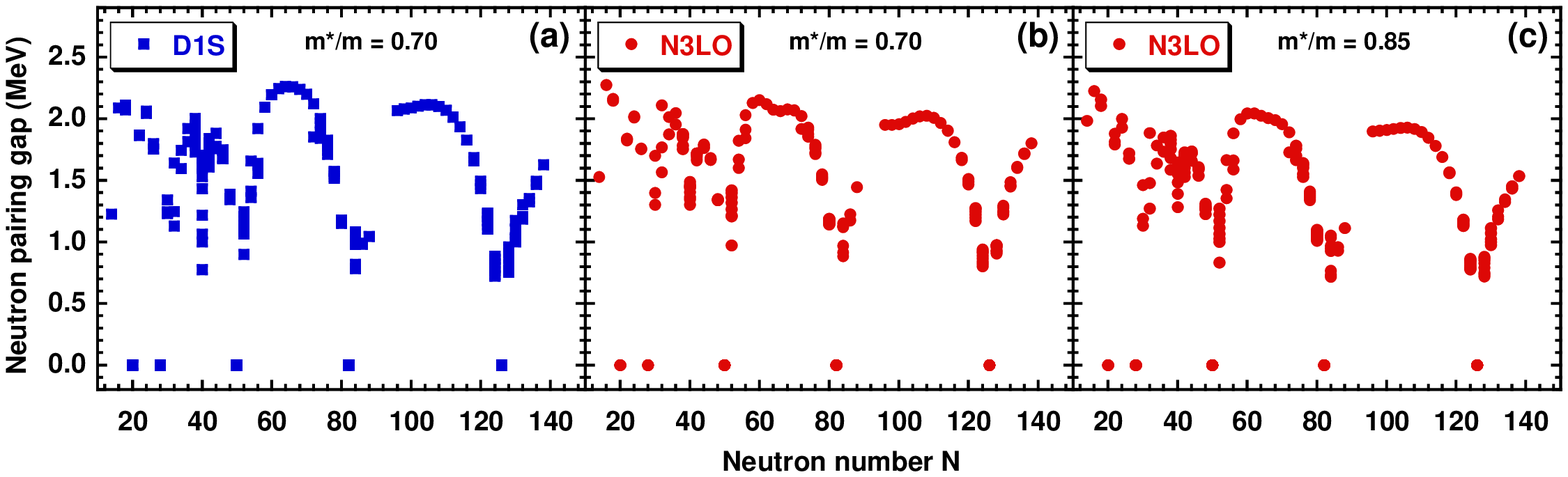}
\caption{Same as in Fig.~\protect\ref{fig:sphres} but for
the average neutron pairing gaps.\label{fig:gapsn}}
\end{figure*}

\begin{figure*}[tbp]
\noindent
\includegraphics[width=\textwidth]{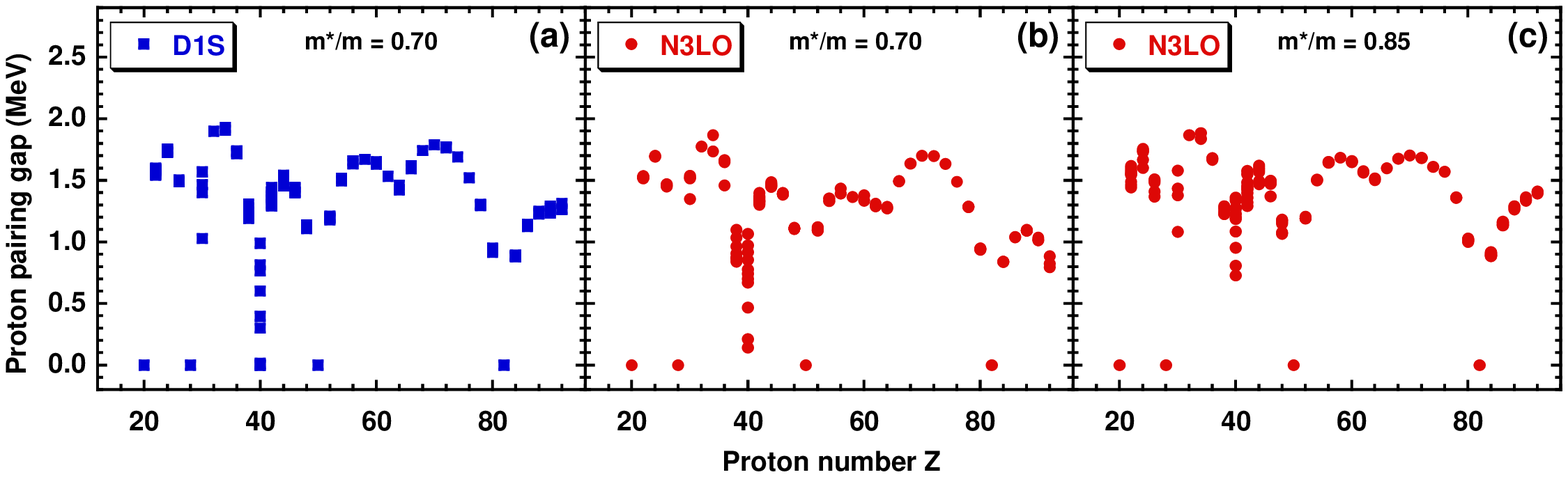}
\caption{Same as in Fig.~\protect\ref{fig:gapsn} but for
the average proton pairing gaps.\label{fig:gapsp}}
\end{figure*}

Finally, in Figs.~\ref{fig:gapsn} and~\ref{fig:gapsp}, we show
calculated average neutron and proton pairing gaps, respectively.
Qualitatively, all three EDFs shown in the figures give very similar results.
A thorough comparison with experimental odd-even mass staggering, along
with parameter adjustments better focused on the pairing channel, will
be the subject of a forthcoming publication.

\subsection{Single particle energies}
\label{sec:Single}

\begin{figure}
\begin{center}
\begin{tabular}{cc}
\includegraphics[height=0.47\linewidth,angle=270,viewport=20 2 500 680,clip]{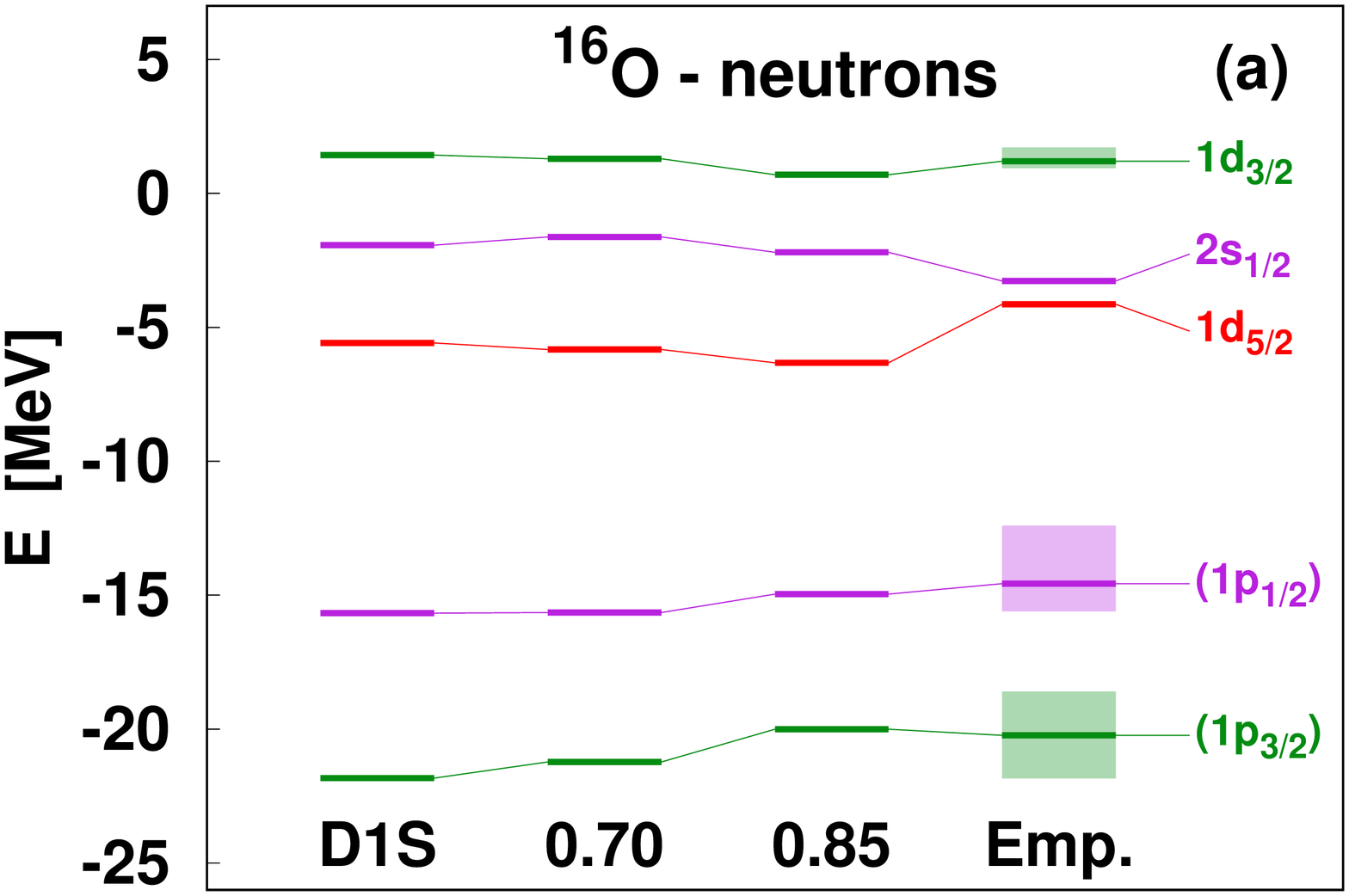} &
\includegraphics[height=0.47\linewidth,angle=270,viewport=20 2 500 680,clip]{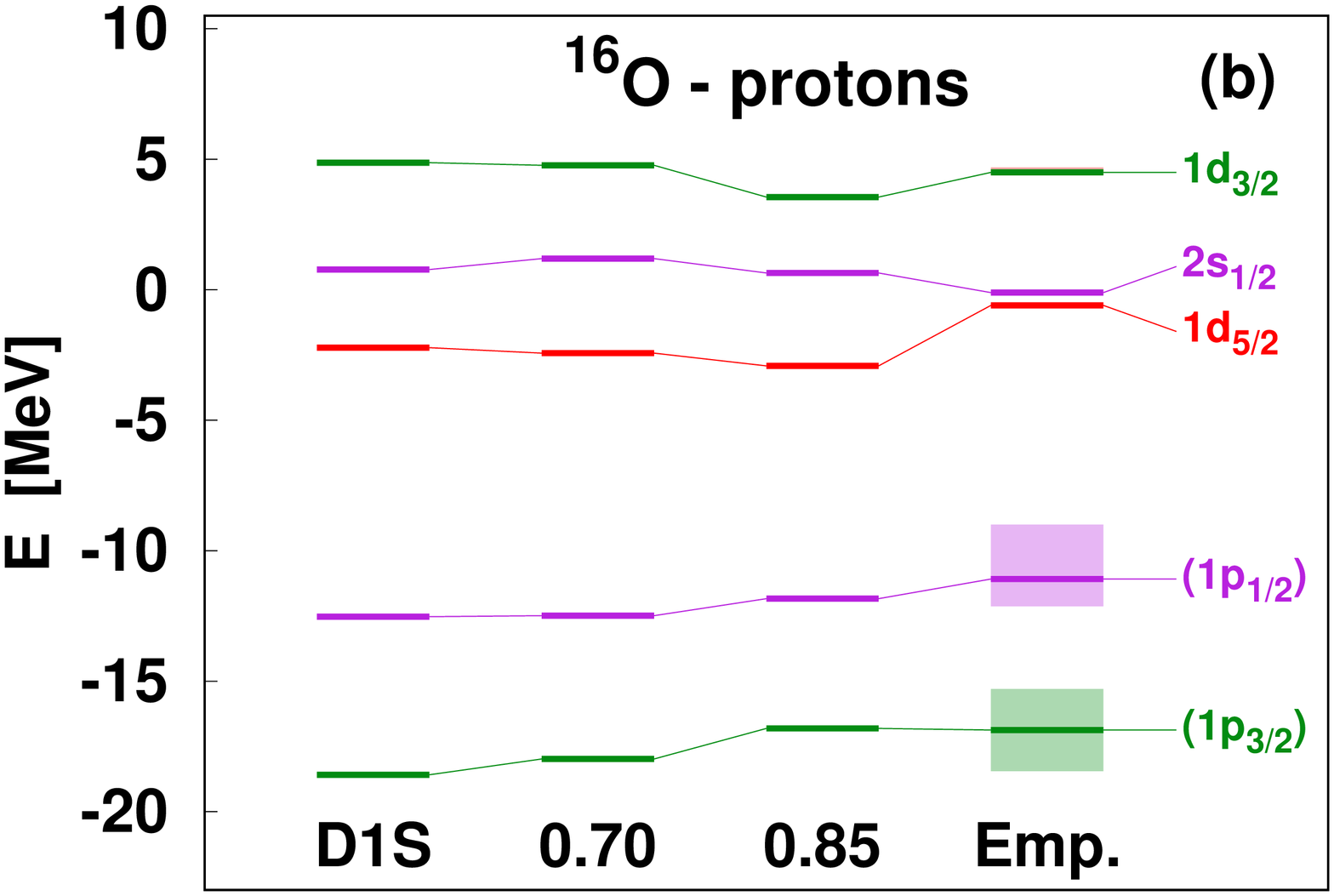} \\
\includegraphics[height=0.47\linewidth,angle=270,viewport=20 2 500 680,clip]{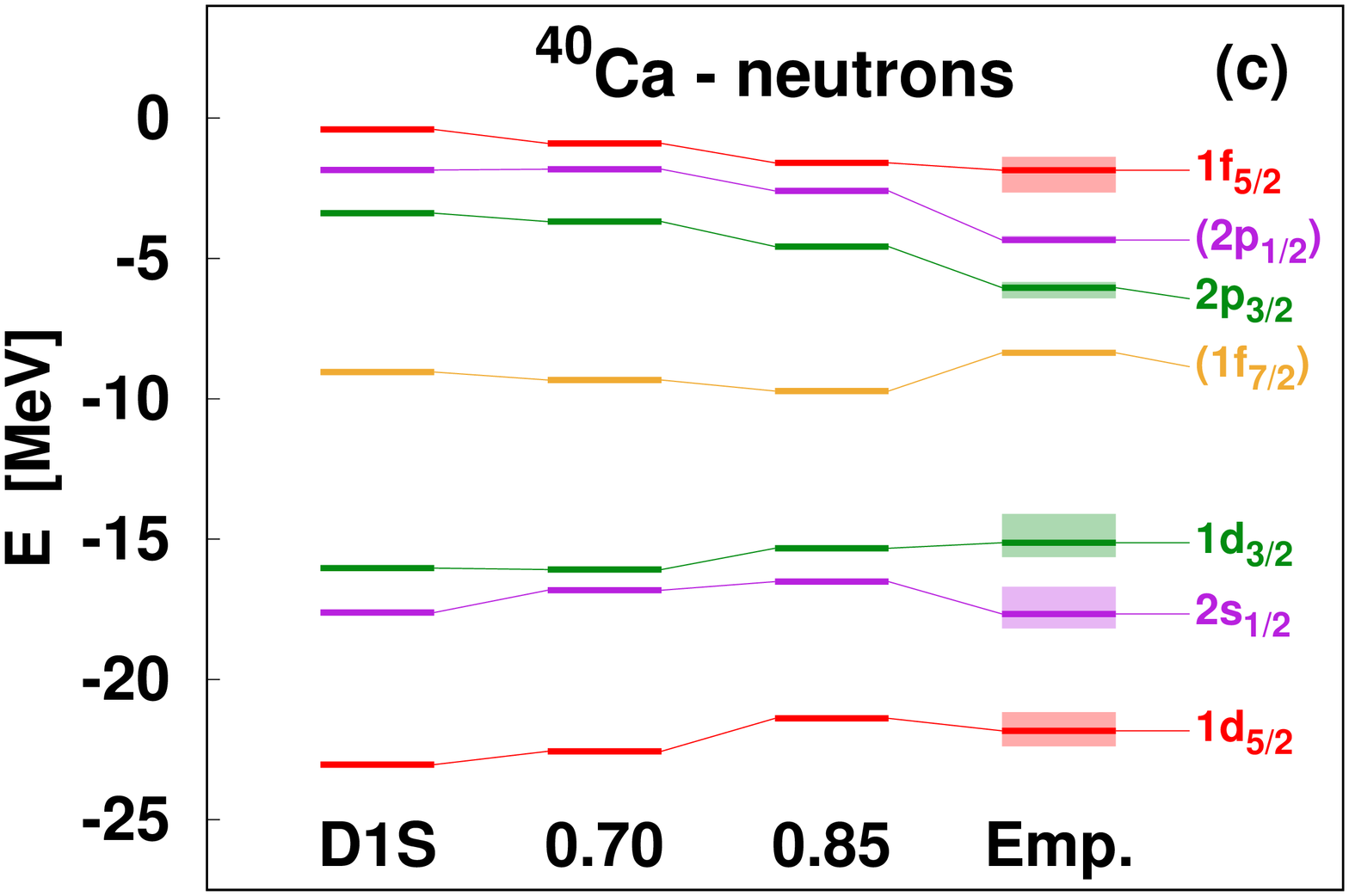} &
\includegraphics[height=0.47\linewidth,angle=270,viewport=20 2 500 680,clip]{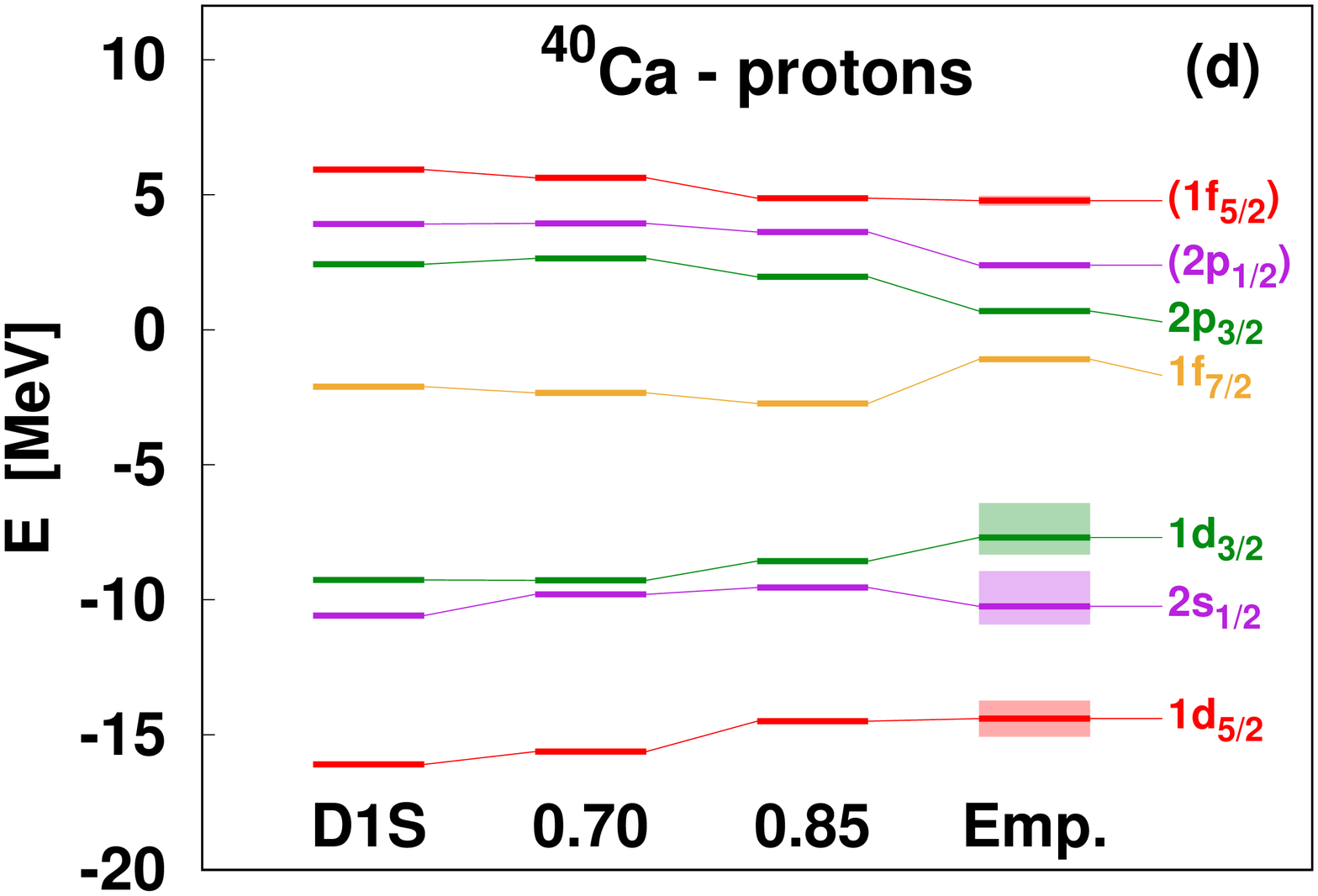} \\
\includegraphics[height=0.47\linewidth,angle=270,viewport=20 2 500 680,clip]{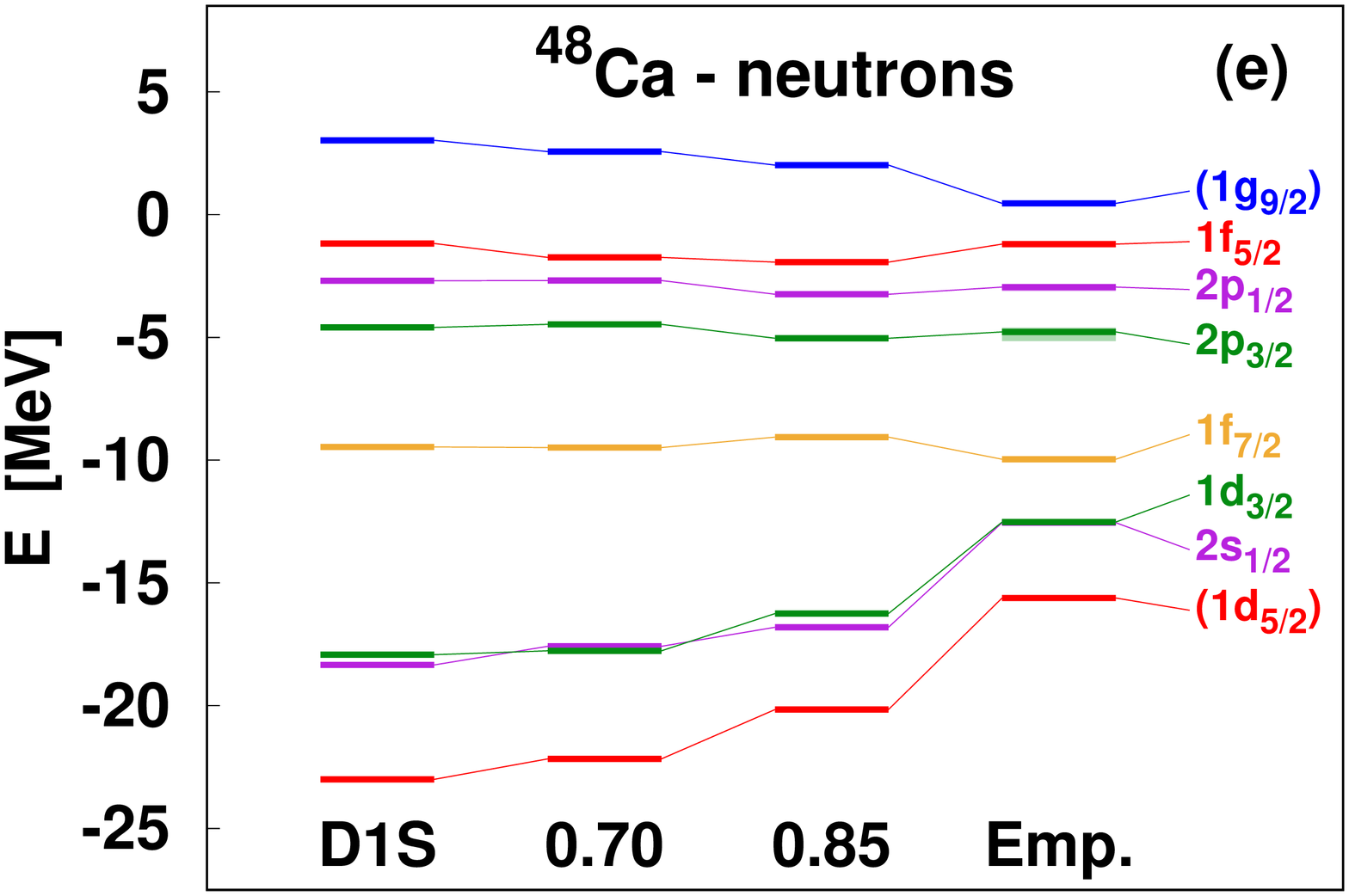} &
\includegraphics[height=0.47\linewidth,angle=270,viewport=20 2 500 680,clip]{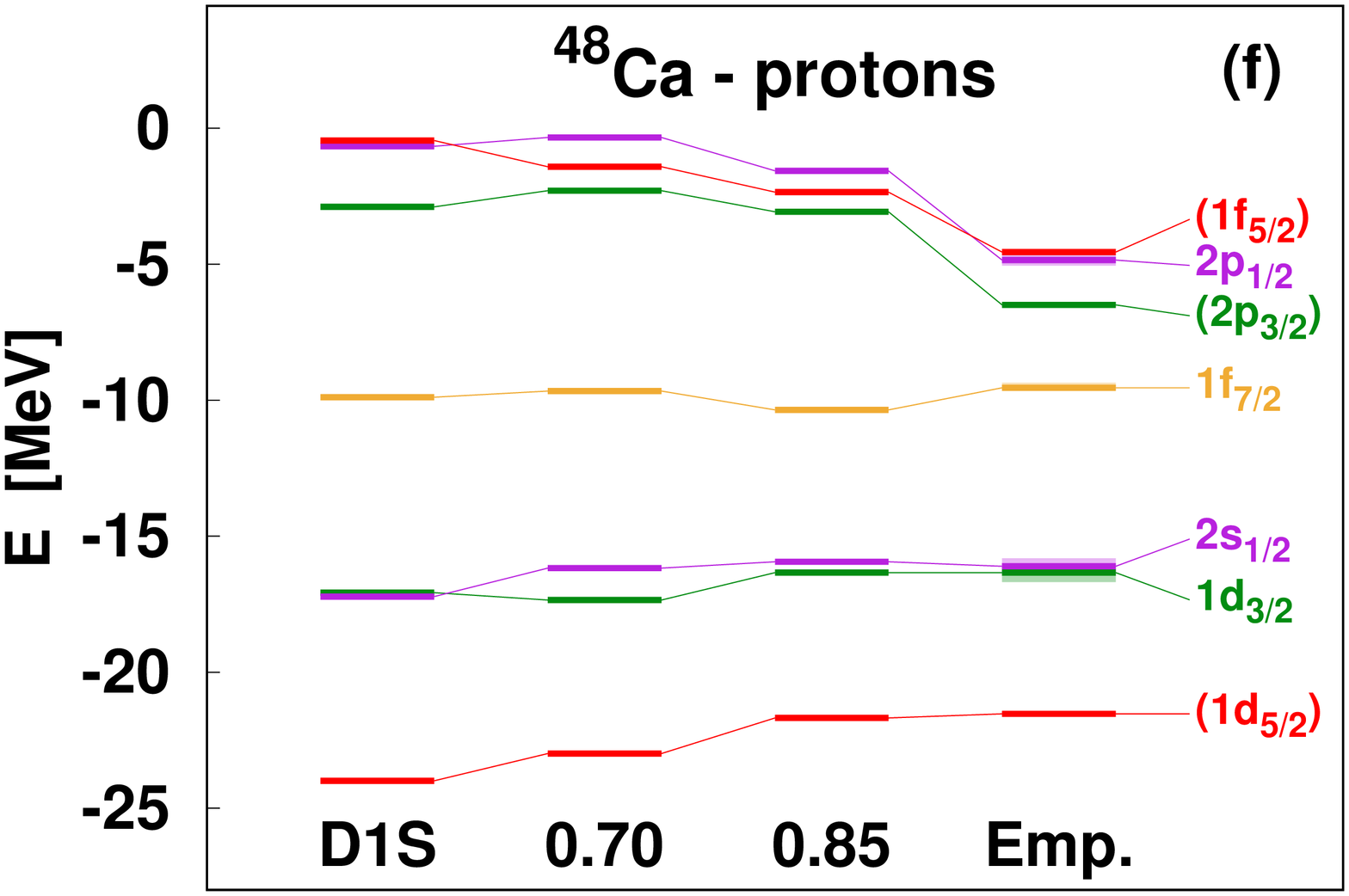}
\end{tabular}
\end{center}
\caption{Neutron (left) and proton (right) single-particle energies in $^{16}$O (top), $^{40}$Ca (middle),
and $^{48}$Ca (bottom), calculated for the D1S~\protect\cite{(Ber91d)},
REG6a\discard{.190617} (N$^3$LO, $m^*/m=0.70$), and REG6d\discard{.190617} (N$^3$LO, $m^*/m=0.85$) functionals.
Empirical values were taken from the compilation of Ref.~\cite{(Tar14b)}.
\label{fig:spe16o40ca48ca}}
\end{figure}

\begin{figure}
\begin{center}
\begin{tabular}{cc}
\includegraphics[height=0.47\linewidth,angle=270,viewport=20 2 500 680,clip]{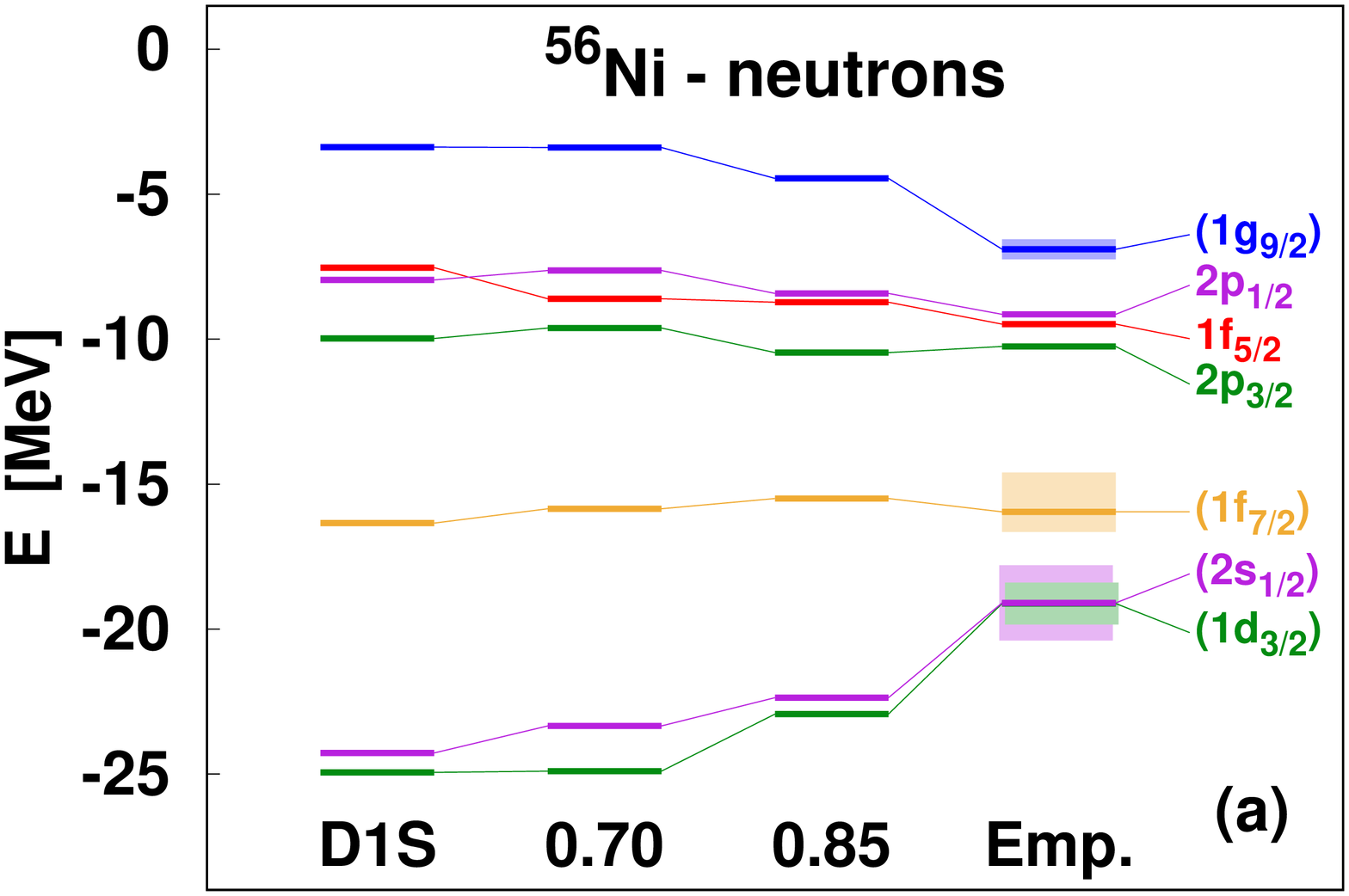} &
\includegraphics[height=0.47\linewidth,angle=270,viewport=20 2 500 680,clip]{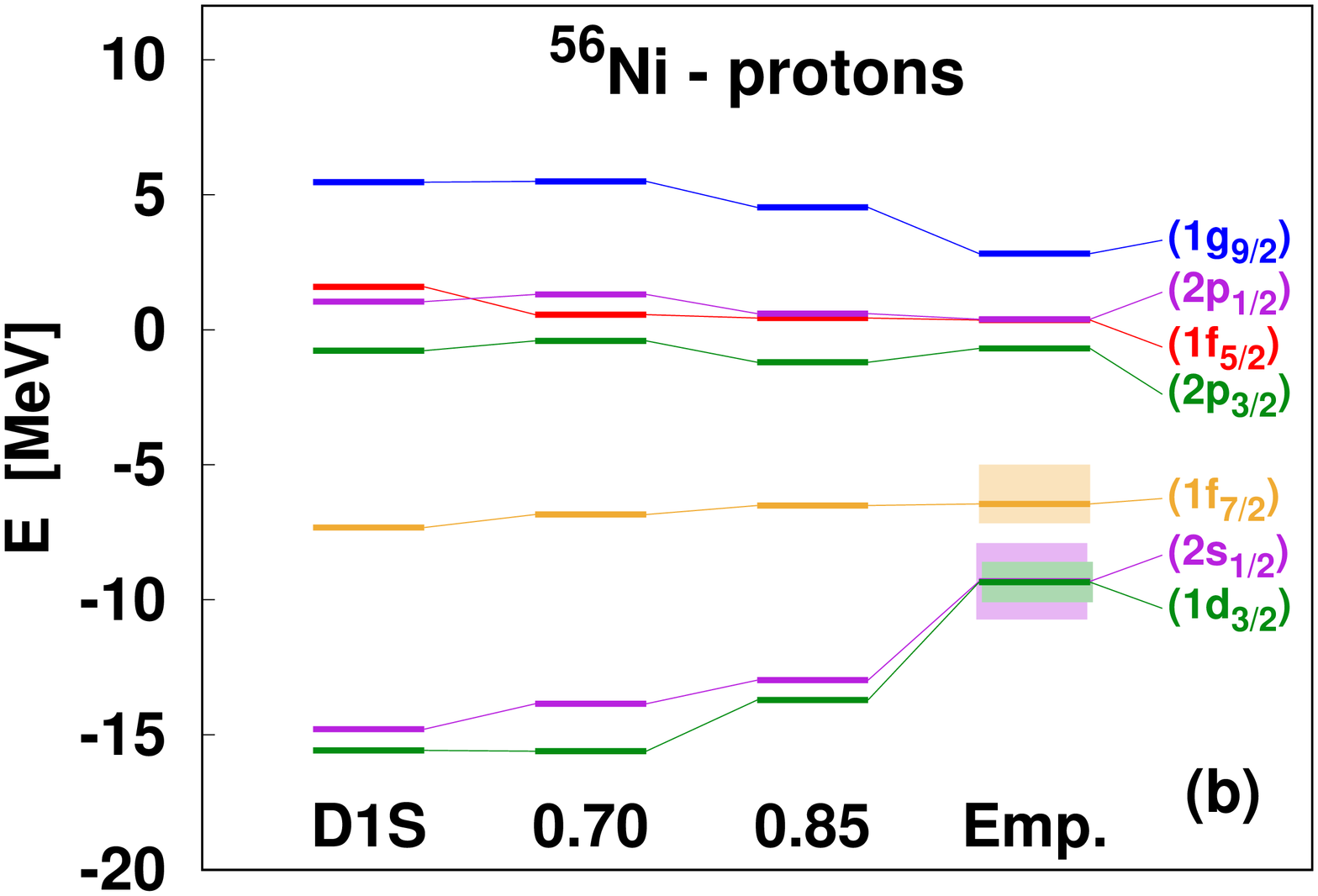} \\
\includegraphics[height=0.47\linewidth,angle=270,viewport=20 2 500 680,clip]{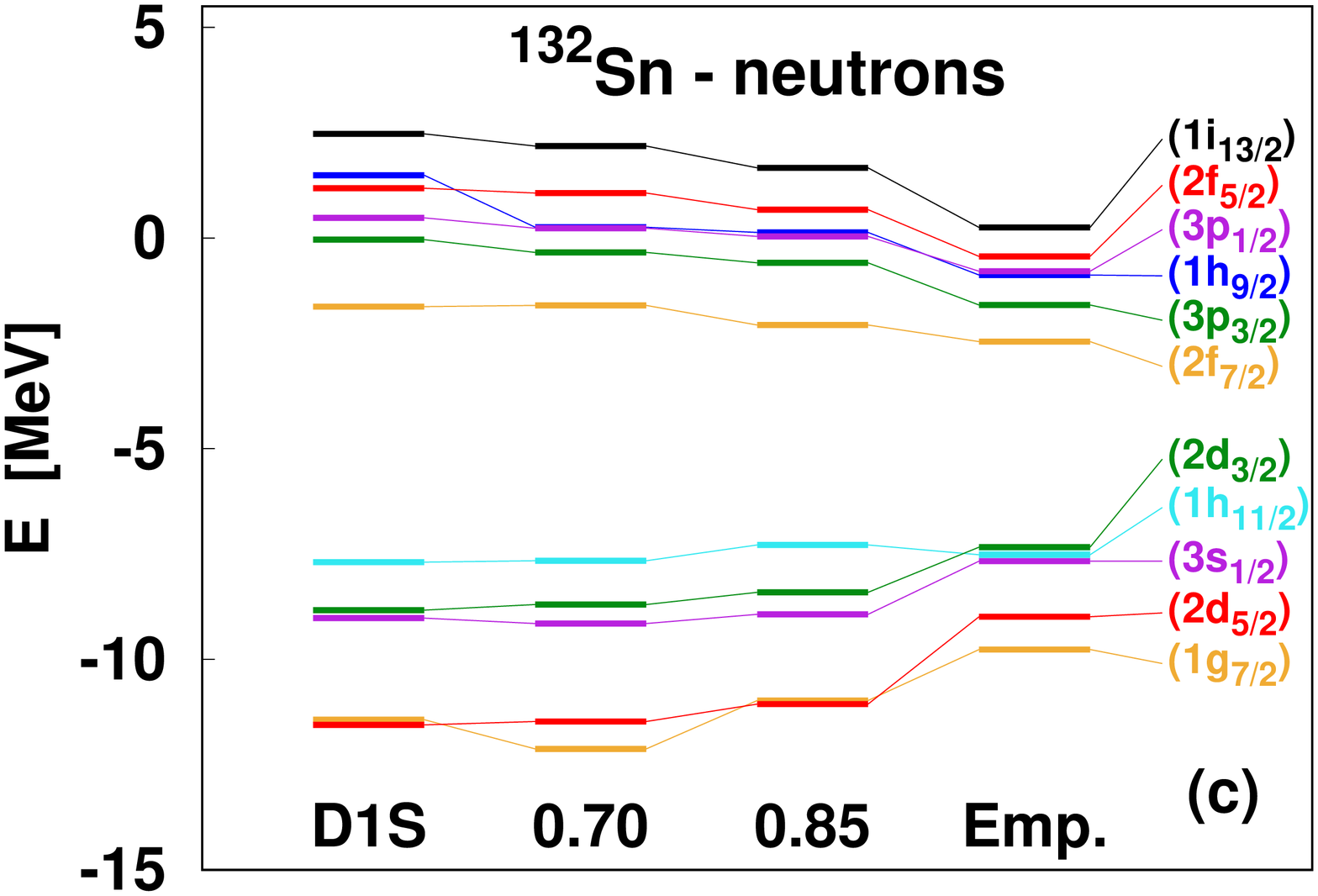} &
\includegraphics[height=0.47\linewidth,angle=270,viewport=20 2 500 680,clip]{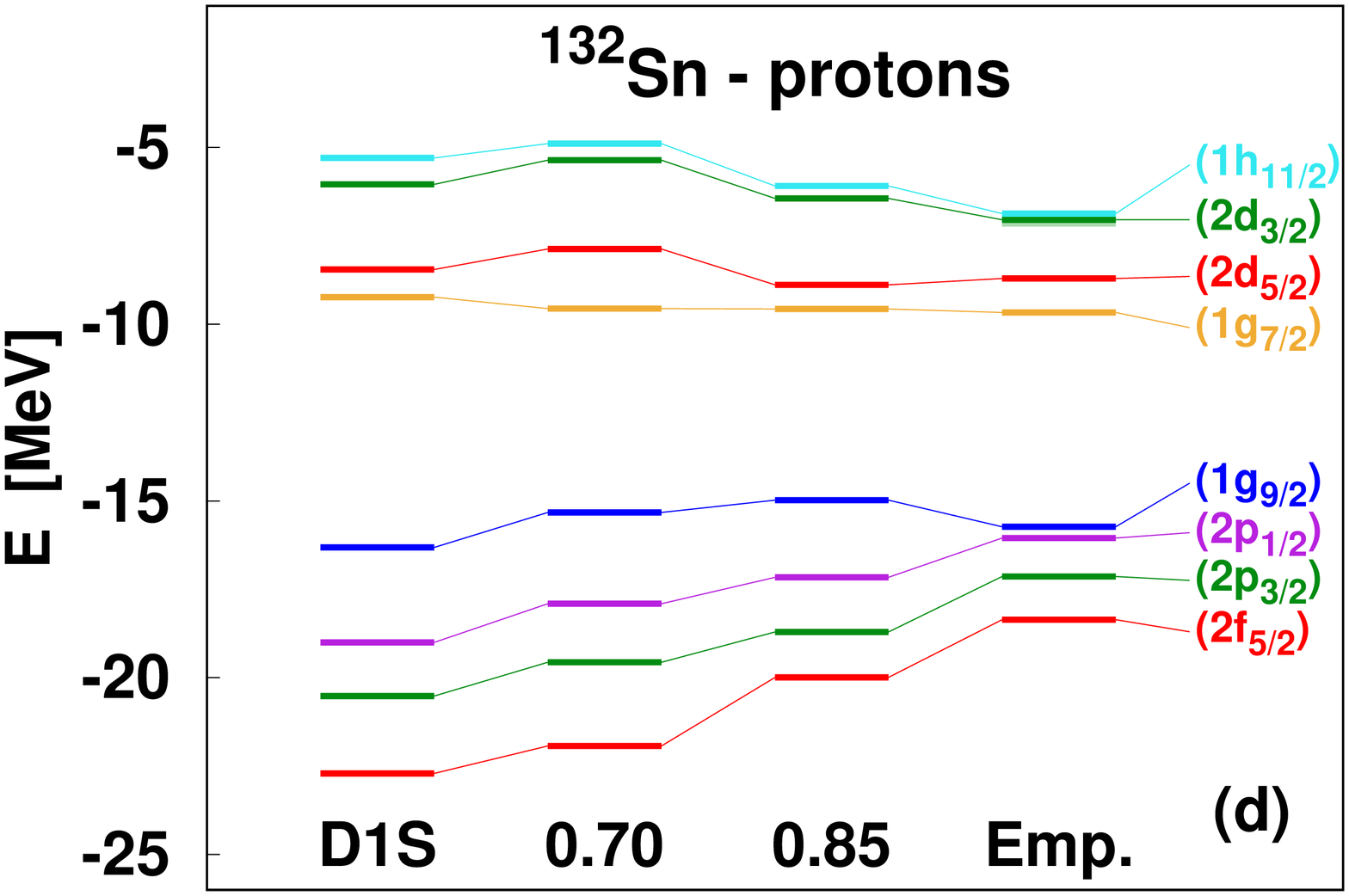} \\
\includegraphics[height=0.47\linewidth,angle=270,viewport=20 2 500 680,clip]{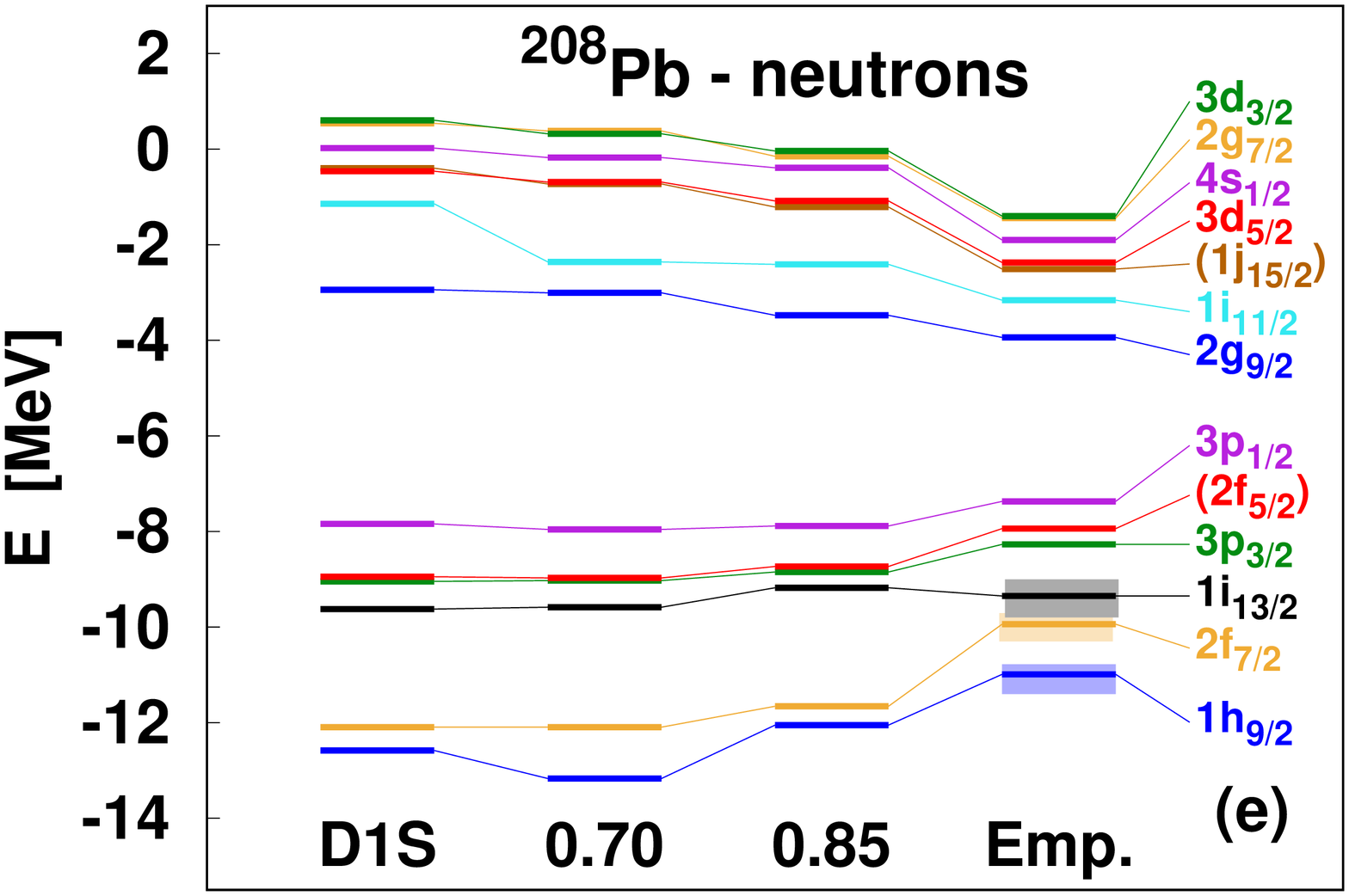} &
\includegraphics[height=0.47\linewidth,angle=270,viewport=20 2 500 680,clip]{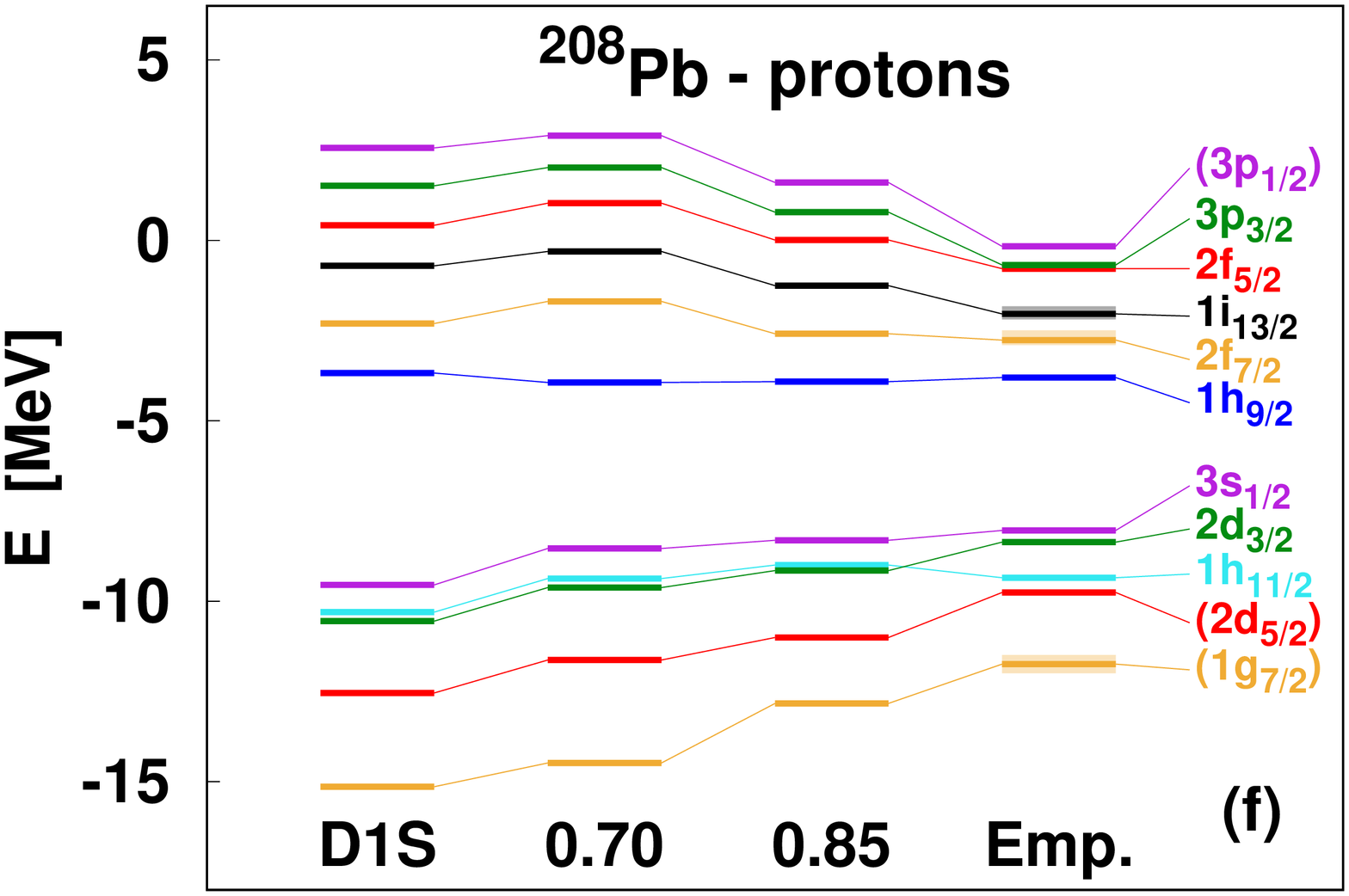}
\end{tabular}
\end{center}
\caption{Same as in Fig.~\protect\ref{fig:spe16o40ca48ca} but for $^{56}$Ni (top),
$^{132}$Sn (middle), and $^{208}$Pb (bottom).
\label{fig:spe56ni132sn208pb}}
\end{figure}

In Figs.~\ref{fig:spe16o40ca48ca} and~\ref{fig:spe56ni132sn208pb}, we show comparison
of single-particle energies calculated in semi-magic nuclei for the D1S~\cite{(Ber91d)},
REG6a\discard{.190617} ($m^*/m=0.70$), and REG6d\discard{.190617} ($m^*/m=0.85$) functionals
with the empirical values taken from the compilation published
in the supplemental material of Ref.~\cite{(Tar14b)}, which contains the single-particle
energies collected within three data sets.
In all panels of Figs.~\ref{fig:spe16o40ca48ca}
and~\ref{fig:spe56ni132sn208pb}, horizontal lines of the rightmost columns represent average
values of the three data sets, whereas
shaded boxes represent spreads between the minimum and maximum values.
Quantum numbers in parentheses indicate single-particle
states with corresponding attributed spectroscopic factors
smaller than 0.8 or unknown.

The spin-orbit interaction corresponding to functional
REG6a\discard{.190617} ($m^*/m=0.70$) is smaller than
that of D1S, which may explain differences between the
single-particle energies of states with large orbital angular
momenta. Differences between the results obtained for functionals
with $m^*/m=0.70$ and $m^*/m=0.85$ mostly amount to a global compression.
Generally speaking, the calculated single-particle energy spacings
are larger than those of the empirical ones, which is typical
for the effective masses being smaller than one.

We note that the comparison between the calculated and empirical
single-particle energies is given here only for the purpose of
illustration. Indeed, both are subjected to uncertainties of
definition and meaning. The calculated ones, which are here
determined as the eigenenergies of the mean-field Hamiltonian, could also be
evaluated from calculated odd-even mass
differences. Similarly, determination of the empirical ones is always
uncertain from the point of view of the fragmentation of the
single-particle strengths. For these reasons, we did not include
single-particle energies in the definition of our penalty
function, see Sec.~\ref{sec:fit}. Nevertheless, positions and
ordering of single-particle energies are crucial for a correct
description of other observables, such as, for example, ground-state
deformations or deformation energies. Therefore, we consider
comparisons presented in Figs.~\ref{fig:spe16o40ca48ca}
and~\ref{fig:spe56ni132sn208pb} to be very useful illustrations of
properties of the underlying EDFs.

\subsection{Deformed nuclei}
\label{sec:Deformed}

\begin{figure}
\begin{center}
\begin{tabular}{c@{\,}c@{\,}c}
\includegraphics[height=0.25\textwidth,angle=0]{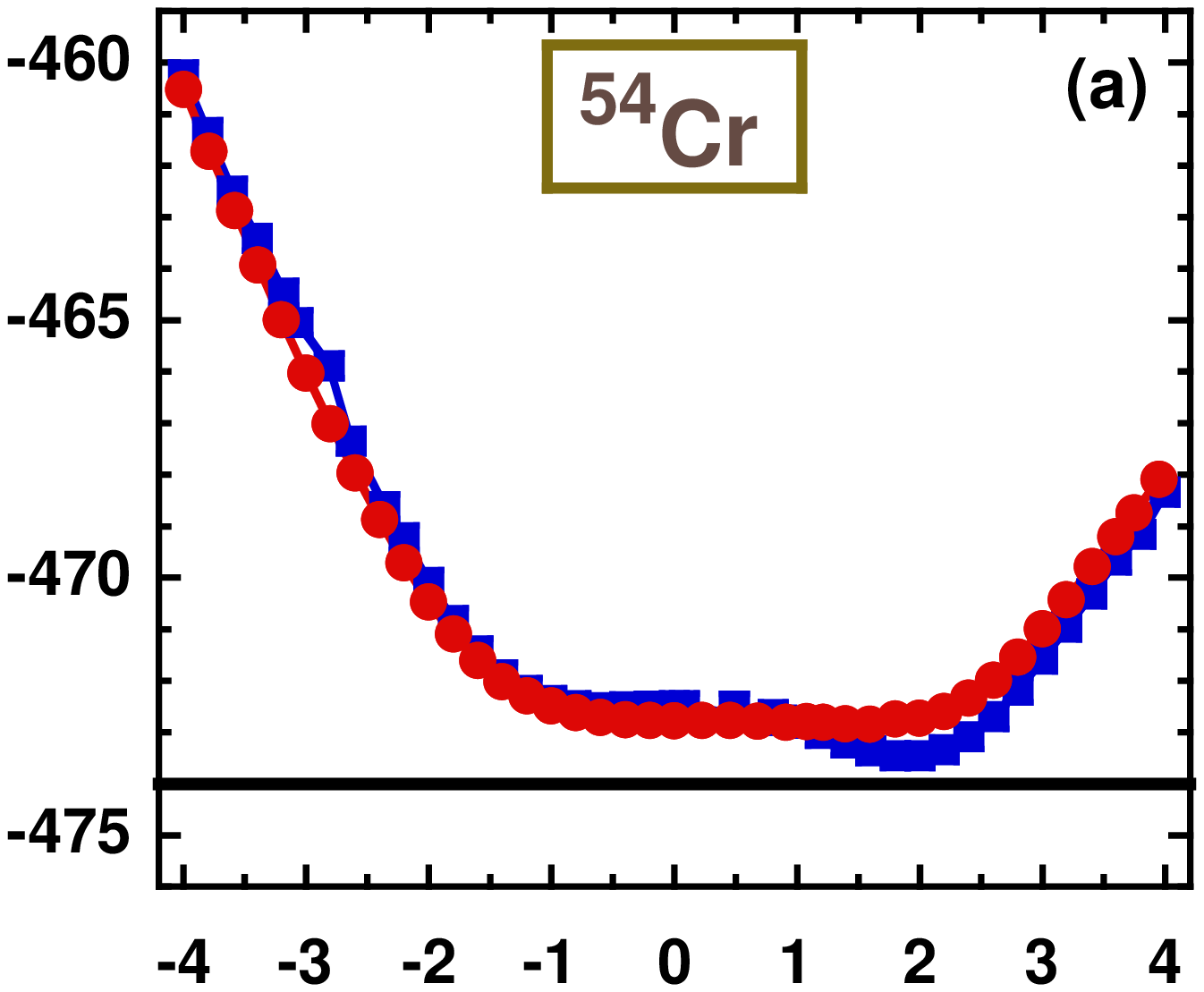} &
\includegraphics[height=0.25\textwidth,angle=0]{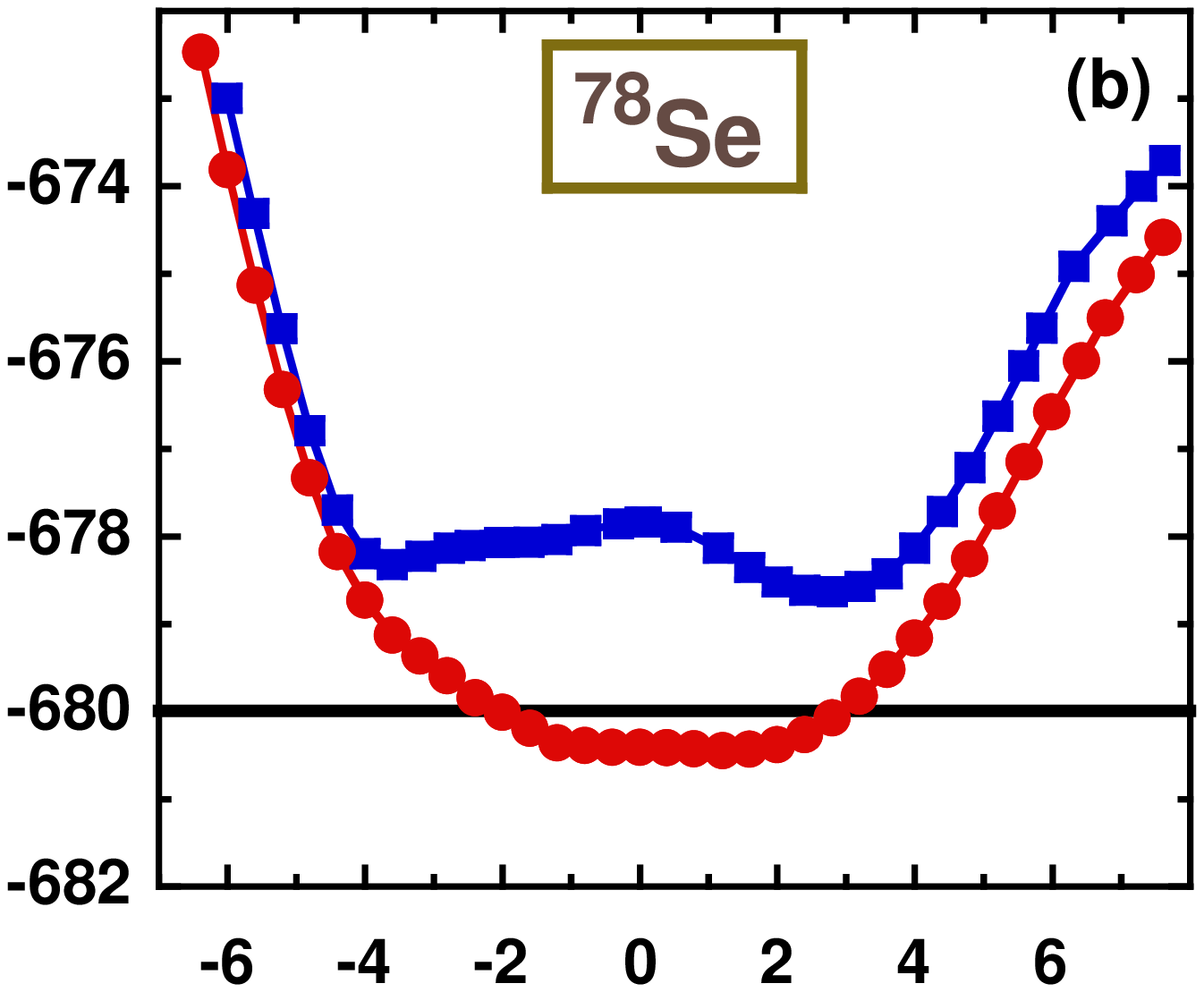} &
\includegraphics[height=0.25\textwidth,angle=0]{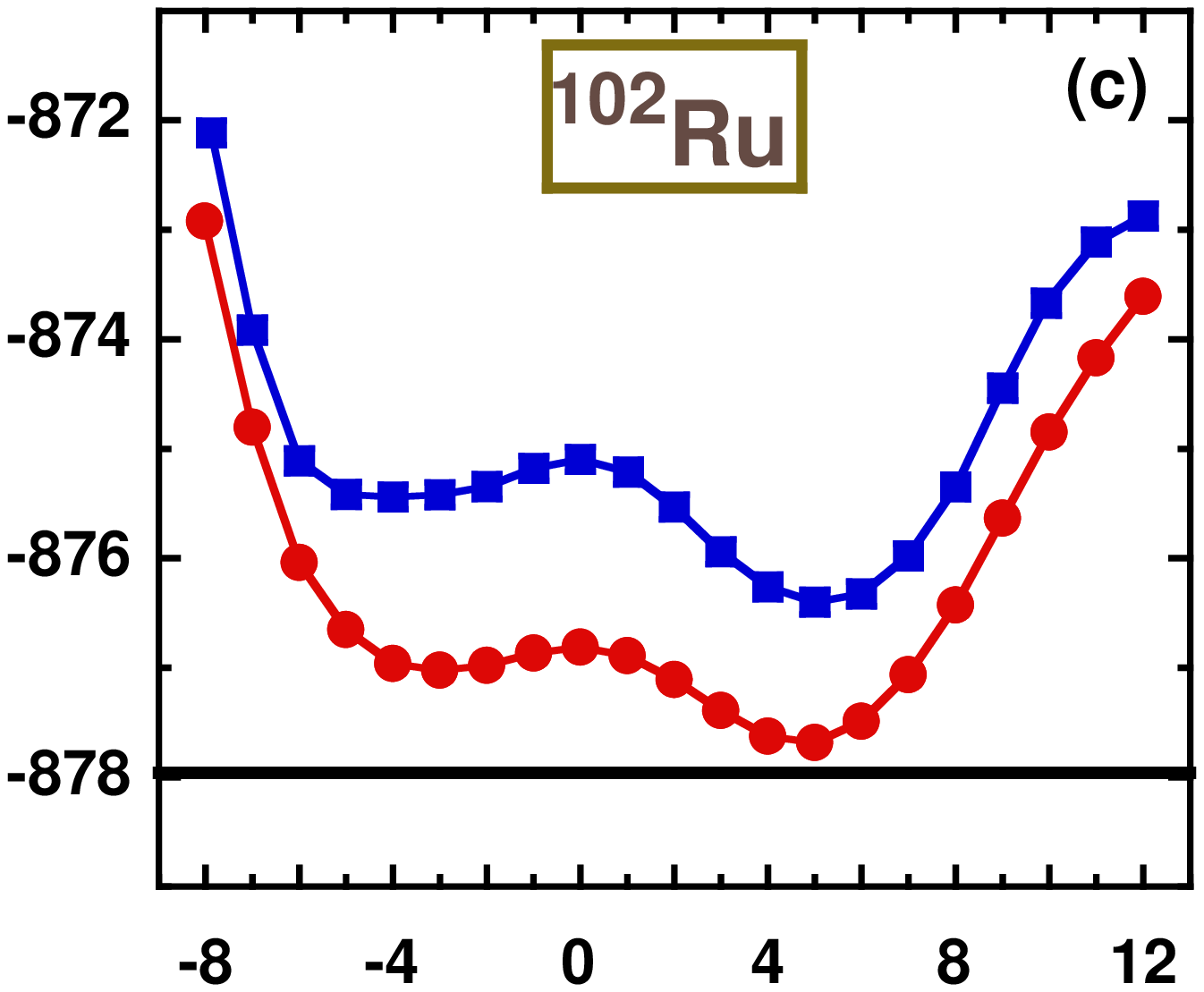} \\
\includegraphics[height=0.25\textwidth,angle=0]{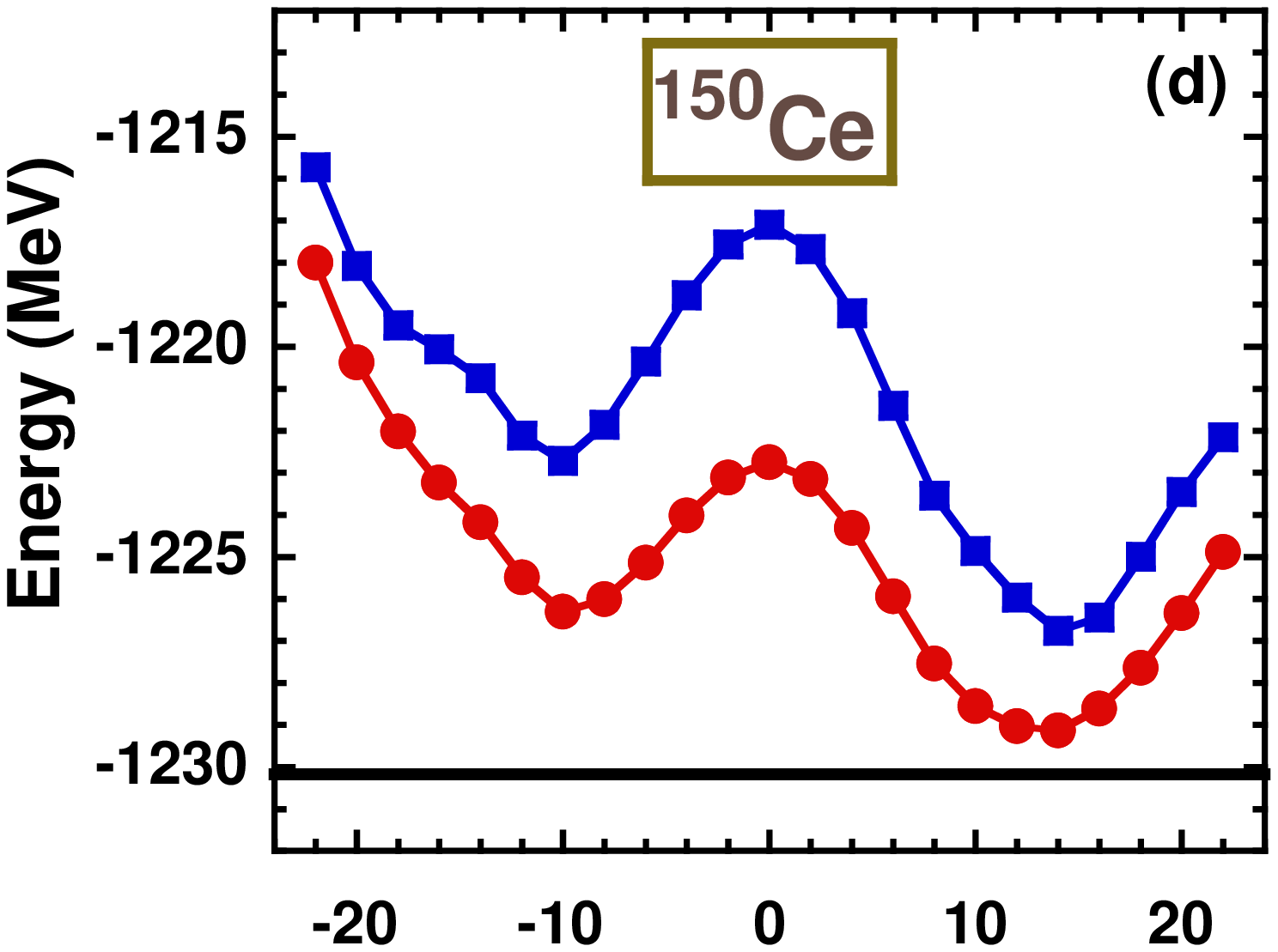} &
\includegraphics[height=0.25\textwidth,angle=0]{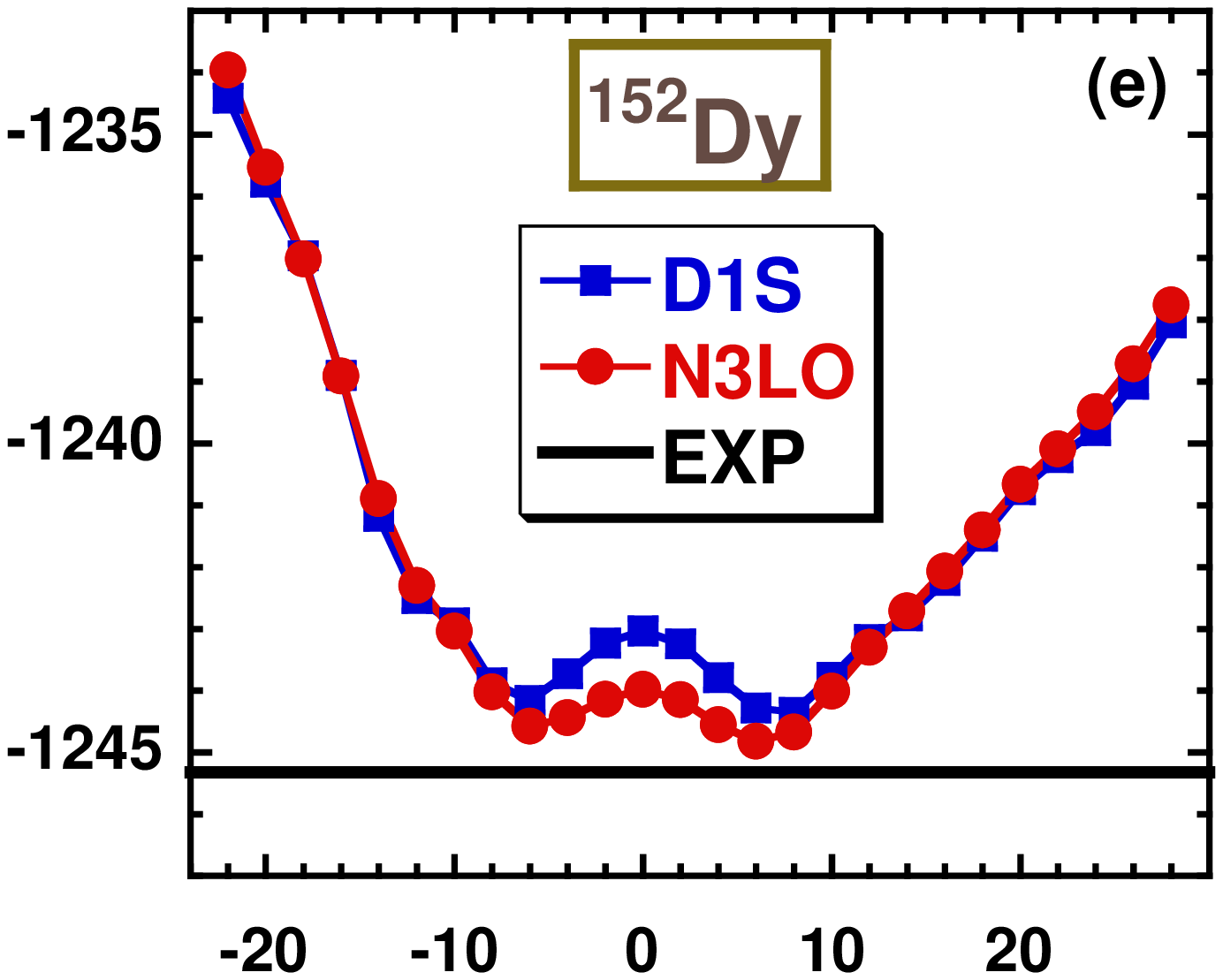} &
\includegraphics[height=0.25\textwidth,angle=0]{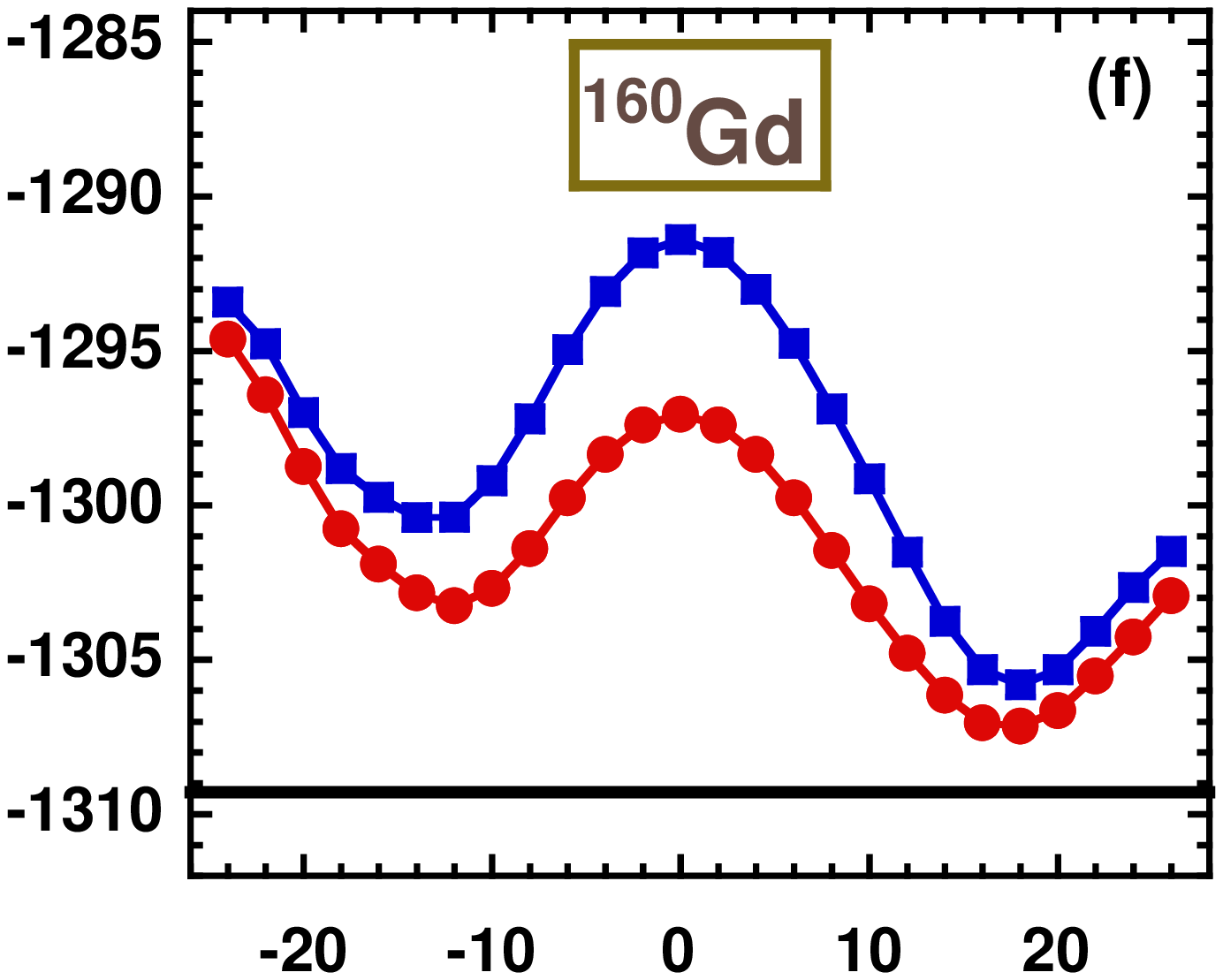} \\
\includegraphics[height=0.28\textwidth,angle=0]{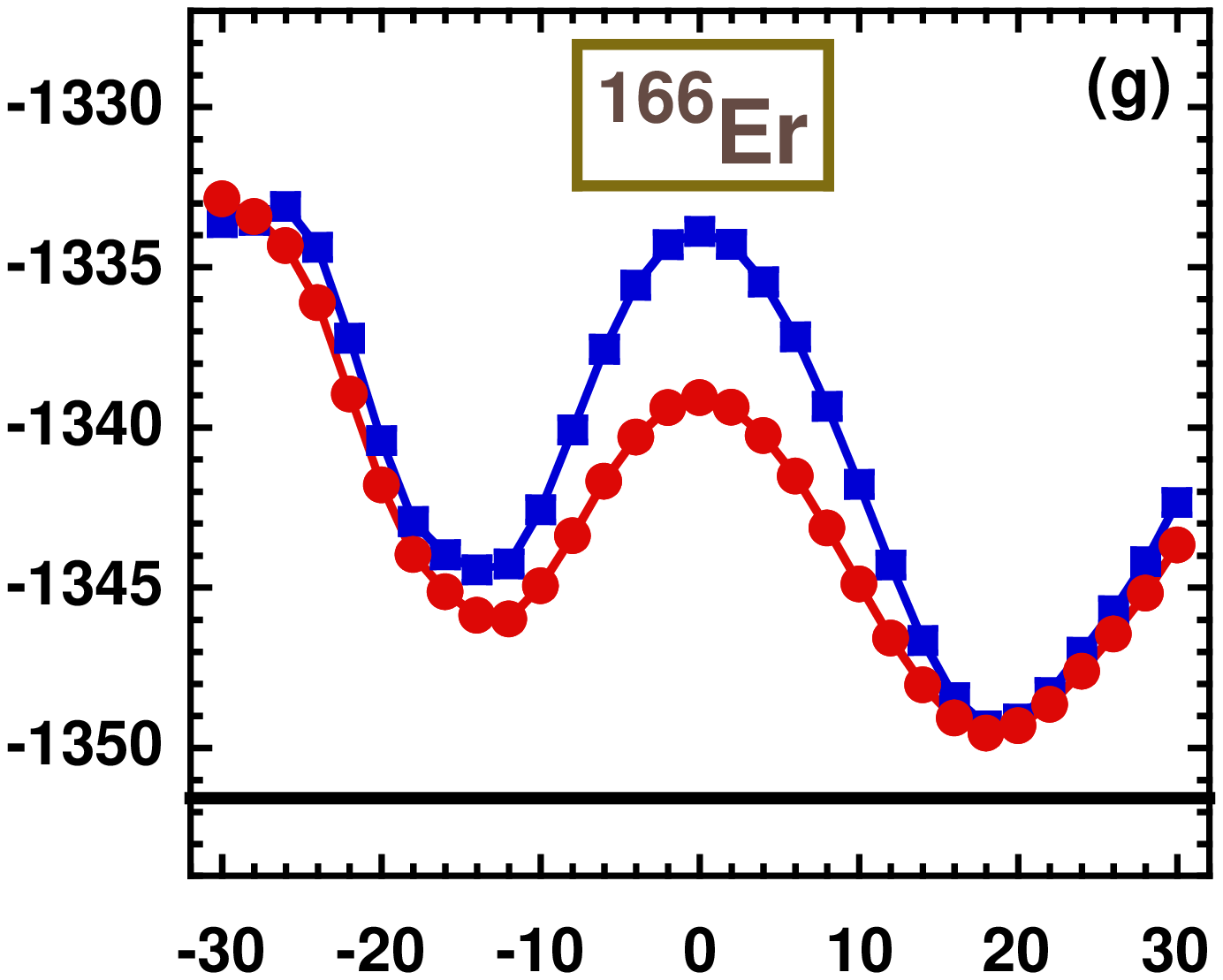} &
\includegraphics[height=0.28\textwidth,angle=0]{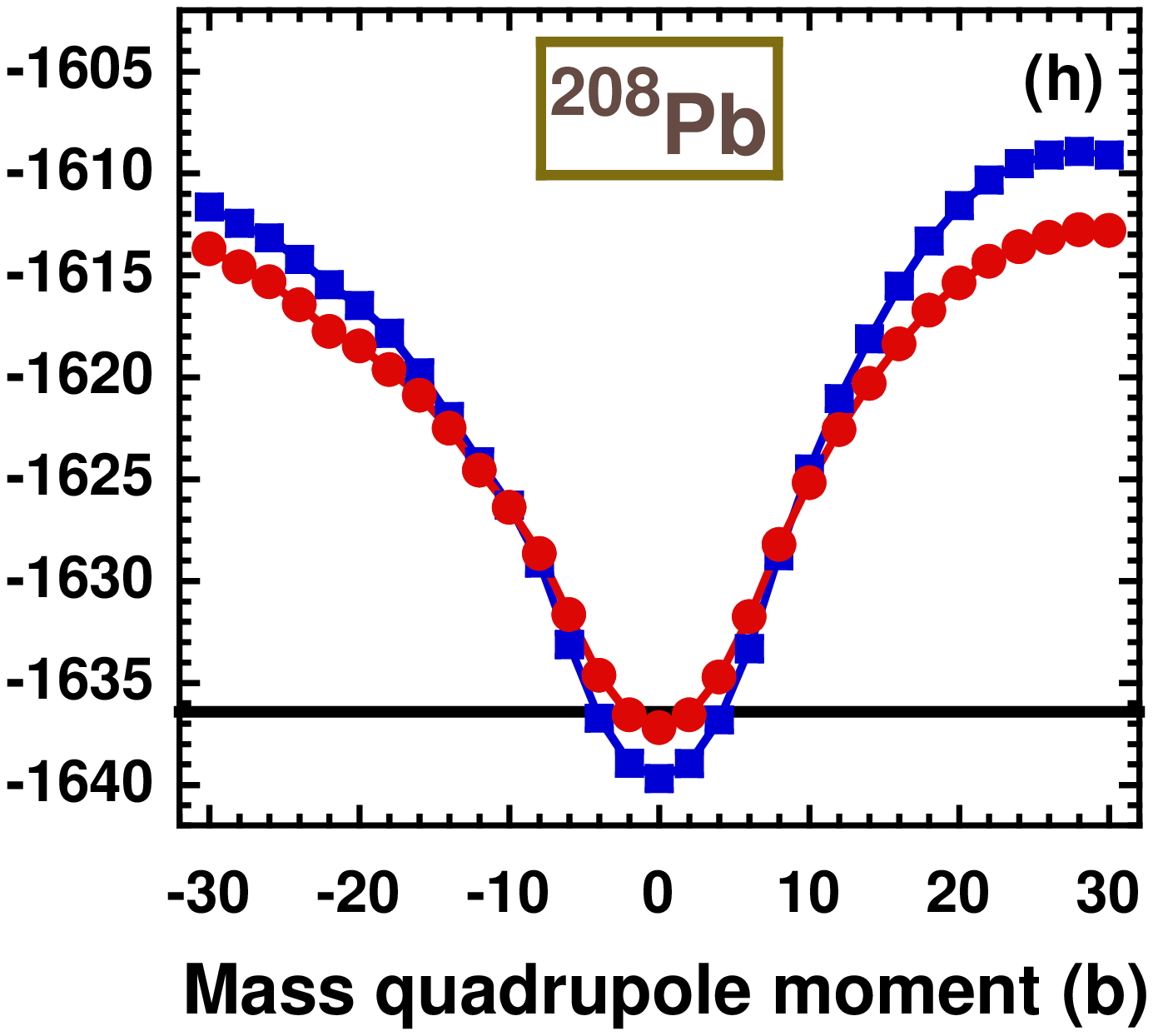} &
\includegraphics[height=0.28\textwidth,angle=0]{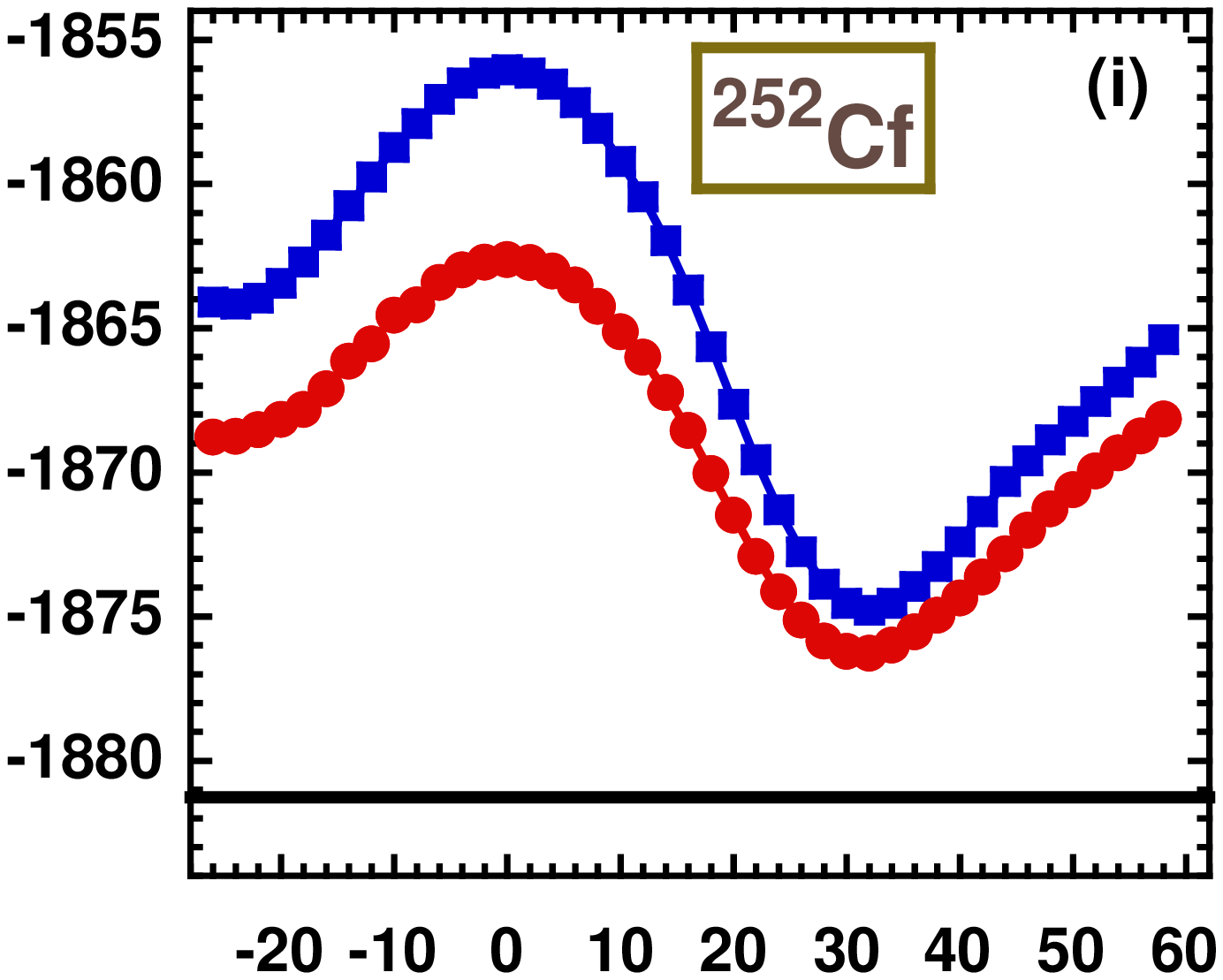} \\
\end{tabular}
\end{center}
\caption{Deformation energies of selected nuclei, calculated for the
D1S~\protect\cite{(Ber91d)} (squares) and N$^3$LO
REG6d\discard{.190617} (circles) functionals, extrapolated to infinite
harmonic-oscillator basis, see~\ref{app:Extrapolation}, plotted in
the absolute energy scale, and compared with the experimental binding
energies~\protect\cite{(Wan17)} (horizontal lines).\label{fig:pes}}
\end{figure}

\begin{figure}
\begin{center}
\begin{tabular}{c@{\,}c}
\includegraphics[height=0.30\textwidth,angle=0]{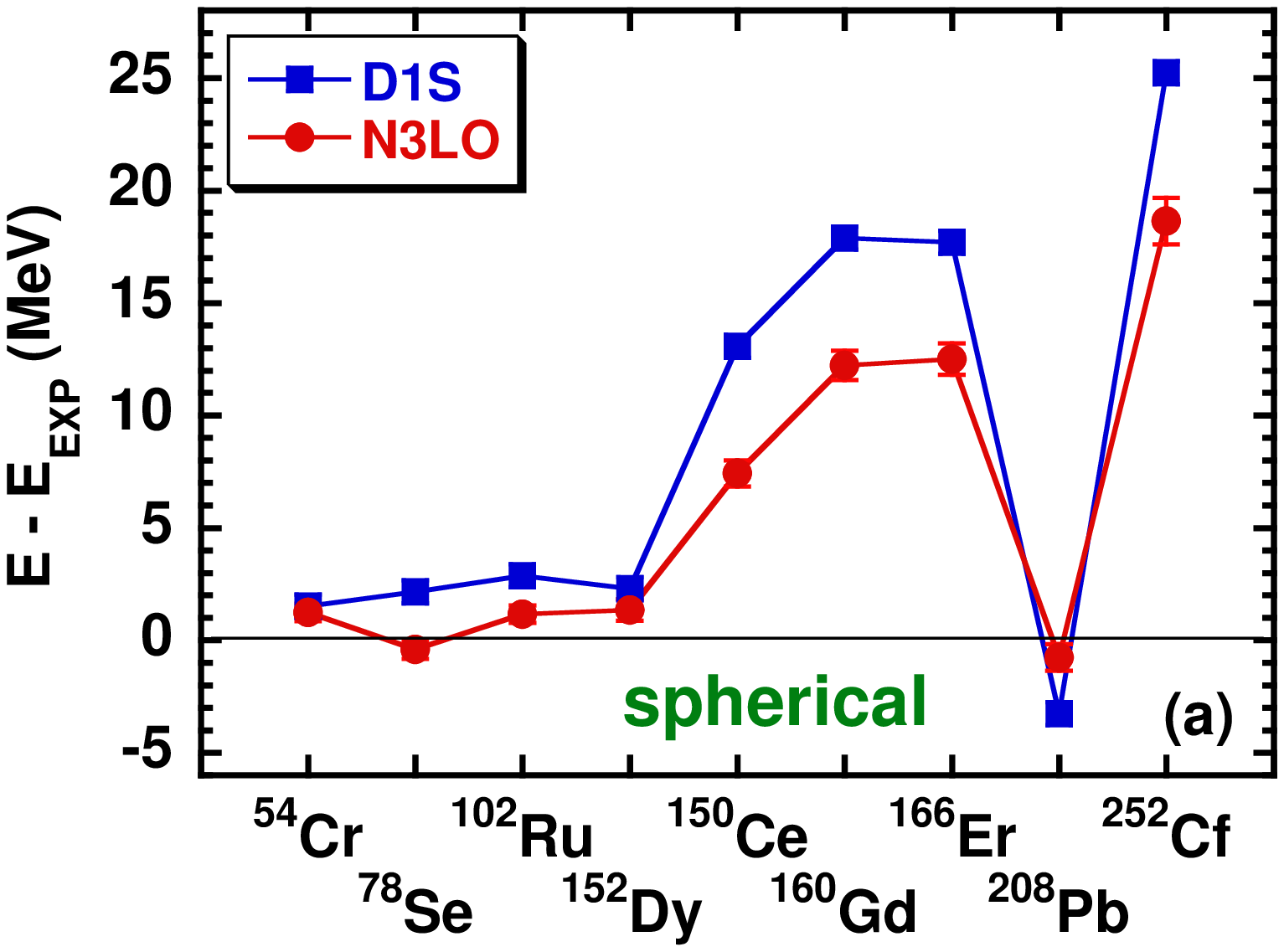} &
\includegraphics[height=0.30\textwidth,angle=0]{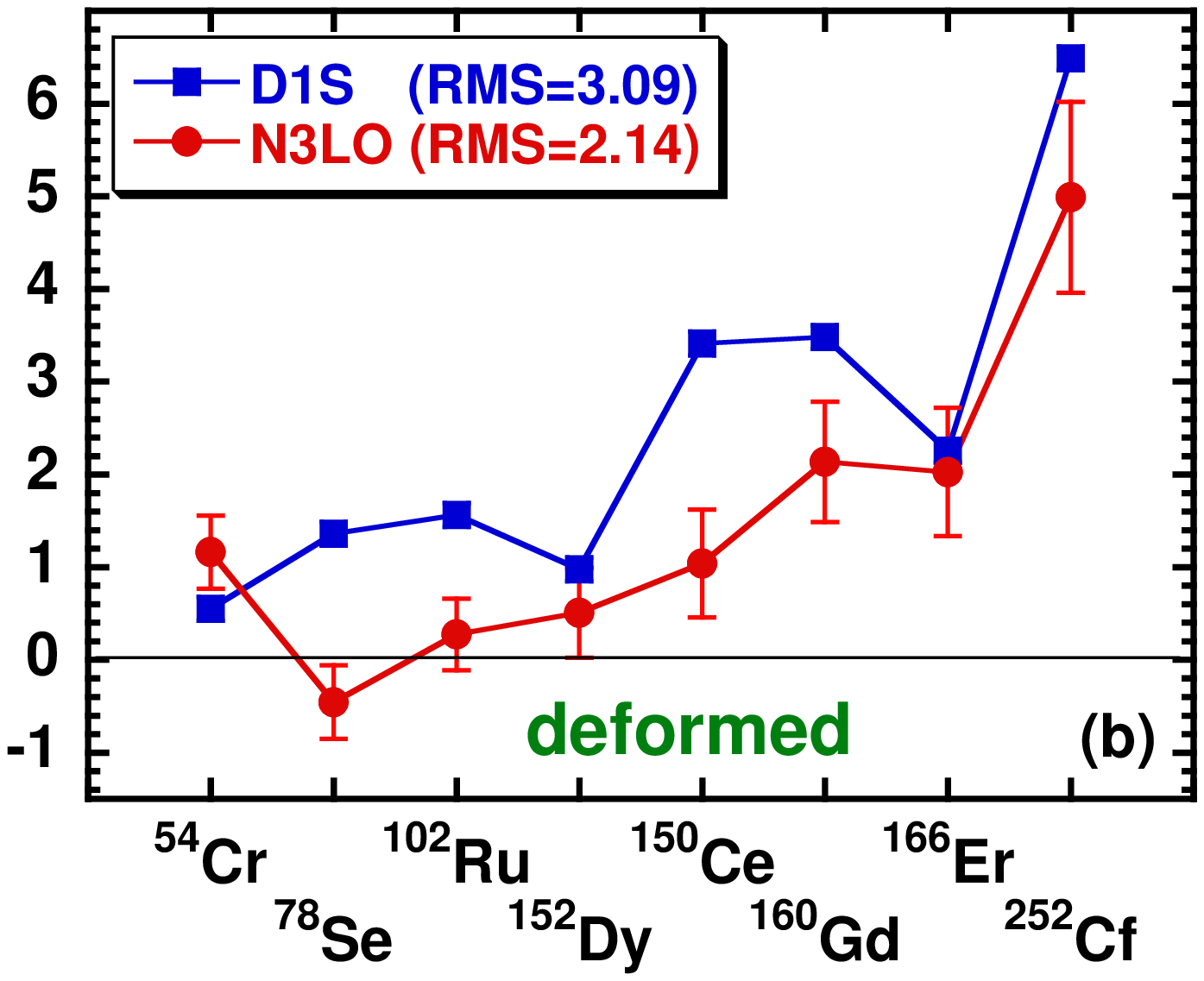}
\end{tabular}
\end{center}
\caption{Same as in Fig.~\protect\ref{fig:pes} but for the
binding-energy residuals calculated at spherical shapes (a) and
deformed minima (b).\label{fig:sph-def}}
\end{figure}

Using the methodology of extrapolating results calculated for
$N_0=16$ harmonic-oscillator shells to infinite $N_0$, presented
in~\ref{app:Extrapolation}, for a set of nine nuclei from $^{54}$Cr
to $^{252}$Cf we determined deformation energies, Fig.~\ref{fig:pes},
and binding-energy residuals, Fig.~\ref{fig:sph-def}. The figures
compare results obtained for the D1S~\protect\cite{(Ber91d)} and
N$^3$LO REG6d\discard{.190617} functionals. For N$^3$LO
REG6d\discard{.190617}, in Fig.~\ref{fig:sph-def}(a) we also show
propagated uncertainties~\cite{(Dob14b)} of spherical energies
determined using the covariance matrix available in the supplemental
material (\suppl). For illustration, the
same propagated uncertainties are also plotted in
Fig.~\ref{fig:sph-def}(b), whereas determination of full propagated
uncertainties of deformed minima is left to a future analysis of
deformed solutions.

In general, the pattern of deformations obtained for both functionals
is very similar. This is gratifying, because deformed nuclei were not included in the
adjustment of either one of the two EDFs. For this
admittedly fairly limited sample of nuclei, the pattern of RMS
binding-energy residuals is fairly analogous to what we have observed
in spherical nuclei, see Sec.~\ref{sec:Binding}, with
REG6d\discard{.190617} giving values that are about 30\% smaller than
those for D1S. It is interesting to see that in several instances,
the two functionals generate absolute energies of the deformed minima
that are more similar to one another than those of the spherical
shapes. In the present study we limit ourselves to presenting results
only for these very few nuclei, whereas attempts of using deformed
nuclei in penalty functions~\cite{(Hav20)} and systematic mass-table
calculations~\cite{(Kor20)} are left to forthcoming publications.

\section{Conclusions}
\label{sec:conclusion}

In this article, we reported on the next step in adjusting parameters
of regularized finite-range functional generators to data. We have
shown that an order-by-order improvement of agreement with data is
possible, and that the sixth order (N$^3$LO) functional describes
data similarly or better than the standard Gogny or Skyrme
functionals.

We implemented adjustments of parameters based on minimizing fairly
complicated penalty function. Our experience shows that a blind
optimization to selected experimental data seldom works. Instead, one
has to implement sophisticated constraints, which prevent wandering of
parameters towards regions where different kinds of instabilities loom.

We consider the process of developing new functionals and adjusting
their parameters a continuous effort to better their precision
and predictive power. At the expense of introducing single
density-dependent generator, here we were able to raise the values of
the effective mass, obtained in our previous study~\cite{(Ben17a)},
well above those that are achievable with purely two-body
density-independent generators~\cite{(Dav18)}. Such a solution is
perfectly satisfactory at the single-reference level. However, for
multi-reference implementations, the density-dependent term must be
replaced by second-order three-body zero-range generators~\cite{(Sad13e)},
or otherwise entirely new yet unknown approach would be required.

Although a definitive conclusion about usefulness of EDFs obtained in
this study can only be drawn after a comparison of observables of
more diverse nature, this class of pseudopotentials looks promising,
even if it can clearly be further improved. In the future, we plan on
continuing novel developments by implementing non-local regularized
pseudopotentials along with their spin-orbit and tensor
terms~\cite{(Rai14a)}. This may allow us to fine-tune spectroscopic
properties of functionals and facilitate precise description of
deformed and odd nuclei.

\appendix

\section{Decomposition of the potential energy in $(S,T)$ channels}
\label{app:st}

The techniques to derive decomposition of the potential energy
per particle, $E^{(S,T)}_\mathrm{pot}/A$, into four spin-isospin $(S,T)$ channels are
the same for finite-range and zero-range pseudopotentials.
Therefore, we do not repeat here the details of the derivation, which
can be found, for example, in Ref.~\cite{(Les06)}.

First, we recall the expression for the auxiliary function $F_0(\xi)$, already introduced
in Ref.~\cite{(Ben14b)},
\begin{equation}
F_0(\xi)=\frac{12}{\xi^3}\left[
\frac{1-\rme^{-\xi^2}}{\xi^3}-\frac{3-\rme^{-\xi^2}}{2\xi}
+\frac{\sqrt{\pi}}{2}\,\mathrm{Erf}\,\xi\right]\,.
\end{equation}
Then, in the symmetric infinite nuclear matter with Fermi momentum $k_F$
and density $\rho_0=2k_F^3/3\pi^2$,
contributions of the finite-range local pseudopotential (\ref{eq:locpot}) at order zero
($n=0$) to $(S,T)$ channels can be expressed as:
\begin{eqnarray}
\frac{E^{(0,0)}_\mathrm{pot}}{A}&=&
\frac{1}{32}\left(W_1^{(0)}-B_1^{(0)}+H_1^{(0)}-M_1^{(0)}\right)\rho_0\left[
1-F_0(k_Fa)\right]\,, \\
\frac{E^{(0,1)}_\mathrm{pot}}{A}&=&
\frac{3}{32}\left(W_1^{(0)}-B_1^{(0)}-H_1^{(0)}+M_1^{(0)}\right)\rho_0\left[
1+F_0(k_Fa)\right]\,, \\
\frac{E^{(1,0)}_\mathrm{pot}}{A}&=&
\frac{1}{32}\left(W_1^{(0)}+B_1^{(0)}+H_1^{(0)}+M_1^{(0)}\right)\rho_0\left[
1+F_0(k_Fa)\right]\,, \\
\frac{E^{(1,1)}_\mathrm{pot}}{A}&=&
\frac{9}{32}\left(W_1^{(0)}+B_1^{(0)}-H_1^{(0)}-M_1^{(0)}\right)\rho_0\left[
1-F_0(k_Fa)\right]\,,
\end{eqnarray}
and those at higher orders $n$ as:
\begin{eqnarray}
\hspace*{-1cm}
\frac{E^{(0,0)}_\mathrm{pot}}{A}&=&-
\frac{1}{32}\left(W_1^{(n)}-B_1^{(n)}+H_1^{(n)}-M_1^{(n)}\right)\rho_0
\left(-\frac{1}{a}\frac{\partial}{\partial a}\right)^{n/2}F_0(k_Fa)\,, \\
\hspace*{-1cm}
\frac{E^{(0,1)}_\mathrm{pot}}{A}&=&~~~~
\frac{3}{32}\left(W_1^{(n)}-B_1^{(n)}-H_1^{(n)}+M_1^{(n)}\right)\rho_0
\left(-\frac{1}{a}\frac{\partial}{\partial a}\right)^{n/2}F_0(k_Fa)\,, \\
\hspace*{-1cm}
\frac{E^{(1,0)}_\mathrm{pot}}{A}&=&~~~~
\frac{1}{32}\left(W_1^{(n)}+B_1^{(n)}+H_1^{(n)}+M_1^{(n)}\right)\rho_0
\left(-\frac{1}{a}\frac{\partial}{\partial a}\right)^{n/2}F_0(k_Fa)\,, \\
\hspace*{-1cm}
\frac{E^{(1,1)}_\mathrm{pot}}{A}&=&-
\frac{9}{32}\left(W_1^{(n)}+B_1^{(n)}-H_1^{(n)}-M_1^{(n)}\right)\rho_0
\left(-\frac{1}{a}\frac{\partial}{\partial a}\right)^{n/2}F_0(k_Fa)\,.
\end{eqnarray}

\section{Estimate of the surface energy coefficient}
\label{app:asurf}

In section~\ref{sec:fit}, we introduced a constraint on the estimate
of the surface energy coefficient $a_\mathrm{surf}^\mathrm{LDM}$,
calculated with a liquid-drop-type formula. In the case of local
functionals (such as Skyrme functionals), to calculate the surface
energy coefficient~\cite{(Jod16)}, several approaches can
be considered, such as the Hartree-Fock (HF)
calculation~\cite{COTE1978104}, approximation of the Extended Thomas
Fermi (ETF) type~\cite{BRACK1985275} or Modified Thomas Fermi (MTF)
type~\cite{KRIVINE1979212}, or within a leptodermous protocol,
which is based on an analysis of calculations performed for very
large fictitious nuclei~\cite{PhysRevC.73.014309}.

Some of these approaches are not usable with the regularized pseudopotentials
considered in this article. Indeed, the ETF and MTF methods can only be used
for functionals that depend on local densities. In principle, the leptodermous
protocol could be used, but it would require a significant expense in CPU time.
Moreover, the HF calculations cannot be considered because the Friedel
oscillations of the density make the extraction of a
stable and precise value of the surface energy coefficient very difficult (see
discussion in Ref.~\cite{(Jod16)} and references therein).

Therefore, for the purpose of performing parameter adjustments, we
decided to use a very simple estimate of the surface energy
coefficient, which is usable with any kind of functional. After
determining the self-consistent total binding energy $E$ of a
fictitious symmetric, spin-saturated, and unpaired $N=Z=40$ nucleus without
Coulomb interaction, we used a simple liquid-drop formula
to calculate the surface energy coefficient,
\begin{equation}
\label{surf}
a_\mathrm{surf}^\mathrm{LDM}=\frac{E-a_v A}{A^{2/3}} ,
\end{equation}
where $A=80$ and $a_v$ is the volume energy coefficient in symmetric infinite
nuclear matter at the saturation point.

Values of $a_\mathrm{surf}^\mathrm{LDM}$ obtained in this way do depends on $A$,
but, at least in the case Skyrme functionals, they vary linearly with
the surface energy coefficients obtained using full HF calculations.
Detailed study of the usability of estimates (\ref{surf}) will be the
subject of a forthcoming publication~\cite{dacosta}.

In section~\ref{sec:fit}, we used the value of
$a_\mathrm{surf}^\mathrm{LDM}=18.5$\,MeV as the target value of the
parameter adjustments. This value is only slightly below the value
obtained for the Skyrme functional SLy5s1 (18.6\,MeV), which is an
improved version of the SLy5 functional with optimized surface
properties~\cite{(Jod16),PhysRevC.99.044315}. This target
value we used was only an educated guess, and it may require
fine-tuning after a systematic study of the properties of deformed
nuclei will have been performed.

\section{Parameters of the pseudopotentials}
\label{app:Parameters}

Parameters of the pseudopotentials used in Sec.~\ref{sec:res},
REG2d\discard{.190617} at NLO, REG4d\discard{.190617} at N$^2$LO, and
REG6d\discard{.190617} at N$^3$LO with $m^*/m\simeq0.85$, and
REG6a\discard{.190617} at N$^3$LO with $m^*/m\simeq0.70$ are reported in
Table~\ref{tab:para} along with their statistical uncertainties. As
it turns out, values of parameters rounded to the significant digits,
which would be consistent with the statistical uncertainties, give
results visibly different than those corresponding to unrounded
values. Therefore, in the Table we give all parameters up to the
sixth decimal figure. Moreover, the statistical uncertainties of
parameters are only given for illustration, whereas the propagated
uncertainties of observables have to evaluated using the full
covariance matrices~\cite{(Dob14b)}. Parameters of other
pseudopotentials derived in this study along with the covariance
matrix corresponding to REG6d\discard{.190617} are listed in the
supplemental material (\suppl).

\begin{table}
\caption{Parameters
$a$ (in fm),
$t_3$ (in MeV\,fm$^4$),
$W_\mathrm{SO}$ (in MeV\,fm$^5$), and
$W_1^{(n)}$,
$B_1^{(n)}$,
$H_1^{(n)}$, and
$M_1^{(n)}$ (in MeV\,fm$^{3+n}$),
of pseudopotentials REG2d\discard{.190617},     REG4d\discard{.190617},
                    REG6d\discard{.190617}, and REG6a\discard{.190617}
with statistical uncertainties.
\label{tab:para}}
\begin{center}
\begin{tabular}{@{}l@{~}r@{~~}r@{~~}r@{~~}r@{}}
\br
                & \multicolumn{1}{c}{REG2d\discard{.190617}}
                & \multicolumn{1}{c}{REG4d\discard{.190617}}
                & \multicolumn{1}{c}{REG6d\discard{.190617}}
                & \multicolumn{1}{c}{REG6a\discard{.190617}}         \\
                & \multicolumn{1}{c}{(NLO)}
                & \multicolumn{1}{c}{(N$^2$LO)}
                & \multicolumn{1}{c}{(N$^3$LO)}
                & \multicolumn{1}{c}{(N$^3$LO)}                     \\
\mr
$a$             & \multicolumn{1}{c}{0.85}
                & \multicolumn{1}{c}{1.15}
                & \multicolumn{1}{c}{1.50}
                & \multicolumn{1}{c}{1.60}                          \\
\mr
$t_3$           & $11516.477663( 0.5)$ & $11399.197904 (0.1)$ & $11509.501921(0.3)$ & $ 9521.936183(0.3) $   \\
$W_\mathrm{SO}$ & $  106.098237( 2.8)$ & $  115.427981 (2.2)$ & $  116.417478(1.9)$ & $  122.713008(1.9) $   \\
\mr
$W_1^{(0)}$     & $-2510.198547( 3.6)$ & $  -689.651657(2.4)$ & $-2253.706132(0.5)$ & $-1478.053786(0.9) $   \\
$B_1^{(0)}$     & $ 1108.303995(10.0)$ & $  -824.881825(6.4)$ & $  740.258749(1.9)$ & $   87.165128(2.4) $   \\
$H_1^{(0)}$     & $-2138.673166( 2.2)$ & $  -247.692094(1.1)$ & $-1794.716098(2.2)$ & $-1031.141021(2.3) $   \\
$M_1^{(0)}$     & $  746.778833( 1.6)$ & $ -1270.827895(2.2)$ & $  282.583629(1.0)$ & $ -362.705492(1.3) $   \\
\mr
$W_1^{(2)}$     & $ -637.749560( 3.7)$ & $  -741.229448(2.0)$ & $-3207.567147(2.1)$ & $-2459.995595(2.2) $   \\
$B_1^{(2)}$     & $  210.327285( 3.1)$ & $   434.961848(2.6)$ & $ 2368.246502(1.2)$ & $ 1412.933291(1.6) $   \\
$H_1^{(2)}$     & $ -892.452162( 2.3)$ & $  -951.018473(1.0)$ & $-3163.516190(0.8)$ & $-2418.882336(0.8) $   \\
$M_1^{(2)}$     & $  379.274480( 7.1)$ & $   615.750351(2.0)$ & $ 2319.605187(0.4)$ & $ 1370.885213(0.6) $   \\
\mr
$W_1^{(4)}$     &                      & $   442.206742(3.8)$ & $  559.364051(1.3)$ & $  835.586806(2.1) $   \\
$B_1^{(4)}$     &                      & $  -972.382568(2.9)$ & $-1398.820389(0.9)$ & $-1594.561771(1.2) $   \\
$H_1^{(4)}$     &                      & $   420.867921(4.9)$ & $  351.670752(1.3)$ & $  774.195095(1.9) $   \\
$M_1^{(4)}$     &                      & $  -953.687931(2.3)$ & $-1197.878374(0.3)$ & $-1535.312092(0.5) $   \\
\mr
$W_1^{(6)}$     &                      &                      & $-1603.038264(2.8)$ & $-1700.366589(3.8) $   \\
$B_1^{(6)}$     &                      &                      & $  828.626124(2.0)$ & $  904.903410(2.8) $   \\
$H_1^{(6)}$     &                      &                      & $-1581.833642(2.0)$ & $-1705.515946(2.4) $   \\
$M_1^{(6)}$     &                      &                      & $  802.445641(1.9)$ & $  914.640430(2.5) $   \\
\br
\end{tabular}
\end{center}
\end{table}

\clearpage

\section{Extrapolation of binding energies of deformed nuclei to infinite harmonic-oscillator basis}
\label{app:Extrapolation}

In this study, results for spherical nuclei were obtained using code {\sc finres}$_4$~\cite{finres}, which
solves HFB equations for finite-range generators on a mesh of points in spherical space coordinates.
Because of the
spherical symmetry, it is perfectly possible to perform calculations with a mesh
dense enough and a number of partial waves high enough for the results to be
stable with respect to any change of the numerical conditions. Results for deformed
nuclei were obtained using the 3D code {\sc hfodd} (v2.92a)~\cite{(Sch17c),(Dob20)}
or axial-symmetry code {\sc hfbtemp}~\cite{hfbtemp}. These two codes solve HFB
equations by expanding single-particle wave functions on harmonic-oscillator bases.
Since for deformed nuclei the amount of CPU time and memory is much
larger than for spherical ones, it is not practically possible to use enough major harmonic-oscillator
shells to reach the asymptotic regime, especially for heavy nuclei.

In order to estimate what would be the converged asymptotic value of the total
binding energy of a given deformed nucleus, we proceeded in the following way.
First, using code {\sc finres}$_4$, we determined the total binding energy $E_\mathrm{sph}$
of a given nucleus at the spherical point. Second,
using codes {\sc hfodd} or {\sc hfbtemp}, for the same nucleus and for a given number of shells $N_0$,
we determined total binding energies
$E_\mathrm{sph}(N_0)$ (constrained to sphericity) and  $E_\mathrm{def}(N_0)$ (constrained to a non-zero deformation).
Third, we assumed that with $N_0\rightarrow\infty$,
the deformation energy $\Delta E_\mathrm{def}(N_0) = E_\mathrm{def}(N_0)-E_\mathrm{sph}(N_0)$
converges much faster to its asymptotic value than either of the two energies.
Fourth, within this assumption, we estimate the asymptotic energies of deformed nuclei as
\begin{equation}
\label{eq:deformation-energy}
E_\mathrm{def}(N_0=\infty) = E_\mathrm{sph} + \Delta E_\mathrm{def}(N_0)
                           = E_\mathrm{def}(N_0) + E_\mathrm{sph} - E_\mathrm{sph}(N_0) .
\end{equation}

\begin{figure}
\begin{center}
\begin{tabular}{c@{~\,}c@{\,~}c}
\includegraphics[height=0.275\textwidth,angle=0]{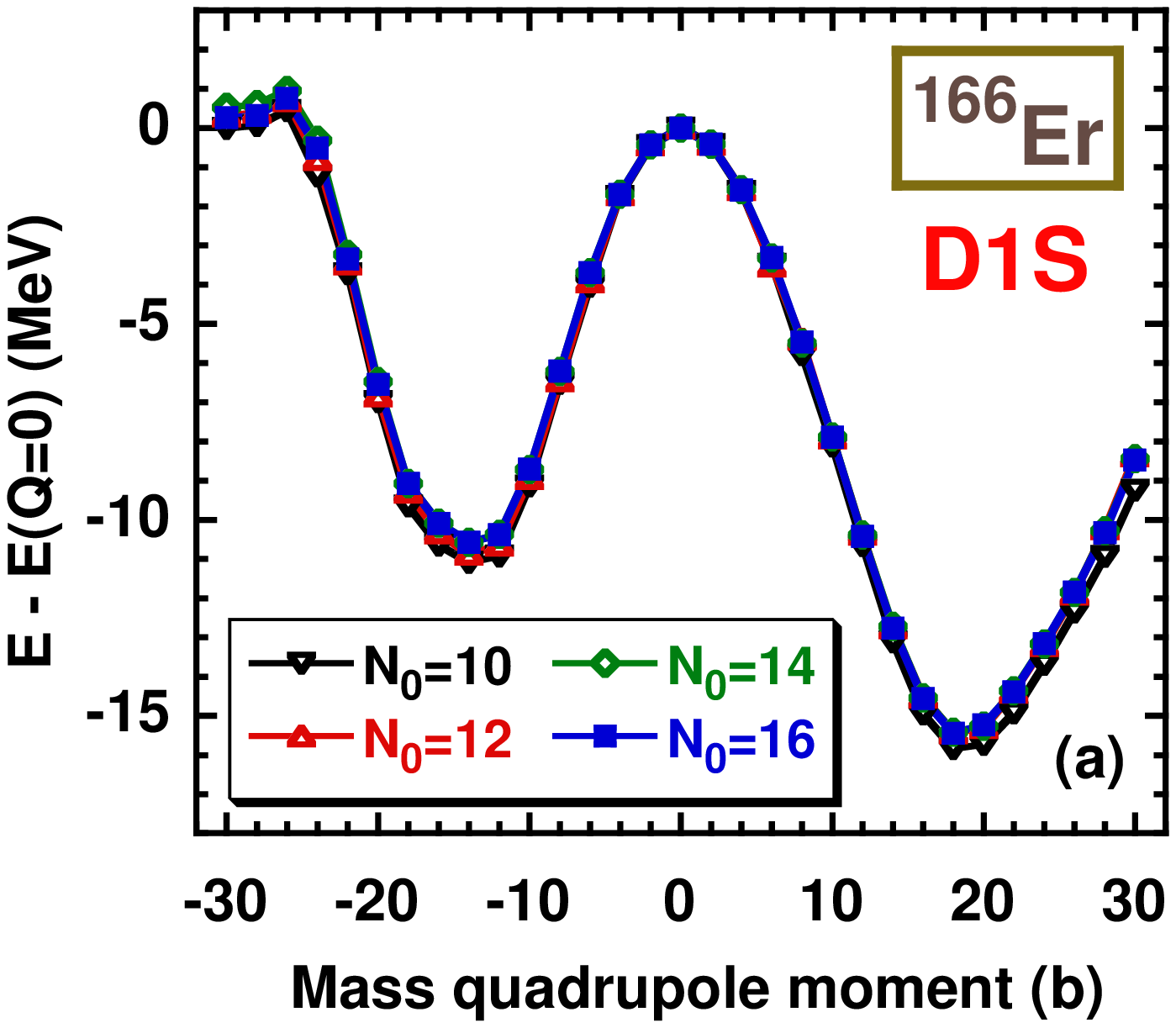} &
\includegraphics[height=0.275\textwidth,angle=0]{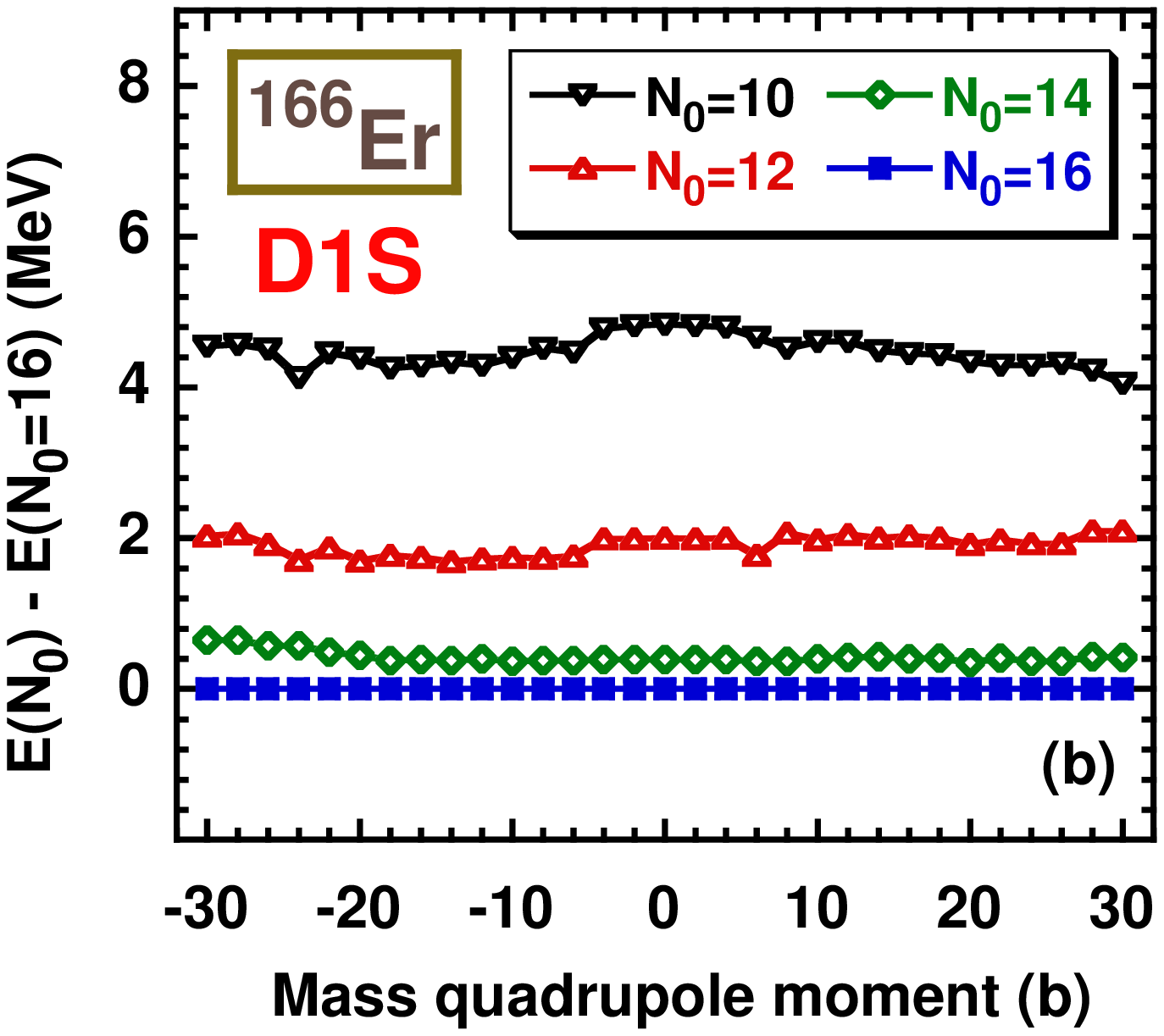} &
\includegraphics[height=0.275\textwidth,angle=0]{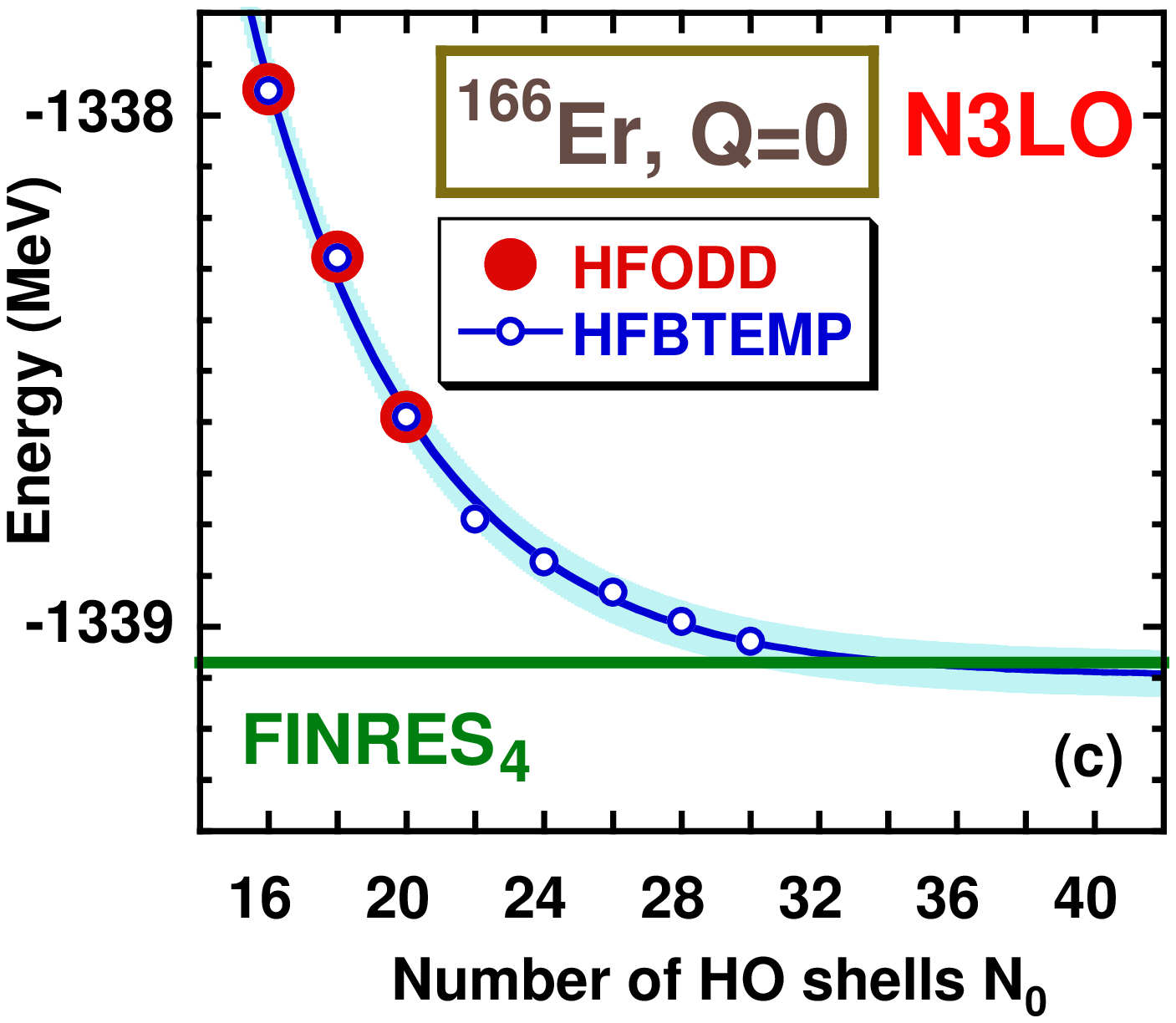}
\end{tabular}
\end{center}
\caption{The relative, panels (a) and (b), and absolute, panel (c),
binding energies of the spherical point in $^{166}$Er, presented in
function of the number $N_0$ of harmonic-oscillator shells used in
codes {\sc hfodd} and {\sc hfbtemp}. In panel (c), the horizontal
line represents the asymptotic value determined using code {\sc
finres}$_4$. \label{fig:extr}}
\end{figure}

In Fig.~\ref{fig:extr}, we present typical convergence pattern that
supports the main assumption leading to estimate
(\ref{eq:deformation-energy}). Fig.~\ref{fig:extr}(a) shows
deformation energies $\Delta E$ of $^{166}$Er calculated for the
numbers of shells between $N_0=10$ and 16 using the D1S
functional~\protect\cite{(Ber91d)}. It is clear that in the scale of
the figure, differences between the four curves are hardly
discernible. In a magnified scale, in Fig.~\ref{fig:extr}(b) we show
total energies $E_\mathrm{def}(N_0)-E_\mathrm{def}(16)$ relative to
that obtained for $N_0=16$. We see that at $N_0=10$ and 12, the
relative energies are fairly flat; however, they also exhibit
significant changes, including sudden jumps related to individual
orbitals entering and leaving the space of harmonic-oscillator wave
functions that are included in the basis with changing deformation.
Nevertheless, already at $N_0=14$, the relative energy becomes very
smooth and almost perfectly flat. This behaviour gives strong support
to applying estimate (\ref{eq:deformation-energy}) already at
$N_0=16$. Such a method was indeed used to present all total binding
energies of deformed nuclei discussed in this article.

Finally, in Fig.~\ref{fig:extr}(c), in the absolute energy
scale we show total binding energies in $^{166}$Er obtained using
codes {\sc hfodd} (up to $N_0=20$, large full circles) and {\sc
hfbtemp} (up to $N_0=30$, small empty circles). Calculations were
constrained to the spherical point and thus the results are directly comparable
with the value of $E_\mathrm{sph}=-1339.069768$ obtained using the
spherical code {\sc finres}$_4$ (horizontal line). These results
constitute a very strong benchmark of our implementations of the
N$^3$LO pseudopotentials in three very different codes. For
$N_0\leq20$, differences between the {\sc hfodd} and {\sc hfbtemp}
total energies (not visible in the scale of the figure) do not
exceed 3\,keV. By fitting an exponential curve to the {\sc hfbtemp}
results (thin line) we obtained the extrapolated asymptotic value of
energy $E_\mathrm{sph}(N_0=\infty)=-1339.097(34)$, which within the
extrapolation error of 34\,keV (shown in the figure by the shaded
band) perfectly agrees with the {\sc finres}$_4$ value.

\vspace*{1cm}

We thank Ph. Da Costa, M. Bender and J. Meyer for valuable discussions during
the development of this project.
This work was partially supported by the STFC Grants No.~ST/M006433/1
and No.~ST/P003885/1, and by the Polish National Science Centre under
Contract No.~2018/31/B/ST2/02220.
We acknowledge the CSC-IT Center for Science Ltd. (Finland) and the
IN2P3 Computing Center (CNRS, Lyon-Villeurbanne, France) for the
allocation of computational resources.

\section*{References}

\bibliographystyle{iopart-num}


\providecommand{\newblock}{}

\end{document}